\documentclass[11pt,epsf]{article}
\usepackage{hyperref}
\usepackage{graphics}
\usepackage{amssymb}
\usepackage{epsf}
\usepackage{epsfig}
\usepackage{rotating}
\usepackage{axodraw}
\usepackage{pstricks}

\newcommand{\Li}[2]{{\mbox{Li}}_{#1}\left(#2\right)}

\newcommand{\veps}{\varepsilon}

\newcommand{\nn}{\nonumber}

\newcommand{\be}{\begin{equation}}
\newcommand{\ee}{\end{equation}}
\newcommand{\bea}{\begin{eqnarray}}
\newcommand{\eea}{\end{eqnarray}}
\newcommand{\ba}{\begin{eqnarray*}}
\newcommand{\ea}{\end{eqnarray*}}
\newcommand{\lbl}[1]{\label{eq:#1}}

\newcommand{\eps}{\epsilon}
\newcommand{\epo}{\;\:.}

\def\slashed{\ds}

\def\a{\alpha}
\def\b{\beta}

\def\g{\gamma}

\def\tr{\mathop{\rm Tr}}
\def\ds#1{#1\kern-1ex\hbox{/}}
\def\dsh{h\kern-1.2ex /}

\def\nn{\nonumber}
\def\beq{\begin{equation}}
\def\eeq{\end{equation}}

\def\beqn{\begin{eqnarray}}
\def\eeqn{\end{eqnarray}}
\def\ba{\begin{eqnarray}}
\def\ea{\end{eqnarray}}

\newcommand{\beqa}{\begin{eqnarray}}
\newcommand{\eeqa}{\end{eqnarray}}
\newcommand{\la}{\lambda}

\newcommand{\ksls}{\not \! k}

\newcommand{\qsls}{\not \! q}
\newcommand{\pslsh}{\not \! p}

\newcommand{\dd}{\!\cdot\!}

\begin{document}
\begin{center}
\vspace{1.cm}
{\bf \large Axions and Anomaly-Mediated Interactions: \\
The Green-Schwarz and Wess-Zumino Vertices at Higher Orders\\ and  ${g-2}$ of the muon }

\vspace{1.5cm}
{\bf $^{a}$Roberta Armillis $^{a,b}$Claudio Corian\`{o}  $^{a}$Marco Guzzi and $^a$Simone Morelli}

\vspace{1cm}

{\it $^a$Dipartimento di Fisica, Universit\`{a} del Salento \\
and  INFN Sezione di Lecce,  Via Arnesano 73100 Lecce, Italy}\\
\vspace{.5cm}
{\it $^b$ Department of Physics and Institute of Plasma Physics \\
University of Crete, 71003 Heraklion, Greece}\\
\vspace{.5cm}

\begin{abstract}
We present a study of the mechanism of anomaly cancellation using only transverse invariant amplitudes on anomaly diagrams  at higher perturbative orders. The method is the realization of the Green-Schwarz (GS) mechanism at field theory level, which restores the Ward identities by a subtraction of the anomaly pole. Some of the properties of the GS vertex are analyzed both in the context of unitarity and of the organization of the related perturbative expansion.
We investigate the role played by the GS and the Wess-Zumino vertices in the anomalous magnetic moment of the muon and in the hyperfine splitting of muonium, which are processes that can be accompanied by the exchange of a virtual anomalous extra Z prime and an axion-like particle.
\end{abstract}
\end{center}
\newpage
\section{Introduction}
Understanding the role played by the Green-Schwarz (GS) and Wess-Zumino (WZ) mechanisms in quantum field theory is important in order to grasp the implications of the chiral gauge anomaly at the level of model building, especially in the search of extra trilinear gauge interactions at future colliders. In recent years several proposals coming either from string theory or from theories with extra dimensions have introduced new perspectives in regard to the various mechanisms of cancellation of the gauge anomalies in effective low energy lagrangeans, which require further investigation in order to be fully understood. These effective models are characterized by the presence of higher dimensional operators
and interactions of axion-like particles. In a class of vacua of string theory this enterprise has some justification, for instance in orientifold models (see \cite{Kiritsis:2003mc, Antoniadis:2000ena, Ibanez:2001nd}), where deviations from the Standard Model may appear in the form of higher dimensional corrections which are not heavily suppressed and which could be accessible at the LHC.

In anomaly-free realizations of chiral gauge theories the trilinear anomalous gauge interactions vanish (identically) in the chiral limit, by a suitable distribution of charges among the fermions of each generation (or inter-generational), showing that residual interactions are proportional to the mass differences of the various fermions. In the GS realization this request is far more relaxed and the mechanism requires only the cancellation, in the presence of anomalous contributions, of the longitudinal component of the anomaly vertex rather than that of the entire triangle diagram. In the WZ case, the cancellation of the anomaly takes place at lagrangean level, rather than at the vertex level, and requires an axion as an asymptotic state, which is a generalization of the Peccei-Quinn (PQ) interaction.

The effective field theory of the WZ mechanism has been analyzed in \cite{Coriano:2005js, Coriano:2007fw, Coriano:2007xg, Coriano:2006xh, Coriano:2008pg, Armillis:2007tb}, together with  its supersymmetric extensions \cite{Anastasopoulos:2008jt} while a string derivation of the GS constructions has been outlined in \cite{Anastasopoulos:2006cz}. Pseudoscalar fields (axion-like particles) - with a mass and a coupling to gauge fields which are left unrelated - have been the subject of several investigations and proposals for their detection either in ground-based experiments \cite{Jaeckel:2006xm} or to explain some puzzling results on gamma ray propagation \cite{DeAngelis:2008sk, DeAngelis:2007dy},
while new solutions of the strong CP problem in more general scenarios have also received attention \cite{Berezhiani:2000gh}. At the same time the search for extra $Z^\prime$  at the LHC
from string models and extra dimensions, together with precision studies on the  resonance to uncover new effects,
has also received a new strength (\cite{Adam:2008ge, Lee:2007qx, Langacker:2008yv, Fuks:2008yp, Feldman:2006wb, Fuks:2007gk}).

If the GS and the WZ mechanisms are bound to play any role at future experiments (see for instance \cite{Kumar:2007zza}) remains to be seen, given the very small numerical impact of the anomaly corrections in the cleanest processes that can be studied, for instance, at the LHC; nevertheless more analysis is needed in order to understand the theoretical implications of ``anomaly mediation'' and of its various realizations, in the form of GS and WZ interactions, in effective models.

Both mechanisms are quite tricky, since they show some unusual features which are not common to the rest of anomaly-free field theories and it is not hard to find in the literature several issues which have been debated for a long time, concerning the consistency of these approaches \cite{Andrianov:1989by, Andrianov:1993qy, Fosco:1993qx}.

For instance, in the GS case, one of them concerns unitarity, due to the claimed presence of extra ``double poles''  \cite{Adam:1997gj} in a certain class of interactions which would render completely invalid a perturbative prescription; another one concerns the physical interpretation of the longitudinal subtraction, realized within the same mechanism, which is usually interpreted as due to the exchange of an axion, which, however, as pointed out in \cite{Coriano:2008pg} is not an asymptotic state.

The first part of our investigation is a study of the organization of the perturbative expansion for anomalous theories in the presence of counterterms containing double poles in virtual corrections and, in principle, in $s/t/u$ channel exchanges. Our point of view and conclusions are in contradiction with those of
\cite{Adam:1997gj}, formulated within axial QED, where the analysis of the anomaly pole counterterm was not taking into account the fact that the subtracted term is an intrinsic part of the triangle (anomaly) diagram, corresponding to one of its invariant amplitudes, in a specific formulation. We will come to a rather detailed discussion of these subtle points.

The picture that emerges from our analysis is that of consistency -rather than of inconsistency- of the GS mechanism at the level of effective field theory. In other words, it should be possible to subtract the longitudinal pole of the anomaly diagram with no further consequences at perturbative level.
The structure of the perturbative expansion in the presence of explicit GS counterterms is worked out in two sections and in an appendix, where we detail the methods for the computations of graphs containing extra poles in the propagators and compare the general features of this expansion to an ordinary expansion.

In any case, in the absence of a direct check of the unitarity equations -which is hard to perform given the rather large order at which these anomalous corrections appear- the problems in perturbation theory can potentially appear in the form of double poles in some ({\em external} ) propagators. A re-examination of several diagrams brings us to conclude that this situation is avoided.

Coming to a direct phenomenological application, we investigate the role of these vertices in the study of the anomalous magnetic moment of the muon. We stress that if the physical mechanism introduced for the cancellation of the anomaly is of WZ type, then a physical axion appears in the spectrum. This is the case if the anomalous extra $Z^{\prime}$ receives its mass both by the Higgs and the St\"uckelberg mechanisms. Both for the GS and the WZ case we outline the role of the anomalous extra $Z^{\prime}$ and of the pseudoscalar exchange up to 2-loop level. A previous analysis of the leading contribution to $g-2$ for intersecting brane models can be found in \cite{Kiritsis:2002aj}.

More recently, the GS vertex has been used in the study of the coupling of the Kaluza-Klein (KK) \cite{Kumar:2007zza, Djouadi:2007eg} excitations of gauge bosons to fermions, where it has been pointed out the possibility to detect these coupling at the LHC, for instance in $t\bar{t}$ production. We find that several of these results are based on a still unsatisfactory understanding of the GS mechanism at theoretical level, and our work is an attempt to clarify some of these points. From our analysis will emerge the correct structure of the broken Ward identities for the GS vertex, which are specific of a non-local theory.
These points will be carefully analyzed in the final section of this work.

\section{The GS and WZ vertices}
The field theory version of the GS mechanism, deprived of all its stringy features, appears in an attempt to cancel the anomaly by introducing a specific non-local counterterm added to the anomalous theory. This attempt had been the cause of serious debates which have questioned the consistency of the approach. The mechanism uses a ghost-like particle, which in string theory is generically identified as an axion - although there it does not appear as an asymptotic state - to restore the broken Ward identities due to the anomaly.
A paradigm for the GS mechanism in field theory is an anomalous version of axial QED in 4-dimensions defined by the lagrangean
\beq
\mathcal{L}_{5\,QED}= \overline{\psi} \left( i \slashed{\partial} + e \slashed{B} \gamma_5\right)\psi - \frac{1}{4} F_B^2
\label{count0}
\eeq
plus the counterterm
\beq
\mathcal{S}_{ct}= \frac{1}{24\pi^2} \langle \partial B(x) \square^{-1}(x-y) F(y)\wedge F(y)
\rangle.
\label{count}
\eeq
Federbush \cite{Federbush:1996cp} proposed to reformulate this  lagrangean in terms of one axion and one ghost-like particle interacting via a Wess-Zumino (WZ) counterterm (see the discussion in \cite{Coriano:2008pg}). An equivalent formulation of the same subtraction counterterm is given in \cite{Andrianov:1989by}, where a transversality constraint ($\partial B=0$) is directly imposed on the lagrangean
via a multiplier. Eq.~(\ref{count}) can be obtained by performing the functional integral over $a$ and $b$ of the following action

\beqa
\mathcal{L} &=& \overline{\psi} \left( i \not{\partial} + e \not{B} \gamma_5\right)\psi - \frac{1}{4} F_B^2 +
\frac{ e^3}{48 \pi^2 M} F_B\wedge F_B ( a + b) \nonumber \\
&& + \frac{1}{2}  \left( \partial_\mu b - M B_\mu\right)^2 -
\frac{1}{2} \left( \partial_\mu a - M B_\mu\right)^2.
\label{fedeq}
\eeqa
The integral on $a$ and $b$ are gaussians and one recovers the non-local contribution in (\ref{count}) after partial integration.
Notice that $b$ has a positive kinetic term and $a$ is ghost-like. Both $a$ and $b$ shift by the same amount under a gauge transformation of $B$
\beq
a\rightarrow a + M \theta,\,\,\,\,\, b\rightarrow b + M \theta
\eeq
where $\theta$ is the gauge parameter.
This second (local) formulation of the pole counterterm contained in (\ref{count}) shows the connection between this action and the WZ mechanism \cite{Coriano:2008pg}. Both actions, in fact, share some similarities, but describe different theories. In particular, the WZ action is obtained by removing the ghost term ($b$) and keeping only the
axion. This second theory is characterized by a unitarity bound \cite{Coriano:2008pg}.  The bound is due to the fact that in effective models containing Wess-Zumino interactions, gauge invariance of the effective action requires a cancellation  between different trilinear vertices: the anomalous vertices and the axion counterterm $\phi F\wedge F$, while for Green-Schwarz vertices the subtraction of the longitudinal component of the anomaly is sufficient to make the effective vertex  gauge-invariant to all orders. In both cases the physical amplitudes are gauge-independent.

In the WZ case the proof of gauge
  independence is rather involved and has been discussed before \cite{Coriano:2007fw}. In the GS case, instead, this is trivial since the vertex is gauge-invariant by construction. Notice that a local counterterm in the form of a Peccei-Quinn term is not sufficient to remove the power-like growth with energy of  a class of amplitude (BIM amplitudes, Bouchiat, Iliopoulos and Meyer) \cite{Coriano:2008pg} that are characterized by anomalous production and anomalous decay of massless gauge bosons in the initial and final states, mediated by the exchange of an anomalous $Z^{\prime}$ in the s-channel. These amplitudes are quite interesting since they evade the Landau-Yang theorem, triggering a $Z\gamma\gamma $ vertex. The phenomenological implications of these amplitudes are discussed in a companion work.
\begin{figure}[t]
\begin{center}
\includegraphics[scale=0.8]{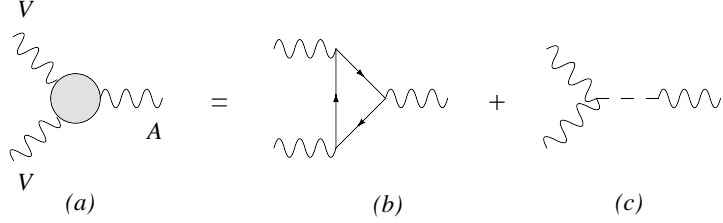}
\caption{\small A  gauge invariant GS vertex of the AVV type, composed of an AVV triangle  and a single counterterm of Dolgov-Zakharov form. Each term is denoted by  $\Delta_{AVV}^{\lambda \mu \nu \, GS}$ (a), $\Delta_{AVV}^{\lambda \mu \nu}$ (b) and $C^{\lambda \mu \nu}_{AVV}$ (c).}
\label{GS_AVV}
\end{center}
\end{figure}
\begin{figure}[t]
\begin{center}
\includegraphics[scale=0.8]{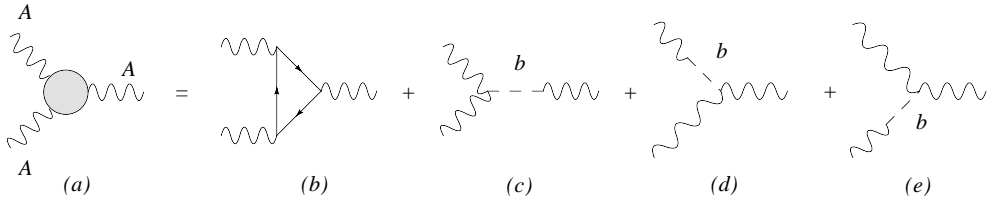}
\caption{\small All the contributions to the GS gauge invariant vertex, for an AAA triangle. The single terms are denoted by $\Delta_{AAA}^{\lambda \mu \nu \, GS}$ (a), $\Delta_{AAA}^{\lambda \mu \nu}$ (b), $C^{\lambda \mu \nu}_{AVV}$ (c), $C^{\mu \nu \lambda}_{AVV}$ (d) and $C^{\nu \mu \lambda}_{AVV}$ (e).}
\label{GS_AAA}
\end{center}
\end{figure}

\subsection{The GS vertex in the $AAA$ and $AVV$ cases}

In our analysis, we denote with $T_{\mu\nu\lambda}$ the 3-point function in momentum space, obtained from the lagrangeans (\ref{count0}) and (\ref{count}). In the case of three axial-vector currents we define the correlator
\begin{eqnarray}
(2\pi)^4\delta^{(4)}(k_1+k_2-k)\Delta_{AAA}^{\lambda\mu\nu}(k,k_1,k_2)
=\int d^4x_1 d^4x_2 d^4x_3e^{i(k_1 x_1 + k_2 x_2 - k x_3)}
\langle J_{\mu}^5(x_1)J_{\nu}^5(x_2)J_{\lambda}^5(x_3)\rangle.
\end{eqnarray}
and a symmetric distribution of the anomaly for the $AAA$ vertex
\footnote{We have used the following notation
$a_n=-\frac{i}{2\pi^2}$ and
$\varepsilon[\mu,\nu,k_1,k_2]=\varepsilon^{\mu\nu\alpha\beta}k_{1\alpha}k_{2\beta}$}
\begin{eqnarray}
\label{anomaly}
&&k_{\lambda} \Delta_{AAA}^{\lambda\mu\nu}(k,k_1,k_2)=
\frac{a_n}{3}\varepsilon[\mu, \nu, k_1, k_2]
\nonumber\\
&&k_{1\mu} \Delta_{AAA}^{\lambda\mu\nu}(k,k_1,k_2)=
\frac{a_n}{3}\varepsilon[\lambda, \nu, k, k_2]
\nonumber\\
&&k_{2\nu} \Delta_{AAA}^{\lambda\mu\nu}(k,k_1,k_2)=
\frac{a_n}{3}\varepsilon[\lambda, \mu, k, k_1].
\end{eqnarray}
In the $AVV$ case, a second vector-like gauge interaction ($A_\mu$) is introduced in Eq.(\ref{fedeq}) for more generality and we have
\begin{eqnarray}
(2\pi)^4\delta^{(4)}(k_1+k_2-k)\Delta_{AVV}^{\lambda\mu\nu}(k,k_1,k_2)
=\int d^4x_1 d^4x_2 d^4x_3e^{i(k_1 x_1 + k_2 x_2 - k x_3)}
\langle J_{\mu}(x_1)J_{\nu}(x_2)J_{\lambda}^5(x_3)\rangle,
\end{eqnarray}
where the anomaly equations are
\begin{eqnarray}
&&k_{\lambda} \Delta_{AVV}^{\lambda\mu\nu}(k,k_1,k_2)=
a_n\varepsilon^{\mu\nu\alpha\beta}k_{1\alpha}k_{2\beta}
\nonumber\\
&&k_{1\mu} \Delta_{AVV}^{\lambda\mu\nu}(k,k_1,k_2)=0
\nonumber\\
&&k_{2\nu} \Delta_{AVV}^{\lambda\mu\nu}(k,k_1,k_2)=0.
\end{eqnarray}
Below we will consider both the $AVV$ and $AAA$ cases.
The GS counterterm that corresponds to the exchange of the massless pole of Eq.~(\ref{count}) takes the following form in momentum space in the $AVV$ case
\bea
C^{\lambda \mu \nu}_{AVV}(k,k_1,k_2) =
C^{\mu \nu}(k_1,k_2) k^\lambda =
-\frac{a_n}{k^2} k^\lambda \epsilon[\mu,\nu,k_1,k_2].
\eea

Similarly, a GS counterterm in the $AAA$ case, with incoming momentum $k$  and outgoing momenta $k_1, k_2$,  is defined as
\beqa
C^{\lambda \mu \nu}_{AAA}(k,k_1,k_2) &=&
\frac{1}{3} \left( C^{\lambda \mu \nu}_{AVV}(k,k_1,k_2)
+ C^{\mu \nu \lambda}_{AVV}(-k_1,k_2,-k)
+ C^{\nu \lambda \mu}_{AVV}(-k_2, -k,k_1)\right)
\nonumber\\
&=&  \frac{1}{3} \Bigg( C^{\mu \nu}(k_1,k_2)k^\lambda
- C^{\nu \lambda}(k_2,-k)k_1^\mu -  C^{\lambda \mu}(-k,k_1)k_2^\nu   \Bigg) \nonumber\\
&=&- \frac{1}{3} \left(
\frac{a_n}{k^2} k^\lambda \epsilon[\mu, \nu, k_1, k_2]
+ \frac{a_n}{k_1^2} k_1^\mu \epsilon[\lambda, \nu, k, k_2]
+ \frac{a_n}{k_2^2} k_2^\nu \epsilon[\lambda, \mu, k, k_1]\right),
\eeqa

and corresponds to the Dolgov-Zakharov form (DZ) of the anomaly diagram \cite{Dolgov:1971ri} (modulo a minus sign).
The re-defined vertex shown in Fig.~\ref{GS_AVV} is written as
\begin{eqnarray}
\label{cance}
\Delta_{AVV}^{\lambda\mu\nu \, GS}(k,k_1,k_2)=
\Delta_{AVV}^{\lambda\mu\nu}(k,k_1,k_2)+C_{AVV}^{\lambda\mu\nu}(k,k_1,k_2)
\end{eqnarray}
and we obtain a similar expression for the $AAA$ vertex (Fig.~\ref{GS_AAA})
just by replacing $AVV$ with $AAA$ in Eq.(\ref{cance}) and taking into account the different form of the counterterms.
These gauge invariant vertices trivially satisfy the Ward identities
\begin{eqnarray}
\label{WI}
k_{\lambda}\Delta^{\lambda\mu\nu\,  GS}(k,k_1,k_2)
=k_{1\mu} \Delta^{\lambda\mu\nu\,  GS}(k,k_1,k_2)
=k_{2\nu} \Delta^{\lambda\mu\nu\,  GS}(k,k_1,k_2)
=0,
\end{eqnarray}
where again $\Delta^{\lambda\mu\nu\,  GS}$  refers either to an $AVV$ or to an $AAA$ correlator.

\subsection{Implications of the GS vertex: vanishing of (real) light-by-light  scattering at 2-loop}
\begin{figure}[t]
\begin{center}
\includegraphics[scale=0.9]{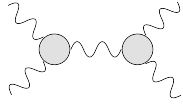}
\caption{\small Amplitude with two full GS vertices and the exchange of an axial-vector current in the s-channel. For on-shell external lines the contributions from the extra poles disappear.}
\label{GSGS}
\end{center}
\end{figure}
To illustrate some of the properties of the GS vertex and its implications, we consider a 2-loop process in which we have two massless vector bosons in the initial and in the final state with the exchange in the s-channel of an axial-vector current (Fig.~\ref{GSGS}). The example that we provide here comes from anomalous axial QED,  but it can be extended to more realistic models with no major variants. In Fig.~\ref{GSGS} both the initial and the final state contain anomalous subdiagrams, but the two GS vertices are defined in such a way to absorb all the longitudinal subtractions terms inside each of the blobs.
The amplitude of such a process is given by
\ba
\label{GS_BIMeq}
{\mathcal M}^{\, \mu \nu \mu^\prime \nu^\prime}=
\Delta_{AVV}^{\lambda \mu \nu \, GS}(- k, - k_1, - k_2) \left(- \frac{i g_{\lambda \lambda^\prime}}{k^2} \right)
\Delta_{AVV}^{\lambda^\prime \mu^\prime \nu^\prime \,  GS}(k,k^\prime_1,k^\prime_2).
\ea
\begin{figure}[t]
\begin{center}
\includegraphics[scale=0.6]{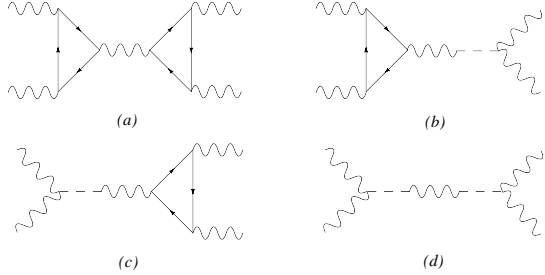}
\caption{\small All the contributions from the GS gauge invariant vertex, in the AVV case, to the amplitude VV$ \rightarrow$ VV via an axial-vector current. }
\label{GS_BIM}
\end{center}
\end{figure}
In the expression above, the propagator is deprived of its longitudinal momentum dependence due a Ward identity. The amplitude in Eq.~(\ref{GS_BIMeq}) can be decomposed into the four sub-amplitudes shown in Fig.~\ref{GS_BIM} after expanding the two GS vertices with Eq.~(\ref{cance})
\bea
{\mathcal M}^{\, \mu \nu \mu^\prime \nu^\prime} =
- \left( \Delta_{AVV}^{\lambda \mu \nu}(- k,- k_1,- k_2)-
k^\la C^{\mu \nu}(-k_1, -k_2)\right) \frac{i g_{\lambda \lambda^\prime}}{k^2}
\left( \Delta_{AVV}^{\lambda^\prime \mu^\prime \nu^\prime}(k, k^\prime_1,k^\prime_2)+
C^{\la^\prime \mu^\prime \nu^\prime}(k^\prime_1, k^\prime_2)\right),
\eea
but only two sub-amplitudes survive (Fig.~\ref{GS_BIM}a and \ref{GS_BIM}b)) because of the Ward identities in Eq.~(\ref{WI}).
\begin{figure}[t]
\begin{center}
\includegraphics[scale=0.7]{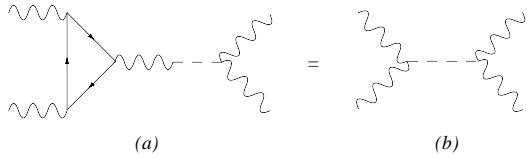}
\caption{\small The sub-amplitude in Fig.~\ref{GS_BIM}b after the contraction $k^\la \Delta_{AVV}^{\lambda \mu \nu}$ which gives the anomaly equation.}
\label{GS_BIM_WI}
\end{center}
\end{figure}
We are left with two contributions which cancel, for on-shell matrix elements. In fact, while off-shell the graph in Fig.~\ref{GS_BIM_WI} spoils unitarity, when instead the four external lines are on-shell  the triangle contribution (the first term) reduces to the DZ form and the cancellation between the two terms is identical. In view of the structure of the anomaly vertex and of the GS vertex given before, this cancellation implies that the anomaly diagram, for on-shell (axial-vector) photons and in the chiral limit, is purely longitudinal (DZ form). A similar result holds also for the AAA case.

It is then natural to look at more general situations when these types of diagrams appear into higher order contributions. In these more general  cases, the anomaly diagram does  not coincide with its DZ form, except for specific kinematical points ($k_1^2=k_2^2=k^2$, off-shell), and there is no identical cancellation of the anomalous trilinear gauge interactions.  We conclude that compared to the identical cancellation of the anomaly by charge assignment on each generation, which eliminates all-together all the trilinear gauge interactions, the GS vertex can be either transversal or vanishing (in the chiral limit) for each given flavour.

As we have previously mentioned, an anomaly vertex with the addition of the pole counterterms
(i.e. deprived of the anomaly pole) has been criticized in previous works in \cite{Adam:1997gj}. There the author brings in as an example a class of amplitudes which are affected by double poles, claiming a unitarity failure of the model. We will argue against this interpretation.
\begin{figure}[t]
\begin{center}
\includegraphics[scale=0.9]{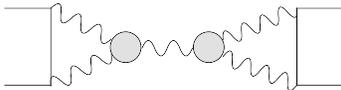}
\caption{\small The embedding of the BIM amplitude with GS vertices in a fermion/antifermion scattering.}
\label{ninefig}
\end{center}
\end{figure}
\begin{figure}[t]
\begin{center}
\includegraphics[scale=0.8]{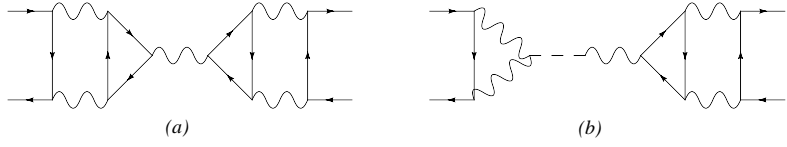}
\caption{\small Contributions to the fermion-antifermion scattering with the GS mechanism. }
\label{antifermion_fermion_GS}
\end{center}
\end{figure}
\subsection{Embedding the GS vertex into higher order diagrams}
When we embed the amplitude into higher order diagrams (see Fig.~\ref{ninefig}), and consider an on-shell fermion-antifermion scattering, according to \cite{Adam:1997gj}, we are forced to move away from a symmetric configurations of the loop momenta and the identical vanishing of the anomaly is not ensured any longer, for the reasons that we have
just  raised above. In particular, according to \cite{Adam:1997gj}, the s-channel exchange is affected
by a double pole.

There is no better way to check the correctness of these conclusions than going through an explicit computation of this amplitude, using some results of the recent literature on radiative corrections.

Expanding the two GS vertices, which are clearly non-zero in this case, we end up with several contributions,  such as a graph with two triangle diagrams, and a specific set of counterterms.
The two contributions involved in the study of fermion-antifermion scattering are shown
in Fig.~\ref{antifermion_fermion_GS}. Notice the presence of the box-triangle diagram ($\mathcal{BT}$) in the first graph, which remains non-trivial to compute even in the chiral limit. The second graph is the only contribution which does not disappear, according to \cite{Adam:1997gj}, and therefore describes a spurious s-channel exchange characterized by a double pole.

The conclusions of \cite{Adam:1997gj} are erroneous for two reasons:
1) in this specific case the double pole cancels in the explicit computation, so it not a good counter example;
2) the cancellation or the presence of double poles should be analyzed together with the anomalous vertex and not separately. This second point will be addressed in the next sections.

To prove point 1) we need the 2-loop structure of the $\mathcal{BT}$ graph, which
is just proportional to its tree-level axial-vector form times a form factor ${\bf G}(s)$,
function of $s=(p_1 + p_2)^2$,
\ba
\bar{v}(p_2) \gamma^\lambda \gamma^5  u(p_1) {\bf G}(s).
\ea
Explicit expressions of the ${\bf G}(s)$ coefficient in the massless case are obtained from \cite{Bernreuther:2005rw} and are given by
\ba
&& {\mathcal Re}{\bf G}(s,m_i=0,m_q=0) =3\log\left(\frac{s}{\mu^2}\right)-9+2\zeta(2)
\nonumber\\
&& {\mathcal Im}{\bf G}(s,m_i=0,m_q=0) =-3,
\ea
where $m_i$ is the mass of the internal fermion with flavor $i$ circulating in the triangle diagram
while $m_q$ is the mass of the fermionic external leg with flavor $q$.

We have for the two sub-amplitudes in Fig.~\ref{antifermion_fermion_GS}
\bea
&&{\mathcal M}_a= - {\bf G}^2(s)\bar{v}(p_2) \gamma^\lambda \gamma^5u(p_1) \frac{i}{k^2} \bar{u}(p^\prime_1)
\gamma_\lambda \gamma^5 v(p^\prime_2)
\nonumber\\
&&{\mathcal M}_b= i \, \int \frac{d^4 k_1}{(2 \pi)^4} \Bigg[ \bar{v}(p_2)
\gamma^\nu  \frac{ \ds p_1 -\ds k_1} {(p_1 -k_1)^2} \gamma^\mu u(p_1)
\frac{1}{k^2_1}  \frac{1}{k^2_2} a_n \epsilon[\mu, \nu, k_1, k_2]  \Bigg]
\frac{1}{(k^2)^2} \bar{u}(p^\prime_1) \ds k \gamma^5 {\bf G}(s) v(p^\prime_2)=0\,,
\nonumber\\
\eea
where ${\mathcal M}_b$ is identically zero, because of the equations of motion satisfied by the external fermion lines ($k = p^\prime_1 + p^\prime_2$).

\begin{figure}[t]
\begin{center}
\includegraphics[scale=0.7]{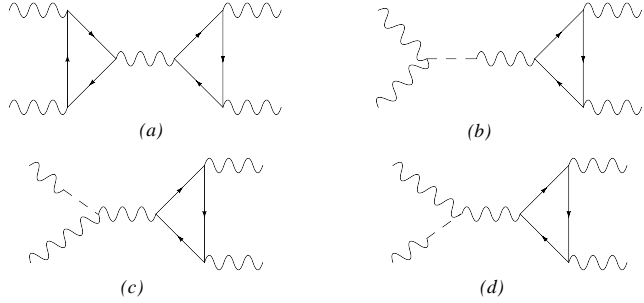}
\caption{\small All the contributions in the symmetric GS vertex. }
\label{four_terms}
\end{center}
\end{figure}

We can easily generalize our analysis to an $AAA$ case, for this
we have to consider Eqs. (\ref{count0},\ref{count}). When an $AAA$ triangle is embedded in the 2-loop fermion-antifermion scattering process
we can formally write the amplitude as follows
\ba
{\mathcal S}=\int\frac{d^4 k_1}{(2\pi)^4}\frac{d^4 k^{\prime}_1}{(2\pi)^4}
\bar{v}(p_2)\gamma^{\nu}\frac{1}{\ds p_1-\ds k_1}\gamma^{\mu}u(p_1)
\frac{1}{k_1^2}\frac{1}{k_2^2}{\mathcal S}^{\, \mu \nu \mu^\prime \nu^\prime}
\bar{u}(p^{\prime}_1)\gamma^{\mu^{\prime}}\frac{1}{\ds k_1^{\prime}-\ds p_1^{\prime}}
\gamma^{\nu^{\prime}}v(p^{\prime}_2)\frac{1}{{k^{\prime}_1}^2}\frac{1}{{k^{\prime}_2}^2}
\ea
where the tensor sub-amplitude is defined as
\bea
{\mathcal S}^{\, \mu \nu \mu^\prime \nu^\prime}&=&
- \, \Delta_{AAA}^{\lambda \mu \nu \, GS}(-k,-k_1,-k_2)
\frac{i g^{\lambda \lambda^\prime}}{k^2}
\Delta_{AAA}^{\lambda^\prime \mu^\prime \nu^\prime \,  GS}(k,k^\prime_1,k^\prime_2)
\nonumber\\
&=&
- [ \Delta_{AAA}^{\lambda \mu \nu}(-k,-k_1,-k_2)
+ C^{\lambda \mu \nu}_{AAA}(-k,-k_1,-k_2) ]   \frac{i}{k^2}
  [ \Delta_{AAA}^{\lambda \mu^\prime \nu^\prime}(k,k^\prime_1,k^\prime_2)
+ C^{\lambda \mu^\prime \nu^\prime}_{AAA}(k,k^\prime_1,k^\prime_2)].
\nonumber\\
\label{S_AAA}
\eea
Using the Ward identity $k^{\lambda} \Delta_{AAA}^{\lambda \mu \nu \, GS}(-k,-k_1,-k_2) = 0$,
we can drop the GS counterterm $C^{\lambda \mu^\prime \nu^\prime}_{AAA}(k,k^\prime_1,k^\prime_2)$.
In this way the sixteen terms which contribute to the sub-amplitude
${\mathcal S}^{\, \mu \nu \mu^\prime \nu^\prime}$ given in
Eq. (\ref{S_AAA}) reduce to the four terms shown in Fig.~\ref{four_terms}.
Therefore the total femion-antifermion scattering amplitude is given by the sum of four terms
${\mathcal S}_a$, ${\mathcal S}_b$, ${\mathcal S}_c$ and ${\mathcal S}_d$ which are defined below
\begin{figure}[t]
\begin{center}
\includegraphics[scale=0.5]{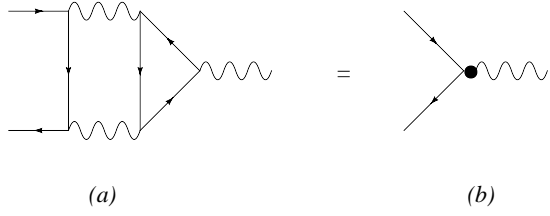}
\caption{\small The axial-vector form factor for the $\mathcal{BT}$ diagram. }
\label{remiddi}
\end{center}
\end{figure}
\begin{figure}[t]
\begin{center}
\includegraphics[scale=0.8]{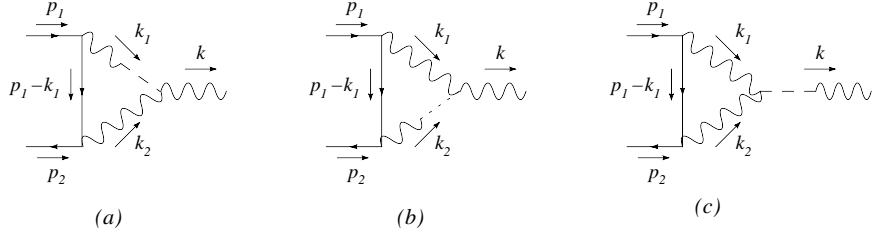}
\caption{\small The one loop GS counterterms included in the $\mathcal{BT}$ as its longitudinal part. }
\label{null2}
\end{center}
\end{figure}
\ba
&&{\mathcal S}_a= - \, \mathcal{ BT}^{ \lambda }_{AAA}  \frac{i}{k^2} \mathcal{ BT}^{ \lambda}_{AAA}
\nonumber\\
&&{\mathcal S}_b= i \, \int  \frac{d^4 k_1}{(2 \pi)^4}   \Bigg( \bar v (p_2)
\gamma_\nu \frac{1}{\ds p_1 - \ds k_1} \gamma_\mu u(p_1) \frac{1}{k_1^2}
\frac{1}{k_2^2}  \Bigg)  \frac{a_n}{3} \frac{k^\lambda}{k^2}
\epsilon[\mu, \nu, k_1, k_2]   \frac{1}{k^2}
\mathcal{ BT}^{ \lambda}_{AAA}
\nonumber\\
&&{\mathcal S}_c= i \, \int\frac{d^4 k_1}{(2 \pi)^4}
\Bigg( \bar v (p_2) \gamma_\nu \frac{1}{\ds p_1 - \ds k_1} \gamma_\mu u(p_1) \frac{1}{k_2^2}
\frac{1}{k_1^2}  \frac{a_n}{3}\frac{k^\mu_1}{k_1^2}
\epsilon[\nu, \lambda, k_2, k]\Bigg) \frac{1}{k^2}
\mathcal{ BT}^{ \lambda}_{AAA}
\nonumber\\
&&{\mathcal S}_d=i \, \int  \frac{d^4 k_1}{(2 \pi)^4} \Bigg( \bar v (p_2) \gamma_\nu \frac{1}{\ds p_1 - \ds k_1} \gamma_\mu u(p_1) \frac{1}{k_1^2}
\frac{1}{k_2^2}  \frac{a_n}{3} \frac{k^\nu_2}{k_2^2}  \epsilon[\lambda, \mu, k, k_1]   \Bigg)   \frac{1}{k^2}
\mathcal{ BT}^{ \lambda}_{AAA}\,,
\ea
where we have defined
\ba
\mathcal{ BT}^{ \lambda}_{AAA}(k,p^\prime_1,p^\prime_2)= - \int\frac{d^4 k^{\prime}_1}{(2\pi^4)}
\Delta_{AAA}^{\lambda \mu^\prime \nu^\prime}(k,k^\prime_1,k^\prime_2)
\bar{u}(p^{\prime}_1)\gamma^{\mu^{\prime}}\frac{1}{\ds k_1^{\prime}-\ds p_1^{\prime}}
\gamma^{\nu^{\prime}}v(p^{\prime}_2)\frac{1}{{k^{\prime}_1}^2}\frac{1}{{k^{\prime}_2}^2}
\ea
and the total amplitude is given by
\bea
{\mathcal S}={\mathcal S}_a +{\mathcal S}_b+{\mathcal S}_c+{\mathcal S}_d.
\eea
In the ${\mathcal S}_b$ sub-amplitude we distribute the anomaly symmetrically on each vertex and using the following Ward identities on the vector currents we obtain
\bea
k^{\lambda}  \mathcal{ BT}^{ \lambda}_{VAV}  =  k^{\lambda}  \mathcal{ BT}^{ \lambda}_{VVA} = 0.
\eea
This allows us to simplify the ${\mathcal S}_b$ expression as follows
\bea
{\mathcal S}_b = i \, \int  \frac{d^4 k_1}{(2 \pi)^4}
\Bigg( \bar v (p_2) \gamma_\nu \frac{1}{\ds p_1 - \ds k_1} \gamma_\mu u(p_1) \frac{1}{k_1^2}
\frac{1}{k_2^2}    \Bigg)  \frac{a_n}{3} \frac{1}{(k^2)^2}
\epsilon[\mu, \nu, k_1, k_2] \bar u(p^\prime_1) \ds k \gamma^5
G(s) v(p^\prime_2)=0\,,   \nonumber\\
\eea
where we have used the result shown in \cite{Bernreuther:2005rw}.

Also the third amplitude ${\mathcal S}_c$ does not contribute to ${\mathcal S}$,
in fact we have
\ba
{\mathcal S}_c = i \, \bar v (p_2) \gamma_\nu u(p_1)
\frac{a_n}{3}   \epsilon[\nu, \lambda, \rho, \sigma] k^\sigma \int  \frac{d^4 k_1}{(2 \pi)^4}   \Bigg(  \frac{k_1^\rho}{(k-k_1)^2 k_1^4 }
\Bigg) \frac{1}{k^2}\mathcal{ BT}^{ \lambda}_{AAA}=0,
\ea
which vanishes by symmetry due to the structure of the tensor integral, which is proportional to $k^{\rho}$. We refer to the
appendices for more details concerning the techniques of evaluation of this and other similar integrals with ``double propagators''.
In the same way also the fourth contribution vanishes, ${\mathcal S}_d =0$. This explicit computations contradicts the
conclusions of \cite{Adam:1997gj} where the same amplitude was conjectured to be affected by double poles in the s-channel.

There are some conclusions to be drawn. The first is that
the replacement of the two anomaly vertices in the amplitude with two GS vertices, in this case,
is irrelevant as far as the fermions are massless.
Equivalently, the longitudinal components of the two anomaly vertices decouple in the graph and the
only invariant amplitudes coming from the anomaly vertices that survive - after the integration on the second loop momentum - are the transverse ones. This is one special case in which the anomaly diagram
is transverse in the virtual corrections just by itself. In other cases this does not happen and the extra poles introduced by the counterterm are sufficient to cancel those generated by the anomaly. For this reason the presence of extra poles in a partial amplitude is not necessarily the sign of an inconsistency.
\begin{figure}[h]
\begin{center}
\includegraphics[scale=0.4]{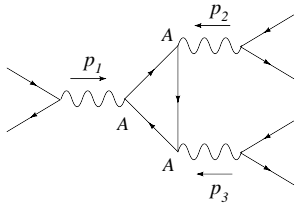}
\caption{\small Typical graph in which the longitudinal anomalous component vanishes.}
\label{fftriangle}
\end{center}
\end{figure}
\subsection{ The vertex in the longitudinal/transverse (L/T) formulation}
The analysis presented above becomes more transparent if we use a special parameterization of the anomaly diagram in which the longitudinal part of the vertex is separated from the transverse one, as done in recent studies of radiative corrections to the anomalous magnetic moment of the gluon \cite{Knecht:2003xy}. This parameterization is more convenient than the usual Rosenberg form \cite{Rosenberg:1962pp}.

While the longitudinal component of the anomaly diagram is given by its DZ form, once this component is subtracted from the general triangle diagram, it leaves behind an anomaly-free vertex which is purely transverse and corresponds to the GS trilinear interaction.
The Ward identities restrict the general covariant decomposition of ${{\Delta^{GS}}}_{\lambda\mu\nu}(k_3,k_1,k_2)$ into invariant functions to three terms (with all the momenta incoming )
\beqa
\label{calw}
{{ \Delta^{GS}}}_{\lambda\mu\nu}(k_1,k_2) &=&
 -\,\frac{1}{8\pi^2}\,\left(
w_T^{(+)}\left(k_1^2,k_2^2,k_3^2\right)\,t^{(+)}_{\lambda\mu\nu}(k_1,k_2)
 +\,w_T^{(-)}\left(k_1^2,k_2^2,k_3^2\right)\,t^{(-)}_{\mu\nu\rho}(k_1,k_2) \right. \nonumber \\
 && \left. + {\widetilde{w}}_T^{(-)}\left(k_1^2,k_2^2,k_3^2\right)\,{\widetilde{t}}^{(-)}_{\mu\nu\rho}(k_1,k_2)
\right),
 \eeqa
with the transverse tensors given by
\beqa
t^{(+)}_{\mu\nu\rho}(k_1,k_2) &=&
k_{1\nu}\,\veps_{\mu\rho\alpha\beta}\ k_1^\alpha k_2^\beta \,-\,
k_{2\mu}\,\veps_{\nu\rho\alpha\beta}\ k_1^\alpha k_2^\beta \,-\, (k_{1}\cdot
k_2)\,\veps_{\mu\nu\rho\alpha}\ (k_1 - k_2)^\alpha
\nonumber\\
&& \quad\quad+\ \frac{k_1^2 + k_2^2 - k_3^2}{k_3^2}\
\veps_{\mu\nu\alpha\beta}\ k_1^\alpha k_2^\beta(k_1 + k_2)_\rho
\nonumber \ , \\
t^{(-)}_{\mu\nu\rho}(k_1,k_2) &=& \left[ (k_1 - k_2)_\rho \,-\, \frac{k_1^2 - k_2^2}{(k_1
+ k_2)^2}\,(k_1 + k_2)_\rho \right] \,\veps_{\mu\nu\alpha\beta}\ k_1^\alpha k_2^\beta
\nonumber\\
{\widetilde{t}}^{(-)}_{\mu\nu\rho}(k_1,k_2) &=& k_{1\nu}\,\veps_{\mu\rho\alpha\beta}\
k_1^\alpha k_2^\beta \,+\, k_{2\mu}\,\veps_{\nu\rho\alpha\beta}\ k_1^\alpha k_2^\beta
\,-\, (k_{1}\cdot k_2)\,\veps_{\mu\nu\rho\alpha}\ (k_1 + k_2)^\alpha \,,
\lbl{tensors}
\eeqa
where, due to Bose symmetry ($k_1,\mu \leftrightarrow k_2,\nu$) we have
\beqa
w_T^{(+)}\left(k_2^2,k_1^2,k_3^2\right)
&=&+w_T^{(+)}\left(k_1^2,k_2^2,k_3^2\right),
\nonumber \\
w_T^{(-)}\left(k_2^2,k_1^2,k_3^2\right) &=& -
w_T^{(-)}\left(k_1^2,k_2^2,k_3^2\right),
~~~
{\widetilde{w}}_T^{(-)}\left(k_2^2,k_1^2,k_3^2\right) = -
{\widetilde{w}}_T^{(-)}\left(k_1^2,k_2^2,k_3^2\right) \,.
\eea
This version of the GS-corrected AVV vertex satisfies the Ward identities on all the three external lines. The explicit expression of these invariant amplitudes can be obtained from
\cite{Jegerlehner:2005fs}
\begin{eqnarray}
{\widetilde{w}}_{T}^{(-)}(k_1^2,k_2^2,k_3^2) &=& -  w_{T}^{(-)}(k_1^2,k_2^2,k_3^2),
\\
k_3^2 \Delta^2  w^{(-)}_{T}(k_1^2,k_2^2,k_3^2)
&=& 8 (x-y)\Delta
+8(x-y)(6xy + \Delta)\Phi^{(1)}(x,y)
\nonumber
\\
&-&4 [18 x y + 6 x^2-6 x + (1+x+y)\Delta)] L_x
\nonumber
\\
&+& 4[18 x y + 6 y^2-6 y + (1+x+y)\Delta)] L_y \, ,
\\
\nonumber
\\
k_3^2 \Delta^2  w^{(+)}_{T}(k_1^2,k_2^2,k_3^2)&=&8[6xy + (x+y)\Delta]\Phi^{(1)}(x,y)
+8\Delta
\nonumber
\\
&-&4 [6 x+ \Delta ](x-y-1) L_x
\nonumber \\
&+&4 [6 y+ \Delta] (x-y+1) L_y
\end{eqnarray}
with
\begin{equation}
L_x = \ln x,~~~~
L_y = \ln y,~~x=\frac{k_1^2}{k_3^2}~~y=\frac{k_2^2}{k_3^2}\epo
\end{equation}
which involves the scalar triangle diagram for general off-shell lines and determines the function $\Phi^{(1)}$ \cite{Usyukina:1992jd} as
\be
\label{Phi1}
\Phi^{(1)} (x,y) = \frac{1}{\lambda} \left\{ \frac{}{}
2 \left( \Li{2}{-\rho x} + \Li{2}{-\rho y} \right)
+ \ln\frac{y}{x} \ln{\frac{1+\rho y}{1+\rho x}}
+ \ln(\rho x) \ln(\rho y) + \frac{\pi^2}{3}
\right\} ,
\ee
where
\be
\label{lambda}
\lambda(x,y) \equiv \sqrt{\Delta} \; \; \; ,
\; \; \; \rho(x,y) \equiv 2 \; (1-x-y+\lambda)^{-1}, ~~\Delta\equiv(1-x-y)^2 - 4 x y \;.
\ee
The full anomaly amplitude is simply obtained by adding the anomaly pole to this expression

\beq
\Delta^{\lambda\mu\nu}= w_L k^\lambda \epsilon[\mu,\nu,k_1,k_2] +\Delta^{GS\, \lambda\mu\nu}
\label{long}
\eeq
with $w_L=1/(8 \pi^2 k^2)$.

In a non-anomalous theory  a specific charge assignment -in the chiral limit- sets to zero the entire trilinear gauge interaction (identically), while in theories characterized by the GS vertex we require the vanishing of the anomalous part (the anomaly pole). The pole is part of the expression of the triangle diagram, which may or may not contribute in certain graphs.
A typical example is shown in Fig.~\ref{fftriangle} which is not sensitive (in the massless fermion limit) to the longitudinal component of the
anomaly, due to the Ward identities satisfied by the fermion antifermion currents on each external photon line (we are considering axial-vector interactions for each photon). In fact, this is a case in which a GS vertex or a complete anomaly vertex give the same contributions.  In this sense, the anomaly, for this graph, is harmless since the external fermion current is conserved, but this situation is not general.  In fact, we will analyze cases in which a similar situation occurs, and others in which the decoupling of the longitudinal part requires the subtraction of $w_L$ from the anomaly diagram. As we have seen above, there are other cases in which the GS vertex is identically vanishing, and this happens in graphs in which the transverse part of the anomaly diagram is zero, as in light-by-light scattering. The anomaly diagrams are purely longitudinal, and their replacement with the GS vertex has to give necessarily zero. We come therefore to discuss point 2) which has been raised in the previous section. We cannot address the issue of double poles {\em only} in the DZ counterterms and forget that the same poles are also present in the triangle anomaly. In other words: the cancellation of the counterterms in specific graphs takes place if and only if the anomaly diagram is harmless.
\subsection{Examples of explicit GS counterterms: anomaly in muon decay}
\begin{figure}[t]
\begin{center}
\includegraphics[scale=0.8]{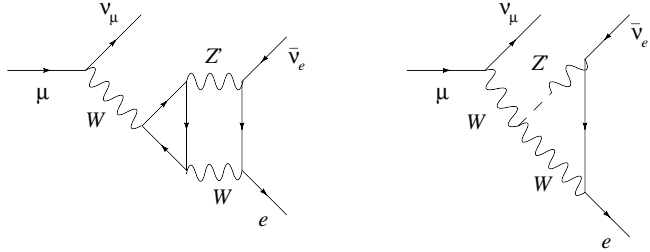}
\caption{\small The muon decay process via a $BT$ diagram. }
\label{muon_decay}
\end{center}
\end{figure}
In general, a given amplitude containing an anomalous extra $Z^{\prime}$ can be harmless if we neglect all the fermion masses and harmful in the opposite case.
An interesting example is shown in Fig.~\ref{muon_decay} which describes a special decay of the muon, mediated by a $WW Z$ vertex. In general, in anomalous extensions of the SM, this amplitude requires a longitudinal subtractions either with the inclusion of a GS or a WZ counterterm. For massless fermions, for instance, the process is anomaly-free. To show this point consider the amplitude for the second diagram in Fig.~\ref{muon_decay} which is given by
\bea
\mathcal{P}&=& i \, \int \frac{d^4 k_2}{(2 \pi)^4} \bar u (p^\prime_1) \g^\mu \frac{1}{\ds p^\prime_1+ \ds k_1} \g^\nu v(p^\prime_2)\, \frac{1}{k_1^2} \left( g^{ \b \nu} - \frac{k_2^\b k_2^\nu}{M_W^2} \right) \frac{1}{k_2^2-M_W^2} \nn \\
&& \frac{a_n}{3} \frac{k_2^\b}{k_2^2} \varepsilon[\alpha, \mu, k, k_1] \frac{1}{k^2}\bar u (p_2) \g^\alpha u(p_1) \nn \\
&=& i \, \bar u (p^\prime_1) \g^\mu v(p^\prime_2)\frac{1}{k^2}\bar u (p_2) \g^\alpha u(p_1)\varepsilon[\alpha, \mu, p^\prime_{12}, \tau] \,\int \frac{d^4 k_2}{(2 \pi)^4} \frac{k_2^\tau}{k_2^2\,(-p^\prime_{12} - k_2)^2}.
\eea
After using the equations of motion for on-shell spinors with $p^\prime_1+p^\prime_2+k_1+k_2=0$ and the tensor integral decomposition in terms of the only momentum in the loop, $p^\prime_{12}=p^\prime_1+p^\prime_2$, it is trivial to verify that the expression vanishes. If we switch-on the external fermion masses, violation of the Ward identities will induce a longitudinal coupling of the anomaly pole on the neutral current, which need an explicit GS subtraction. We will come back
to address the structure of the anomalous contributions away from the chiral limit below, when we will analyze the contribution of similar diagrams to $g-2$ of the muon.

\subsection{Self-energy and gauge invariance}
\begin{figure}[t]
\begin{center}
\includegraphics[scale=0.8]{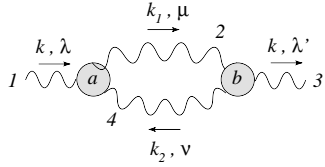}
\caption{\small Self energy  amplitude. }
\label{self}
\end{center}
\end{figure}
Anomalous contributions, in these models, appear also in the running of the coupling, though at a rather large order.
Also in this case transversality of the self-energy is ensured by construction, being the GS vertex transverse
by definition, however, the separation of the vertex into anomaly graph and GS counterterms illustrates how the cancellation of all the double poles takes place.

The corrections appear at 3-loop level and are shown in Fig.~\ref{self}. We denote with $a$, $b$ the GS vertices and assign a number on each line of all the vertices. For instance, in
Fig.~\ref{self} vertex $a$ shares lines 2 and 4 with vertex $b$. The anomaly diagrams are denoted by
$\Delta_a$ and $\Delta_b$, respectively, and are separated into the linear combinations $AVV, VAV$ and $VVA$  as in the previous example, carrying partial anomalies. The pole counterterm can be ``emitted'' by the
vertex either toward the initial or the final state of the diagram along the numbered line. For instance $C_{a2}$ denotes the DZ counterterm that is generated by vertex $a$ with a mixing/ double pole term generated on line $2$. Other trivial cancellations are obtained due to the orthogonality relations between DZ counterterms associated to different lines when they are contracted together.

The expression of the integrand in the amplitude is given by
\bea
{\mathcal M}_{self}^{\lambda \lambda^\prime} &=& - \Delta_{AAA, a}^{\lambda \mu \nu \, GS} (k, k_1, -k_2)
\frac{ g^{\mu \mu^\prime} }{ k_1^2 }  \frac{ g^{ \nu \nu^\prime} }{(k_1-k)^2} \Delta_{AAA, b}^{\lambda^\prime \mu^\prime \nu^\prime  \, GS}( - k, - k_1, k_2)    \nonumber\\
&=&  - \left[  \Delta_{AAA,a} + C_{AAA,a}  \right] ^{\lambda \mu \nu } \frac{1}{k_1^2 (k_1-k)^2}
  \left[  \Delta_{AAA,b} + C_{AAA,b}  \right]^{\lambda^\prime \mu \nu },
\eea
which is given explicitly by
\beqa
C^{\lambda^\prime \mu \nu}_{AAA, b}(-k,-k_1,k_2) &=& \frac{1}{3} \left(C^{\lambda^\prime \mu \nu}_{b3}(-k,-k_1,k_2)
+ C^{\mu \nu \lambda^\prime}_{b2}(k_1,k_2,k)  +  C^{\nu \lambda^\prime \mu}_{b4}(-k_2,k,- k_1)   \right)    \nonumber\\
&=& C^{\mu \nu}_{b3}(- k_1,k_2)k^{\lambda^\prime} + C^{\nu \lambda^\prime}_{b2}(k_2,k)k_1^\mu
 + C^{\lambda^\prime \mu}_{b4}(k,- k_1) k_2^\nu .
\eeqa
\beqa
C^{\lambda \mu \nu}_{AAA, a}(k,k_1,-k_2) &=&
\frac{1}{3} \left( C^{\lambda \mu \nu}_{a1}(k,k_1,-k_2) + C^{\mu \nu \lambda}_{a2}(-k_1,-k_2,-k)
+  C^{\nu \lambda \mu}_{a4}(k_2, -k,k_1)   \right)    \nonumber\\
&=&   C^{\mu \nu}_{a1}(k_1,-k_2)k^\lambda + C^{\nu \lambda}_{a2}(-k_2,-k)k_1^\mu
 +  C^{\lambda \mu}_{a4}(-k,k_1)k_2^\nu
\eeqa
so using the Ward identities
\bea
k^{\mu}_1 \Delta_{AAA, b}^{\lambda^\prime \mu \nu \, GS}( - k, - k_1, k_2)=
k^{\nu}_2 \Delta_{AAA, b}^{\lambda^\prime \mu \nu \, GS}( - k, - k_1, k_2)=0
\eea
we can reduce the 16 contributions of the amplitude ${\mathcal M}_{self}^{\lambda \lambda^\prime}$ to $8$ terms.

\begin{figure}[t]
\begin{center}
\includegraphics[scale=0.7]{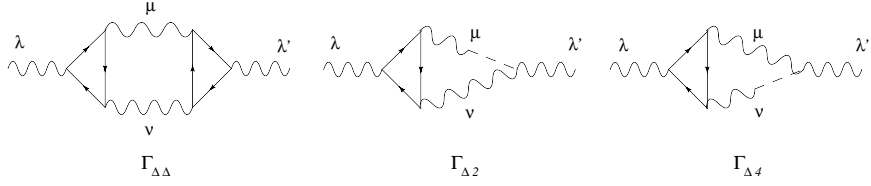}
\caption{\small Contributions to the self-energy amplitude. }
\label{self_terms}
\end{center}
\end{figure}

The GS counterterms $C_{a1}$ and $C_{b3}$  are non-zero for off-shell external photons and are needed to remove the longitudinal poles from the anomaly diagram, while the remaining counterterm contributions are those shown in Fig.~\ref{self_terms}. The latter are transverse just by themselves, as we are going to show. They are given by
\bea
{\mathcal M}_{self}^{\lambda \lambda^\prime} &=&
 -  \Delta_{AAA,a} ^{\lambda \mu \nu } \frac{1}{k_1^2 (k_1-k)^2}
  \left[  \Delta_{AAA,b} + C_{b2} + C_{b4}  \right]^{\lambda^\prime \mu \nu }     \nonumber\\
  &=& -\left(  \Delta_{AAA,a}^{\lambda \mu \nu}  \frac{1}{k_1^2 (k_1-k)^2}   \Delta_{AAA,b}^{\lambda^\prime \mu \nu}
  +   \Delta_{AAA,a}^{\lambda \mu \nu}  \frac{1}{k_1^2 (k_1-k)^2} C_{b2}^{ \mu \nu \lambda^\prime}
 +  \Delta_{AAA,a}^{\lambda \mu \nu}  \frac{1}{k_1^2 (k_1-k)^2}    C_{b4}^{\nu \lambda^\prime \mu } \right) \nonumber\\
 &=&  \Gamma_{\Delta \Delta}^{\lambda \lambda^\prime} + \Gamma_{\Delta 2}^{\lambda \lambda^\prime}
 + \Gamma_{\Delta 4}^{\lambda \lambda^\prime}.
\eea

The amplitude $\Gamma_{\Delta2}$ can be cast in this form using dimensional regularization in $D$ dimensions
\bea
\Gamma^{\lambda \lambda^\prime}_{\Delta 2} &=&
- \int \frac{d^4 k_1}{(2 \pi)^4} \Delta^{\lambda \mu \nu}_{AAA,a}(k, k_1, -k_2)
\frac{1}{k^2_1(k_1-k)^2} C^{\mu \nu \lambda^\prime}_{b2}(k_1, k_2,k)  \nonumber\\
&=&   \frac{a_n}{3} \epsilon[\lambda, \nu, \alpha, k]   \frac{a_n}{3} \epsilon[\nu, \lambda^\prime, \beta, k] \int \frac{d^4 k_1}{(2 \pi)^4}
\frac{k_1^\alpha k_1^\beta}{k_1^4 (k_1 -k)^2}       \nonumber\\
&=& - \frac{1}{2} \left( \frac{a_n}{3} \right)^2 \epsilon[\lambda, \nu, \alpha, k] \epsilon[\nu, \lambda^\prime, \beta, k]
g^{\alpha \beta} \frac{1-D}{s} Bub^{D+2}(s)    \nonumber\\
 &=& \left( \frac{a_n}{3} \right)^2 (k^\lambda k^{\lambda^\prime} - k^2 g^{\lambda \lambda^\prime})
 (1-D) \frac {Bub^D(s)}{8 \pi (3-2 \eps )} \nn \\
 &=&  C\, (k^\lambda k^{\lambda^\prime} - k^2 g^{\lambda \lambda^\prime})  Bub^D(s),
\eea
 where the explicit expressions of the two master integrals $Bub^D(s)$ and $Bub^{D+2}(s)$ can be found in the Appendix in Eqs.(\ref{BubDs}, \ref{BubDp2s}) and
 \bea
 C= \left( \frac{a_n}{3} \right)^2 \frac{1-D}{8 \pi (3-2 \eps )}.
 \eea
\begin{figure}[t]
\begin{center}
\includegraphics[scale=0.75]{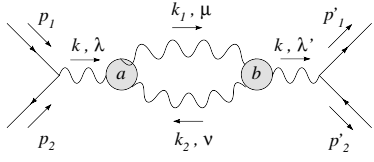}
\caption{\small A self-energy amplitude embedded in a physical process of fermion-antifermion scattering with on-shell external lines. }
\label{self_ffbar}
\end{center}
\end{figure}
If we include the same amplitude in a fermion-antifermion scattering, see Fig.~\ref{self_ffbar}, we obtain
\bea
\mathcal{S}_{\Delta2} &=& - \bar v(p_2) \gamma^\lambda u(p_1) \frac{1}{k^2}
 \left( \frac{k^\lambda k^{\lambda^\prime}}{k^2} - g^{\lambda \lambda^\prime} \right) k^2 C \,Bub^D(s)\,  \frac{1}{k^2} \bar u(p^\prime_1)
\gamma^{\lambda^\prime}  v(p^\prime_2)    \nonumber\\
&=& \bar v(p_2) \gamma^\lambda u(p_1) \frac{1}{k^2}
C \, Bub^D(s) \, \bar u(p^\prime_1)
\gamma^{\lambda}  v(p^\prime_2),
\eea
with
\beq
Bub^{D}(s) = \frac{i \pi^{D/2}}{(2 \pi)^D}\, \mu^{2 \eps} \, \biggl(\frac{e^{\g}}{4 \pi}\biggr)^\eps\frac{c_\Gamma}{ \eps (1-2 \eps)} (s)^{- \eps} (-1)^{\eps},
\label{BubDs0}\\
\eeq
where we have used the equations of motion for the on-shell spinors ($k=p_1+p_2=p^\prime_1 + p^\prime_2$).
The transversality of the pole counterterm comes as a surprise, since while the total amplitude with GS vertices is transverse by construction, the anomalous contribution, in principle, is not expected separately to be so. The computation shows that internal double poles, those due to the GS counterterms, give contributions which are also transversal. This shows once more that there are no apparent inconsistencies in the perturbative expansion of the theory.
\section{Higher order diagrams}
Having worked out several examples in which either the extra poles appear explicitly or cancel by themselves, signalling a harmless anomaly, we now
move to discuss more complex cases, where these techniques will be systematized.

We have two ways to apply the GS vertex at higher order. We could use its explicit form -in terms of its transverse invariant amplitudes- or we could use it in the form "anomaly diagrams plus counterterms". This second form is the most useful one.
The presence of higher poles in the counterterms, which balance those -not explicit- in the anomaly
diagrams, can be treated perturbatively as a field theory of a higher perturbative order. We will illustrate below
two cases from which one can easily infer the general features of the perturbative expansion with these types of graphs. It should be clear that the cancellation of all the poles from the external lines takes place only on-shell, but this is not a problem since we are interested in S-matrix elements.

\subsection{3-point functions}
\begin{figure}[th]
\begin{center}
\includegraphics[scale=0.75]{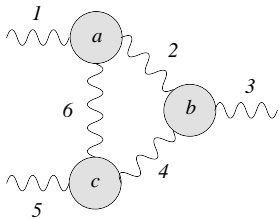}
\caption{\small A 4-loop-amplitude given by three GS vertices with on-shell external lines. }
\label{threeblobs}
\end{center}
\end{figure}

For this reason we consider the 4-loop-diagram shown in Fig.~\ref{threeblobs} with three symmetric GS vertices of the AAA type connected together, which is given by
\bea
\mathcal{M}^{\la \rho \tau}= i\, (\Delta + C_1 + C_2 + C_6)_a^{\la \mu \nu} \frac{1}{k_2^2}\,
(\Delta + C_2 + C_3 + C_4)_b^{\mu \rho \sigma} \, \frac{1}{k_4^2}
(\Delta + C_4 + C_5 + C_6)_c^{\sigma \tau \nu} \frac{1}{k_6^2},
\eea
where, as done in the previous sections, $\Delta$ denotes an $AAA$ triangle amplitude with a symmetric anomaly distribution on each vertex and  $C_i$ a single GS counterterm with the derivative coupling on the i-th line.
At this stage we start simplifying the term $(\Delta + C_2 + C_3 + C_4)_b$ as $(GS)_b$ and in a similar way the $c$ blob using the Ward identities
\bea
C_{a2} (GS)_b = C_{a6} (GS)_c= C_{b4} (GS)_c =0,
\eea
and then omit the GS counterterms $C_{a1}$, $C_{b3}$, $C_{c5}$ in which the transversality conditions
\bea
\eps_{1\la} k_1^{\la} = \eps_{5\tau} k_5^{\tau} = \eps_{3\rho} k_3^{\rho} =0
\eea
 act on the derivative coupling, which allow to reduce the $\mathcal{M}$ amplitude to the six contributions
\bea
\mathcal{M}^{\la \rho \tau} &=& i\, (\Delta)_a^{\la \mu \nu} \frac{1}{k_2^2}\,
(\Delta  + C_2)_b^{\mu \rho \sigma} \frac{1}{k_4^2}\,
(\Delta + C_4 + C_6)_c^{\sigma \tau \nu} \frac{1}{k_6^2}\,  \nn \\
&=& \frac{i}{k_2^2 \, k_4^2 \, k_6^2}
\left(\Delta_a \Delta_b \Delta_c + \Delta_a \Delta_b C_{c4} + \Delta_a \Delta_b C_{c6}
+ \Delta_a C_{b2} \Delta_c + \Delta_a C_{b2} C_{c4} + \Delta_a C_{b2} C_{c6} \right)^{\la \rho \tau} \nn \\
&=& \left( \Delta_a \Delta_b\Delta_c + \Gamma_4 + \Gamma_6 + \Gamma_2 + \Gamma_{24} + \Gamma_{26}\right)^{\la \rho \tau},
\label{Mthree}
\eea
where the notation $\Gamma_i$ and $\Gamma_{ij}$ refers to the line corresponding to the counterterm in Fig.~\ref{threeblobs}.
\begin{figure}[ht]
\begin{center}
\includegraphics[scale=0.65]{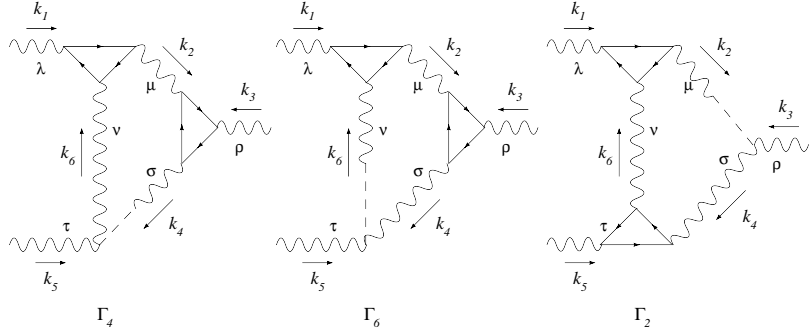}
\caption{\small The $\Gamma_4$,  $\Gamma_6$ and  $\Gamma_2 $ contributions taken from $\mathcal{M}$ at 4-loop-level.}
\label{DDC3}
\end{center}
\end{figure}
At this point we consider the $\Delta_a \Delta_b C_{c4}$ contribution represented in Fig.~\ref{DDC3} with a counterterm on the line $4$  denoted by $\Gamma_4^{\la \rho \tau}$
\bea
{\Gamma}_4^{\la \rho \tau} &=& i  \int \frac{d^4k_4}{(2 \pi)^4}
\bigg[ \Delta_a^{\lambda \mu \nu } (k_1, k_2, -k_6) \frac{1}{ k_2^2 }
\Delta_b^{\mu \rho \sigma} (k_2, -k_3, k_4)
\frac{ 1 }{k_4^2} \frac{a_n}{3 \,k_4^2}
k_4^{\sigma} \epsilon[\tau, \nu , k_5, k_6] \frac{1}{ k_6^2 }\bigg],
\eea
in which we substitute the Rosenberg parametrization for the triangle amplitude $\Delta_a^{\lambda \mu \nu } (k_1, k_2, -k_6)$ given by
\bea
\Delta_a^{\lambda \mu \nu } (k_1, k_2, -k_6) &=&
A_1 \, \epsilon[k_2,\mu, \nu, \la] -
A_2 \, \eps[k_6,\mu, \nu, \la] -
A_3 \, k_2^\nu \, \eps[k_2, k_6, \mu, \la] \nn \\
&+& A_4 \, k_6^\nu \, \eps[k_2, k_6, \mu,\la] -
A_5 \, k_2^\mu \, \eps[k_2, k_6, \nu, \la] +
A_6 \, k_6^\mu \, \eps[k_2, k_6, \nu, \la],
\label{Rosen1}
\eea
and  the anomaly equation
\bea
k_4^{\sigma} \Delta_b^{\mu \rho \sigma} (k_2, -k_3, k_4) = - \frac{a_n}{3} \eps [\mu, \rho, k_2, k_3].
\label{an_eq}
\eea
We choose $k_1$ and $k_5$ as independent momenta, so we have
\bea
k_3=-(k_1+k_5), \qquad \qquad k_2=k_1+k_4+k_5, \qquad \qquad k_6=k_4+k_5,
\eea
with  the on-shell conditions  $k_1^2=k_5^2=k_3^2=0$.
A direct computation of the $\Gamma_4^{\la \rho \tau}$ amplitude shows the complete cancellation of  the spurious double pole relative to the  $k_4$ momentum.

In an analogous way we can consider the $\Delta_a \Delta_b C_{c6}$ term or $\Gamma_6$ in Eq.~(\ref{Mthree}), that is
\bea
{\Gamma}_6^{\la \rho \tau} &=& i \int \frac{d^4k_6}{(2 \pi)^4}
\bigg[ \Delta_a^{\lambda \mu \nu } (k_1, k_2, -k_6) \frac{1}{ k_2^2 }
\Delta_b^{\mu \rho \sigma} (k_2, -k_3, k_4)
\frac{1}{k_4^2} \frac{a_n}{3 \,k_6^2}
\, k_6^\nu \varepsilon[\sigma, \tau, k_4, k_5] \frac{1}{ k_6^2 }\bigg]
\eea
and the $\Delta_a C_{b2} \Delta_c $ term or $\Gamma_2$
\bea
{\Gamma}_2^{\la \rho \tau} &=& i \int \frac{d^4 k_2}{(2 \pi)^4}
\bigg[ \Delta_a^{\lambda \mu \nu } (k_1, k_2, -k_6) \frac{1}{ k_2^2 }
\frac{a_n}{3 \,k_2^2}  k_2^\mu \varepsilon[\rho, \sigma, k_3, k_4] \frac{ 1}{k_4^2}
\Delta_c^{\sigma \tau \nu} (k_4, -k_5, k_6) \frac{1}{ k_6^2 }\bigg],
\eea
for which the conditions in Eq.~(\ref{Rosen1}) and (\ref{an_eq})  have to be modified in a suitable form.   After the expansion of the tensor integrals in terms of the two external momenta $k_1$ and $k_5$ we can conclude that also in this case the double poles don't contribute to the physical on-shell amplitude.\\
\begin{figure}[ht]
 \begin{center}
 \includegraphics[scale=0.6]{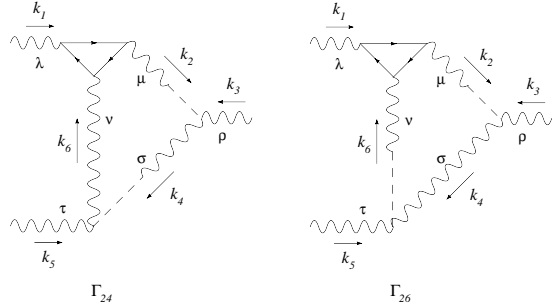}
 \caption{\small Representation of $\Gamma_{24}$ and  $\Gamma_{26}$, the two terms with double poles on the internal lines. }
 \label{DCC}
 \end{center}
\end{figure}
In a similar way we can show the vanishing of the last contributions $\Delta_a\, C_{b2}\, C_{c4}$ ($\Gamma_{24}$) and
 $\Delta_a\, C_{b2}\, C_{c6}$ ($\Gamma_{26}$) shown in Fig.~\ref{DCC} due to antisymmetry
 \bea
{\Gamma}_{24}^{\la \rho \tau} = i \int \frac{d^4k_2}{(2 \pi)^4}
\bigg[ \Delta_a^{\lambda \mu \nu } (k_1, k_2, -k_6) \frac{1}{ k_2^2 }
\frac{a_n}{3 \,k_2^2}  k_2^{\mu} \eps[\rho, \sigma, k_3, k_4]
\frac{1}{k_4^2} \frac{a_n}{3 \,k_4^2}
k_4^{\sigma} \eps[\tau, \nu, k_5, k_6]
\frac{1}{ k_6^2 }\bigg] =0.
 \eea

In the $\Delta_a C_{b2} C_{c6}$ case one obtains the same result after using the anomaly equation, so that
 \bea
{\Gamma}_{26}^{\la \rho \tau} = i \int \frac{d^4k_2}{(2 \pi)^4}
\bigg[ \Delta_a^{\lambda \mu \nu } (k_1, k_2, -k_6) \frac{1 }{ k_2^2 }
\frac{a_n}{3 \,k_2^2}  k_2^{\mu} \eps[\rho, \sigma, k_3, k_4]
\frac{1}{k_4^2} \frac{a_n}{3 \,k_6^2}
\, k_6^\nu \, \eps[\sigma, \tau, k_4, k_5]
\frac{1}{ k_6^2 }\bigg],
 \eea
where the contraction
\bea
k_2^{\mu} \Delta_a^{\lambda \mu \nu } (k_1, k_2, -k_6)= -\frac{a_n}{3} \eps[\nu, \la, k_6, k_1]
\eea
gives $\Gamma_{26}=0$ for antisymmetry.
In conclusion, the amplitude $\mathcal{M}$ at 4-loop-level, composed by three GS symmetric vertices, is not affected by unphysical massless poles arising from the derivative coupling present on some internal lines.
As a result of this analysis it is clear that there are far more cancellations than expected in some of these complex diagrams, due to the structure of the pole counterterms. In fact each DZ counterterm induces  a
Ward identity on an attached triangle diagram and brings in antisymmetric $\epsilon$-tensors into the integrand. This is enough, in many cases, to cause a diagram to vanish by symmetry/antisymmetry of the integrand.

\subsection{Higher point functions: general strategies and examples}
\begin{figure}[ht]
\begin{center}
 \includegraphics[scale=0.7]{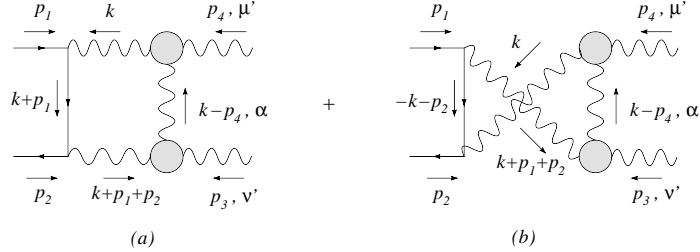}
 \caption{\small Representation of the total process $f \bar f \rightarrow AA$ at 3-loop level via two GS vertices, with $A$ as a generic gauge boson with axial-vector couplings. In diagrams $a)$ and $b)$ we show the amplitudes ${\mathcal M}_a$ + exch. for the tree level process. }
\label{twoblobs1}
\end{center}
\end{figure}
In this subsection we analyze a rather complex example, which is fermion-antifermion annihilation into
two photons at 3-loop level, as shown in Fig.~\ref{twoblobs1}. We will detail our approach, showing how the reduction into typical master integrals of higher orders  takes place for these types of theories.

The total amplitude of the process can be written as
\ba
{\mathcal S}^{\mu^{\prime}\nu^{\prime}} =
- \int \frac{d^4 k}{(2 \pi)^4} \left[  \bar v(p_2) \gamma^\nu \frac{1}{\ds k + \ds p_1 } \gamma^\mu u(p_1)
+\bar v(p_2) \gamma^\mu \frac{1}{- \ds k - \ds p_2}  \gamma^\nu u(p_1)   \right]
\frac{g^{\nu \tau}}{(k + p_1 + p_2)^2}
\frac{g^{\mu \sigma}}{k^2} {\mathcal M}^{\sigma \mu^\prime \nu^\prime \tau}_{c} \nn \\
\ea
where we have defined the sub-amplitude
\ba
{\mathcal M}^{\sigma \mu^\prime \nu^\prime \tau}_{c} =
\Delta_{AAA}^{\alpha \mu^\prime \sigma \, GS} (k -p_4, -p_4, k) \frac{-i}{(k - p_4)^2}
\Delta_{AAA}^{\alpha \nu^\prime \tau  \, GS}(-k + p_4, -p_3, -k-p_1-p_2).
\ea

\begin{figure}[h]
\begin{center}
\includegraphics[scale=0.9]{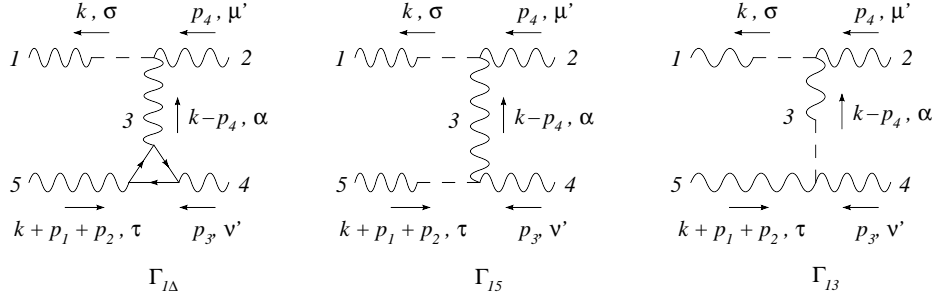}
\caption{\small Amplitudes $\Gamma_{1\Delta}$, $\Gamma_{15}$ and $\Gamma_{13}$ involved in the cancellation of the internal poles. }
\label{internal_poles}
\end{center}
\end{figure}

Using the Ward identity
$(k-p_4)^\alpha \Delta_{AAA}^{\alpha \nu^\prime \tau  \, GS}(-k + p_4, -p_3, -k-p_1-p_2) = 0$
we drop the GS counterterm in $\Delta_{AAA}^{\alpha \mu^\prime \sigma \, GS}$
and we reduce the sub-amplitude ${\mathcal M}^{\sigma \mu^\prime \nu^\prime \tau}_{c}$ to the six contributions
\bea
{\mathcal M}^{\sigma \mu^\prime \nu^\prime \tau}_{c}
&=&   \left[   \Delta_{AAA}^{\alpha \mu^\prime \sigma }(k -p_4, -p_4, k)
+  C^{\sigma \alpha \mu^\prime}(-k, -k+p_4, -p_4)\right]
\nonumber\\
&\times& \frac{-i}{(k-p_4)^2}
\left[\Delta_{AAA}^{\alpha \nu^\prime \tau }(-k + p_4, -p_3, -k-p_1-p_2)
+ C^{\alpha \nu^\prime \tau}(-k + p_4, -p_3, -k-p_1-p_2)  \right.   \nonumber\\
&& \left.
+ C^{\tau \alpha \nu^\prime}(k+p_1+p_2, k-p_4, -p_3)  \right] \nn \\
&=& (\Gamma_{\Delta \Delta} + \Gamma_{\Delta 3} + \Gamma_{\Delta 5}
+\Gamma_{1 \Delta } + \Gamma_{13} + \Gamma_{15})^{\sigma \mu^\prime \nu^\prime \tau},
\eea
where we have used the transversality conditions
\bea
\varepsilon_{\mu^\prime} p_4^{\mu^\prime} = \varepsilon_{\nu^\prime} p_3^{\nu^\prime}=0.
\eea
Some of these contributions are easily shown to vanish, such as $\Gamma_{1 \Delta}$ shown in Fig.~\ref{internal_poles}, which is defined as
\bea
&&\Gamma_{1\Delta}^{\sigma \mu^\prime \nu^\prime \tau} =
\frac{a_n}{3 k^2} k^{ \sigma}
\epsilon[\alpha, \mu^\prime, k-p_4, p_4] \frac{-i}{(k-p_4)^2}
\frac{1}{3} \Delta^{\alpha \nu^\prime \tau}_{AVV}(-k+p_4, -p_3, -k-p_1-p_2).
\eea
If we embed $\Gamma_{1\Delta}$ in a fermion-antifermion annihilation process (Fig.~\ref {twoblobs1})we obtain a new amplitude, $\mathcal S_{1\Delta}$, given by
\bea
{\mathcal S}^{\mu^\prime \nu^\prime}_{1\Delta}
&=&    - \int \frac{d^4 k}{(2 \pi)^4} \left[  \bar v(p_2) \gamma^\nu \frac{1}{\ds k + \ds p_1 } \gamma^\mu u(p_1)
+\bar v(p_2) \gamma^\mu \frac{1}{- \ds k - \ds p_2}  \gamma^\nu u(p_1)   \right]
\frac{g^{\nu \tau}}{(k + p_1 + p_2)^2}
\frac{g^{\mu \sigma}}{k^2}   \nonumber\\
&& \times \frac{a_n}{3 k^2}\varepsilon[\a, \mu^\prime,k-p_4, p_4] k^{\sigma}
\frac{-i}{(k-p_4)^2}
\frac{1}{3} \Delta^{\alpha \nu^\prime \tau}_{AVV}(-k+p_4, -p_3, -k-p_1-p_2)      \nonumber\\
&=&   - \int \frac{d^4 k}{(2 \pi)^4} \left[\bar v(p_2) \gamma^\nu \frac{1}{\ds k + \ds p_1} \ds k u(p_1)
+  \bar v(p_2) \ds k \frac{1}{- \ds k - \ds p_2}  \gamma^\nu u(p_1)   \right]
 \frac{1}{k^2} \frac{1}{(k+p_1+p_2)^2}      \nonumber\\
&& \times  \frac{a_n}{3 k^2}\varepsilon[\a, \mu^\prime,k-p_4, p_4] \frac{-i}{(k-p_4)^2}
\frac{1}{3} \Delta^{\alpha \nu^\prime \tau}_{AVV}(-k+p_4, -p_3, -k-p_1-p_2)  = 0,
\eea
where we have used the equations of motion for the on-shell spinors.
We focus now our attention on the 2 terms shown in Fig.~\ref{internal_poles} called $\Gamma_{13}$ and $\Gamma_{15}$ which both exhibit internal poles.
In a straightforward way we find that $\Gamma_{13}$ vanishes
\bea
{\Gamma}^{\sigma \mu^\prime \nu^\prime \tau}_{13}  &=&
C^{\sigma \alpha \mu^\prime}(-k, -k+p_4, -p_4)  \frac{-i}{(k-p_4)^2}
C^{\alpha \nu^\prime  \tau}(-k+p_4, -p_3, -k - p_1 - p_2)     \nonumber\\
&=& -i \, \frac{a_n}{3 k^2} k^{\sigma}
\epsilon[ \alpha, \mu^\prime, k-p_4, p_4]   \frac{a_n }{3 (k-p_4)^4} (k -p_4)^\alpha \epsilon[\nu^\prime, \tau, p_3,  k +p_1+p_2]  = 0,
\eea
for symmetry, while the amplitude $\Gamma_{15}$ can be written as
\bea
{\Gamma}^{\sigma \mu^\prime \nu^\prime \tau}_{15}   &=&
C^{\sigma \alpha \mu^\prime}(-k, -k+p_4, -p_4)  \frac{-i}{(k-p_4)^2}
C^{\tau \alpha \nu^\prime}(k+p_1+p_2, k-p_4, -p_3)    \nonumber\\
&=& C^{\alpha \mu^\prime}(k-p_4, p_4) k^{\sigma} \frac{-i }{(k-p_4)^2}
C^{\alpha \nu^\prime} (k-p_4, p_3) (k+p_1+p_2)^{\tau} .
\eea
The vanishing of the internal poles coming from $\Gamma_{15}$ is obtained by considering both
contributions of Fig.~\ref{twoblobs1}, obtaining
\bea
{\mathcal S}^{\mu^\prime \nu^\prime}_{15}&=&
- \left( {\mathcal M}^{\mu \nu}_{a}  +  {\mathcal M}^{\mu \nu}_{b} \right)
\frac{g^{\mu \sigma}}{k^2} \frac{g^{\nu \tau}}{(k+p_1+p_2)^2}
{\Gamma}^{\sigma \mu^\prime \nu^\prime \tau}_{15}  \nonumber\\
&=&  - \int \frac{d^4 k}{(2 \pi)^4} \left[\bar v(p_2) (\ds k+\ds p_1+ \ds p_2) \frac{1}{\ds k + \ds p_1} \ds k  u(p_1)
+  \bar v(p_2) \ds k \frac{1}{-\ds k - \ds p_2} ( \ds k+ \ds p_1+ \ds p_2)  u(p_1)   \right] \nn \\
&\times& \frac{1}{k^2} \frac{1}{(k+p_1+p_2)^2}  C^{\alpha \mu^\prime}  \frac{-i}{(k-p_4)^2} C^{\alpha \nu^\prime}= 0.
\eea
At this point we are ready to isolate the only non vanishing contribution to the amplitude, which is expressed in terms of the two components
\bea
\Gamma^{\sigma \mu^\prime \nu^\prime \tau}_{\Delta3} &=&
\frac{1}{3} \Delta^{\alpha \mu^\prime \sigma}_{AVV} (k-p_4, -p_4,k) \frac{-i}{(k - p_4)^2}
C^{\alpha \nu^\prime \tau} (-k+p_4, p_3, k+ p_1+p_2)   \nonumber\\
&=& \frac{a_n}{3} \varepsilon[\mu^\prime, \sigma, -p_4, k] \frac{-i}{(k-p_4)^2}
\frac{a_n}{3(k-p_4)^2} \varepsilon[\nu^\prime, \tau, p_3, k+p_1+p_2]
\eea
and
\bea
{\mathcal S}^{\mu^\prime \nu^\prime}_{\Delta3}&=&
- \int \frac{d^4 k}{(2 \pi)^4}  \left( {\mathcal M}^{\mu \nu}_{a}+  {\mathcal M}^{\mu \nu}_{b} \right)
\frac{g^{\mu \sigma}}{k^2} \frac{g^{\nu \tau}}{(k+p_1+p_2)^2}
{\Gamma}^{\sigma \mu^\prime \nu^\prime \tau}_{\Delta3}    \nonumber\\
&=& -  \int \frac{d^4 k}{(2 \pi)^4} \left[\bar v(p_2) \gamma^\nu \frac{1}{\ds k + \ds p_1} \gamma^\mu u(p_1)
+  \bar v(p_2) \gamma^\mu \frac{1}{-\ds k - \ds p_2 }  \gamma^\nu u(p_1)   \right]
 \frac{g^{\nu \tau}}{(k+p_1+p_2)^2} \frac{g^{\mu \sigma}}{k^2}    \nonumber\\
&& \times \frac{a_n}{3} \varepsilon[\mu^\prime, \sigma, -p_4, k]
\frac{a_n}{3(k-p_4)^4} \varepsilon[\nu^\prime, \tau, p_3, k+p_1+p_2]
\eea
which can't be simplified any further. Also in this case, the presence of explicit "extra poles" in one of the amplitude, brings us to erroneous conclusions if we would claim a failure of unitarity in the process. In fact, hidden in the anomaly diagrams are longitudinal couplings that cancel those of the counterterms by construction, being the GS vertices transverse.

\begin{figure}[t]
\begin{center}
\includegraphics[scale=0.9]{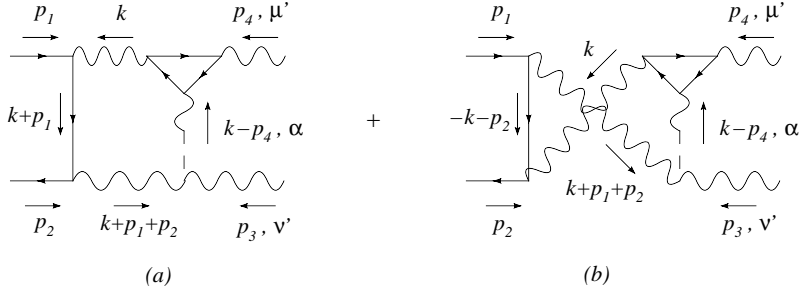}
\caption{\small  Box-like contribution coming from the process show in Fig.~\ref{twoblobs1}. }
\label{box_babis}
\end{center}
\end{figure}

\begin{figure}[t]
\begin{center}
\includegraphics[width=3cm]{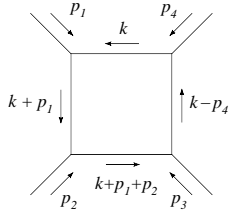}
\caption{\small Representation of the momentum parametrization for the box-like amplitude. }
\label{box_topology}
\end{center}
\end{figure}

The evaluation of the counterterm amplitude follows a standard approach in perturbation theory at higher order.
To show how this takes place, consider the graph in Fig.~\ref{box_babis}.
We introduce the notation $p_3=-p_{124}=-p_1-p_2-p_4$ for momentum conservation, as shown in Fig.~\ref{box_topology}, and a simple computation gives for the direct contribution
\bea
{\mathcal S}^{\mu^\prime \nu^\prime}_{\Delta3}&=&
-\int \frac{d^4 k}{ (2 \pi)^4 }   {\mathcal M}^{\mu \nu}_{a}
\frac{g^{\mu \sigma}}{k^2} \frac{g^{\nu \tau}}{(k+p_1+p_2)^2}   {\Gamma}^{\sigma \mu^\prime \nu^\prime \tau}_{\Delta3}    \nonumber\\
&=& i   \int \frac{d^4 k}{ (2 \pi)^4 }  \bar v(p_2) \gamma^\nu  \frac{ (k + p_1)_\rho}{(k +p_1)^2}
\gamma^\rho \gamma^\mu u(p_1) \frac{1}{ k^2 }  \frac{1}{ (k +p_1 +p_2 )^2 }  \nonumber\\
&& \frac{a_n}{3}  \varepsilon[ \mu^\prime, \mu, - p_4, k ]
\frac{a_n}{3 ( k- p_4)^4} \varepsilon[\nu^\prime, \nu, -p_{124}, k +p_1+p_2]     \nonumber\\
&=&  i \frac{a^2_n}{9} \epsilon[\mu^\prime, \mu, - p_4, \alpha] \epsilon[\nu^\prime, \nu, - p_{124}, \beta]
\bar v(p_2) \gamma^\nu \gamma^\rho \gamma^\mu u(p_1)      \nonumber\\
&&   \int \frac{d^4 k}{ (2 \pi)^4 }  \frac{1}{k^2 (k+p_1)^2 (k+p_1+p_2)^2 (k-p_4)^4}  (k+p_1)^\rho k^\alpha (k+p_1+p_2)^\beta   \nonumber\\
&=&  i T^{\rho \mu^\prime \nu^\prime \alpha \beta}    \int \frac{d^4 k}{ (2 \pi)^4 }  \frac{1}{A_1 A_2 A_3 A^2_4}
(k^\rho k^\alpha k^\beta + k^\rho k^\alpha p_1^\beta + k^\rho k^\alpha p_2^\beta
+p_1^\rho k^\alpha k^\beta + p_1^\rho k^\alpha p_1^\beta + p_1^\rho k^\alpha p_2^\beta )       \nonumber\\
&\equiv& i T^{\rho \mu^\prime \nu^\prime \alpha \beta}
 \left(  {\bf J}^D_{\rho \alpha \beta} + p_1^\beta {\bf J}^D_{\rho \alpha} + p_2^\beta {\bf J}^D_{\rho \alpha}
  + p_1^\rho {\bf J}^D_{\alpha \beta} + p_1^\rho p_1^\beta {\bf J}^D_{\alpha} + p_1^\rho p_2^\beta {\bf J}^D_{\alpha}     \right)
\label{SDelta3}
\eea
where $A_i$ denote, in ordered sequence, the propagators. This amplitude can be computed explicitly, as we illustrate in the appendix.

 There are some interesting
aspects that emerge in the evaluation of these integrals already at 1-loop level. First of all, the combination of the anomaly vertex and of the pole counterterm introduces a Ward identity which trivializes one of the momentum integration, removing the triangle subdiagram from the counterterm graph. The original
two-loop diagram is then reduced to a single one-loop integration but with propagators of higher powers.
The expansion that follows shares therefore the characteristics of an ordinary perturbative expansion
of {\em  higher order},  in which higher powers of the propagators appear quite naturally.

If, in an ordinary perturbative expansion at 2-loop level and higher, we combine the integration by parts and
the usual tensor decomposition of the integrals, trading loop integrals for  higher powers of the propagators, as shown in the appendix, we end up with a perturbative expansion with propagators of arbitrary powers. Therefore, unsurprisingly, the formulation of ordinary perturbative field theories at higher order can be based on a perturbative expansion containing propagators of higher powers. Anomalous field theories treated with the GS prescription are not, from this perspective, that exceptional.

\section{WZ and GS interactions, anomalous magnetic moment of the muon and muonium}
Having clarified some of the subtle issues characterizing a perturbative expansion with GS vertices,
we move to discuss the role played by the GS and the WZ mechanism in $g-2$ of the muon and in
muonium. This is the case where the L/T decomposition of the anomaly amplitude shows its direct relevance and the role of the GS and WZ vertices can be easily worked out.

Our aim is not to proceed with a complete study of these corrections, some of which require a separate study, but to highlight the role played by the two mechanisms in the context of specific
processes which can be accurately quantified in future studies. The possibility of searching for anomalous extra $Z^\prime$ and axions in precision measurements of several observables is challenging but realistic.
\subsection{The GS case}
\begin{figure}[t]
\begin{center}
\includegraphics[scale=0.8]{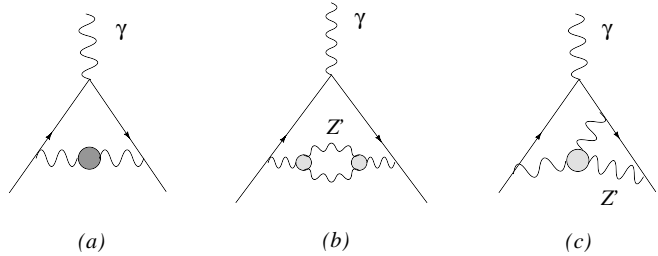}
\caption{\small Higher order contributions to the muon magnetic moment in the GS case.}
\label{magn1}
\end{center}
\end{figure}

\begin{figure}[t]
\begin{center}
\includegraphics[scale=0.8]{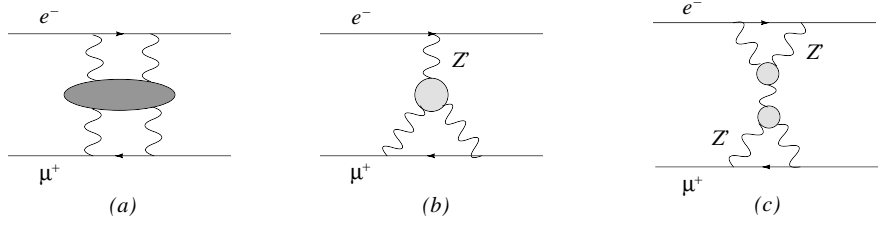}
\caption{\small Hadronic contributions (a), higher order anomalous contributions (b) and light-by-light contributions (c) to the hyperfine splitting in muonium.}
\label{magn2}
\end{center}
\end{figure}

We show in Figs.~\ref{magn1} and \ref{magn2} some of the lowest order GS contributions to the anomalous magnetic moment of the muon and to the hyperfine splitting of muonium.  Some of the recent theoretical attention to $a_\mu\equiv g-2$ has been focused on the study of effects at 2-loop level and higher, such as those shown in Fig.~\ref{magn1}a and
~\ref{magn1}d. The first indicates generically the hadronic contributions coming from self-energy insertions in the lowest order vertex. Of these types are also the corrections coming
from the self-energy graphs involving GS vertices. The corrections are tiny, being of order $g^8$
and their computation involves a 4-loop graph with ordinary propagators (the 2-triangle diagram of  Fig.~\ref{self_terms}) and 2-loop graphs related to the pole counterterms that we have studied in the analysis of
the self-energy. Clearly, the underlying lagrangean should allow an anomalous extra $Z^\prime$ in the spectrum. Working models of this type have been studied previously, and include several anomalous $U(1)'s$, such as in the case of intersecting
branes. The presence of a physical axion that mixes with the Higgs sector (called ``axi-Higgs'' in \cite{Coriano:2005js}) via a kinetic St\"uckelberg term (and eventually a Peccei-Quinn breaking term) makes these models quite attractive. The axi-Higgs is massless in the first case and massive in the second case. Models with
an axi-higgs are constructed using only WZ interactions and not GS interactions.
\begin{figure}[t]
\begin{center}
\includegraphics[scale=0.8]{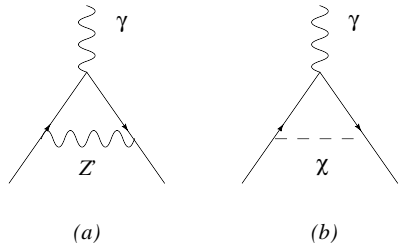}
\caption{\small Leading order corrections to the anomalous magnetic moment of the muon.}
\label{magn_LO}
\end{center}
\end{figure}
\begin{figure}[t]
\begin{center}
\includegraphics[scale=0.8]{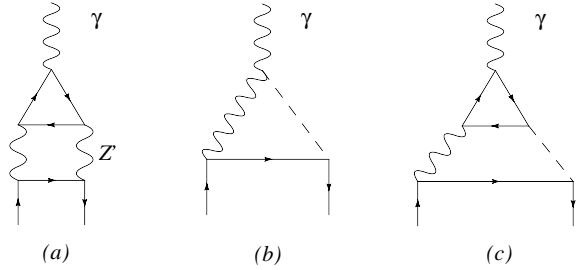}
\caption{\small Higher order corrections to the anomalous magnetic moment of the muon with a WZ vertex.}
\label{magn_WZ}
\end{center}
\end{figure}
\begin{figure}[t]
\begin{center}
\includegraphics[scale=0.8]{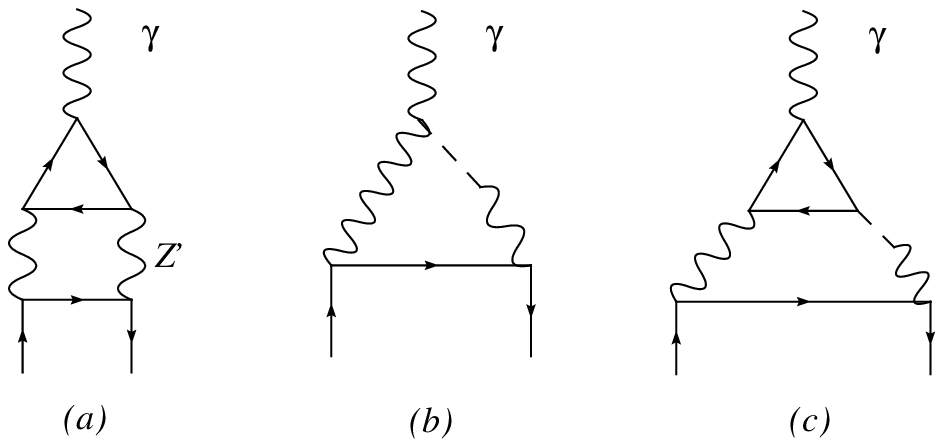}
\caption{\small As in Fig.~\ref{magn_WZ} for a GS vertex.}
\label{magn_GS}
\end{center}
\end{figure}
\subsection{Anomalous corrections to $a_\mu$}
The evaluation of the anomalous corrections to $a_\mu$ in the GS and WZ cases are quite different. In the WZ case there is a larger set of contributing graphs due to the interaction of the axion with the fermions and involve the exchange of an axi-Higgs (massless or massive), which is proportional to the fermion mass. The simplest corrections due to the presence of an anomalous extra $Z^\prime$ are shown in Fig.~\ref{magn_LO}. These do not involve the anomaly diagram and are the leading ones. They have been computed in \cite{Kiritsis:2002aj}. Higher order corrections are those shown in Fig.~\ref{magn_WZ}, also involving  a physical axi-Higgs.

It is convenient to describe in some detail the structure of the perturbative expansion at higher orders to emphasize the differences between the two mechanisms.

The structure of the expansion can be grasped more easily if we work in the chiral limit
(all the fermions are taken to be massless) and focus our attention, for example, on graph (b) in Fig.~\ref{magn1}
since in this case there is no direct point-like interaction of the axion with the fermion.
If we decide to cancel the anomaly with WZ counterterms, we know that we can draw
a counterterm diagram in which the axion is emitted and absorbed by the gauge line.
In this case it is clear that the anomaly is potentially harmful and only a direct
computation is able to show if the counterterm is zero or not. In this specific diagrams we
know that explicit pole counterterms are needed, as we have shown in the previous sections.
If we consider diagram (c), however, the application of this argument shows immediately that the anomaly,
in this case, is harmless, since there is no axion counterterm of WZ type that we can draw.
A similar result is obtained for diagram (a) in Fig.~\ref{magn_WZ}. Also in this case we are unable to draw a WZ counterterm in which the axion is attached only to gauge lines. Therefore this diagram is also well defined even in the presence of an anomaly diagram, since its longitudinal part cancels automatically due to the topology of the graph. In these last two diagrams the gauge lines have to be attached in all possible ways to the muon lines for this to happen. Diagram (c) appears in the massive case, but it is not a counterterm.

Coming to the GS case in the massive fermion case, the anomaly diagram developes a mass-dependence in
the residue of the anomaly pole, shown in graphs (c) of Fig.~\ref{magn_GS}. As we are going to show in the next section,  this is not an anomaly  counterterm. The only counterterm is still given only by diagram (b).
More details will be given below and in the final section.

\begin{figure}[t]
\begin{center}
\includegraphics[scale=0.9]{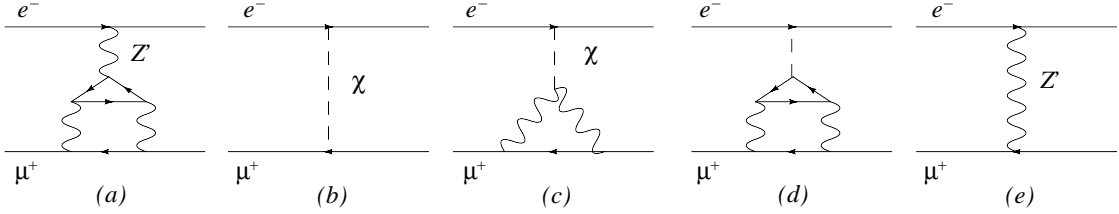}
\caption{\small  Leading contributions to the hyperfine splitting in muonium and denoted as $\mathcal N_i$, with i=a,b,c,d in the WZ case.}
\label{muonium_others}
\end{center}
\end{figure}

\begin{figure}[h]
\begin{center}
\includegraphics[scale=0.9]{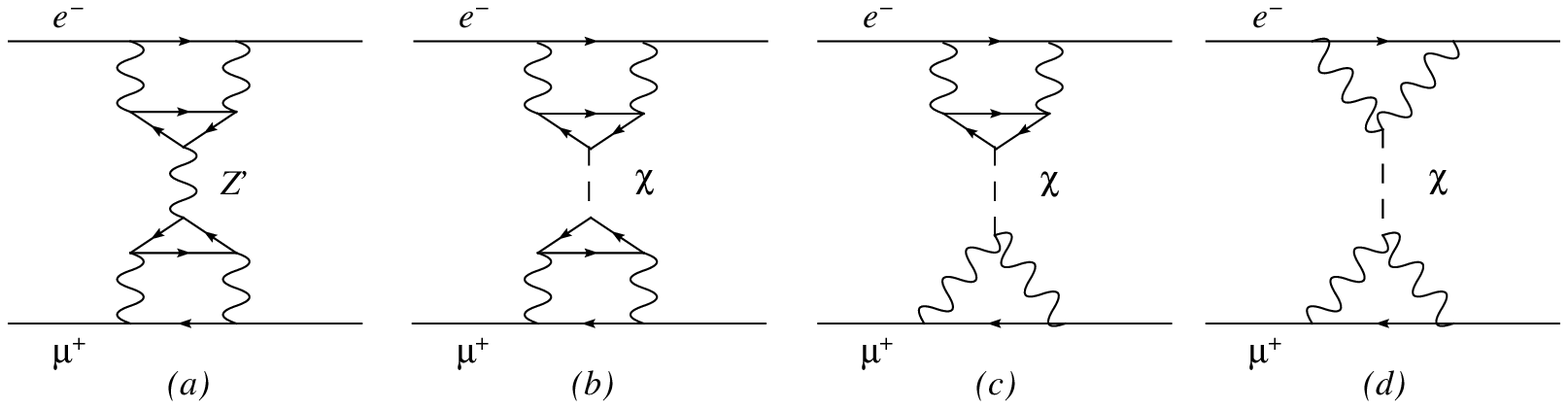}
\caption{\small Explicit (higher order) expansion of the diagram in Fig.~\ref{magn2}c  for a trilinear WZ vertex. Each amplitude is denoted as $\mathcal{M}_i$ with i=a,b,c,d in the text with $p_1$, $p_1^\prime$ as incoming momenta and $p_2$, $p_2^\prime$ as outcoming ones.}
\label{LBL_WZ}
\end{center}
\end{figure}

\section{2-loop Contributions to $g-2$: Anomalous Diagrams}
In this section we briefly analyze the general structure of these corrections for both mechanisms when anomaly diagrams are present. The analysis that we follow is close to the discussions for $g-2$ presented in \cite{Czarnecki:2002nt, Knecht:2002hr, Kukhto:1992qv}, adapted to our case. Most of the physical discussion carried out in these papers, in the case of the muon anomalous magnetic moment, has to do with the identification of the effects due to chiral symmetry breaking in the computation of the anomaly diagrams, which are related both to perturbative and to non-perturbative effects, treated within the operator product expansion. In our case we will be interested only in the perturbative contributions with the GS and WZ vertices.  We will point out the differences
compared to those previous studies while reviewing their derivation in order to be self-contained.  In the GS case the anomaly diagram, corrected by the pole subtraction, does not satisfy any longer the Vainshstein relation \cite{Vainshtein:2002nv} between the longitudinal ($w_L(q^2)$) and transverse ($w_T(q^2)$) component of the anomaly vertex

\beq
w_L(q^2)=2 w_T(q^2),
\eeq
which is obtained in a specific kinematical limit of the anomaly diagram. In particular, in the chiral limit, the
longitudinal component $w_L$ of the GS vertex is zero. Away from the chiral limit a pole $O(1/q^4)$ reappears, multiplied by additional contributions proportional to the fermion mass squared $(m_f^2)$, but it is not an anomaly pole.  The separation between $L$ and $T$ components, away from the chiral limit, for $m_f\neq 0$,
can be done in several ways. In \cite{Knecht:2002hr} this is obtained by isolating the anomalous pole contribution from the rest. After the subtraction of the pole term, the new anomaly-free vertex is still not transverse and satisfies a broken Ward identity. The truly transverse component ($\tilde{w}^T$) is isolated by acting with a specific projection on the vertex, as we shall see below. This assumes a special form in the limit
in which one of the photons is on-shell ($k^2\to 0$) and soft ($k\to 0$). It can be expressed in terms of a set of scalar diagrams which come from the rank-2 tensor decomposition of the fermionic triangle ($C_{ij}$), and which are well-known in the literature. The explicit expressions of these integrals, which are for instance given in  \cite{Kniehl:1989qu}, are singular in the soft/on-shell photon limit that is needed in order to extract their contribution to $g-2$. This is the reason
for  the re-analysis of these contributions using the operator product expansion (OPE), which in this case follows the approach of  \cite{Czarnecki:2002nt, Knecht:2002hr}. In our case, both for the GS and WZ vertices, the OPE analysis would be similar, and can be performed on the two currents carrying large momentum ($q^2\to \infty$), therefore we omit it. In the WZ case the pseudoscalar exchanges involve a goldstone and of a physical axion, this second one being not present in the SM.

We start our analysis by stating our  conventions.
 The coupling of the extra neutral current to the fermions is given by
\beq
-\frac{ig_2}{4\cos\theta_W}\bar{\psi_i}\left(g_V^{Z,Z'}\gamma^{\mu}
+g_A^{Z,Z'}\gamma^{\mu}\gamma^{5}\right)\psi_i V_{\mu}
\eeq
where the vector boson $V_{\mu}$ stays for the $Z$ or the $Z^\prime$ and
the vector and the axial-vector couplings can be written as
\ba
&&\frac{-i g_2}{4 c_w}\gamma^{\mu} {g_V}^{Z^{\prime},j}=\frac{-i g_2}{c_w}
\frac{1}{2}\left[ -\varepsilon c_w^2 T_3^{L,j}
+\varepsilon s_w^2(\frac{\hat{Y}^{j}_L}{2}+\frac{\hat{Y}^{j}_R}{2})
+\frac{g_z}{g_2}c_w(\frac{\hat{z}_{L,j}}{2}+
\frac{\hat{z}_{R,j}}{2})\right]\gamma^{\mu}
\nonumber\\
&&\frac{-i g_2}{4 c_w}\gamma^{\mu}\gamma^{5} {g_A}^{Z^{\prime},j}=\frac{-i
g_2}{c_w} \frac{1}{2}\left[ \varepsilon c_w^2 T_3^{L,j}
+\varepsilon s_w^2(\frac{\hat{Y}^{j}_R}{2}-\frac{\hat{Y}^{j}_L}{2})
+\frac{g_z}{g_2}c_w(\frac{\hat{z}_{R,j}}{2}-
\frac{\hat{z}_{L,j}}{2})\right]\gamma^{\mu}\gamma^{5}.
\ea
Here $j$ is an index which represents the quark or the lepton and we have
set $\sin\theta_W=s_w,\cos\theta_W=c_w$ for brevity.
We denote with $\hat{z}_{R,L}$ the charges of the fermions
under the extra anomalous $U(1)$ and with
$g_z$ the coupling constant of the anomalous gauge interaction \cite{Coriano:2008wf}.

The electroweak vertex that we need to compute in order to take into account the corrections to the anomalous magnetic moment of the muon, due to
the exchange of an extra anomalous $Z^\prime$, in analogy with the discussion presented in \cite{Knecht:2002hr},  is given by
\ba
&&\langle\bar{\mu}(p')\vert
V_{\rho}^{em}(0)\vert\mu(p)\rangle=
\bar{u}(p')\Gamma_{\rho}(p',p)u(p)=
\nonumber\\
&&\int\frac{d^4q}{(2\pi)^4}\,
\frac{-i}{q^2}\,\frac{-i}{(p'-p-q)^2-M_{Z}^2}(-ie)(-ie)
\left(\frac{ig_2}{4\cos\theta_{W}}\right)
\left(\frac{-ig_2}{4\cos\theta_{W}}\right)\times
\nonumber\\
&&\bar{u}(p')\left[\gamma^{\mu}\frac{i}{\pslsh'\ -\qsls-m_{\mu}}g_{A,\mu}^{Z'}\gamma^{\nu}\gamma_{5}+
g_{A,\mu}^{Z'}\gamma^{\nu}\gamma_{5}\frac{i}{\pslsh\ +\qsls-m_{\mu}}
\gamma^{\mu}\right]u(p)\times
\nonumber\\
&&\int d^4x\,e^{iq.x}\int d^4y\, e^{i(p'-p-q).y}
\langle 0\vert
T\{A_{\mu}^{\gamma}(x)Z_{\nu}^{\prime}(y)A_{\rho}^{\gamma }(0)\}\vert
0\rangle\,,
\ea
where
\beq
A_{\mu}^{\gamma}(x)=\bar{q}(x)\gamma_{\mu}\, Q_f\, q(x)\,,\qquad
Z_{\nu}^{\prime}(y)=\bar{q}(y)\gamma_{\nu}\gamma_{5}\, g_{A,f}^{Z'}\,q(y)
\eeq
are the fermion currents of quarks, and $g_{A,f}^{Z^\prime}$ refers to a quark of flavor $f$.
The most general CP invariant expression for a vertex function
satisfying the current conservation is defined by
\ba
\Gamma^{\mu}=-i e\bar{u}(p_2)\left[F_1(q^2) \gamma^{\mu}
+ F_2(q^2)\frac{q_{\alpha}}{4 m_{\mu}}\sigma^{\alpha\mu}
+F_3(q^2)\frac{\left(q^{\mu}\ds{q}-\gamma^{\mu}q^2\right)\gamma^5}{4 M_W^2}
\right]u(p_1)
\ea
where the coefficients $F_i(q^2)$ are the form-factors and $q=p_2-p_1$, $M_W$ and $m_\mu$ denote the mass of the $W$ and of the muon, respectively.
Taking the limit  $q^2\rightarrow 0$ in the Pauli form-factor we obtain the value
of the anomalous magnetic moment
\beq
a=\frac{g-2}{2}=F_2(0),
\eeq
and using the equation of motion we obtain \cite{Kukhto:1992qv}
\ba
\Gamma_{\mu}=a a_{\mu} && a_{\mu}=i e \bar{u}(p_2)\frac{1}{2 m_{\mu}}(p_1+p_2)_{\mu}u(p_1).
\ea
We can use a project operator to extract $F_{2}(0)$
\beq
F_{2}(0)=\lim_{k^2\rightarrow 0}\tr\left\{(\pslsh+m_{\mu})\Lambda_{2}^{\rho}(p',p)(\pslsh'+m_{\mu})
\Gamma_{\rho}(p',p)
\right\}\,,
\eeq
where $(p'=p+k)$
\beq
\Lambda_{2}^{\rho}(p',p)=\frac{m_{\mu}^2}{k^2}\frac{1}
{4m_{\mu}^2-k^2}\gamma^{\rho}
-\frac{m_{\mu}}{k^2}\frac{2m_{\mu}^2+k^2}{(4m_{\mu}^2-k^2)^2}
(p+p')^{\rho}
\eeq
is the projector on the Pauli form factor.

The triangle contribution is obtained from the 1-loop correlator of the electroweak currents
\beq
(2\pi)^4\delta(p'-p-q) \Delta^{\mu\nu\rho}_{VAV}(q,k)=\int d^4x\,e^{iq.x}\int d^4y\, e^{i(p'-p-q).y}
\langle 0\vert T\{A_{\mu}^{\gamma}(x)Z_{\nu}^{\prime}(y)A_{\rho}^{\gamma }(0)\}\vert
0\rangle
\eeq
with $p'$ the incoming photon four-momentum.
The corresponding tensor structure of the triangle in the $k^2\rightarrow 0$ limit for
a fermion of flavor $f$ is given by \cite{Kukhto:1992qv}, obtained from the Rosenberg representation
\cite{Rosenberg:1962pp}
\ba
&&\Delta^{\mu\nu\rho}_{VAV}(q,k)\longrightarrow \frac{g}{\pi^2\cos\theta_W} g_{A,f}^{Z'}~e^2 Q_f^2~
q^{\alpha}q^{\beta} S^{\mu\nu\rho}_{\alpha\beta}(k)\int_0^{1}d x\frac{x(1-x)}{x(1-x)q^2-m_{f}^2}
\nonumber\\
&&S^{\mu\nu\rho}_{\alpha\beta}(k)=-2 k_{\tau}\varepsilon^{\tau\lambda\mu\rho}
\left(g_{\alpha\lambda}g^{\nu}_{\beta} - g_{\alpha\beta}g^{\nu}_{\lambda}\right)
+g_{\alpha\lambda}\varepsilon^{\lambda\tau\nu\rho}\left(k_{\beta}g^{\mu}_{\tau}-g^{\mu}_{\beta}k_{\tau}\right),
\ea
where $k=p'-p$. This expression, in the GS case, is simply modified by the subtraction of the longitudinal pole due to the anomaly.
The tensor $\Delta^{\mu\nu\rho}_{VAV}(q,k)$
in momentum space is affected by the longitudinal (anomaly) pole, similarly to the case of axial QED discussed above, in the form of a longitudinal $w_L(q^2)$ contribution \cite{Jegerlehner:2005fs}.
In fact the asymptotic behavior at large $Q^2=-q^2$ is given by \cite{Czarnecki:2002nt}
\ba
w_L^{f}(Q^2)=\frac{g_2}{\cos\theta_W} g_{A,f}^{Z'}e^2 Q_f^2
\left[ \frac{1}{2 \pi^2 Q^2}-2\frac{m_f^2}{\pi^2 Q^4}\log{\frac{Q^2}{m_f^2}}+O\left(\frac{1}{Q^6}\right)\right].
\label{carn}
\ea
and in the GS case it becomes
\ba
{w_{L}}^{f}(Q^2)\vert_{GS}=-\frac{g_2}{\pi^2\cos\theta_W} g_{A,f}^{Z'}e^2 Q_f^2
\left[ 2\frac{m_f^2}{Q^4}\log{\frac{Q^2}{m_f^2}}+O\left(\frac{1}{Q^6}\right)\right].
\ea
Following \cite{Knecht:2002hr}
 we can always write
\beq
\Delta^{f}_{\mu\nu\rho}(q,k)=\Delta_{\mu\nu\rho}(q,k)_{\mbox{\rm\tiny
anomaly}}+\tilde{\Delta}^{f}_{\mu\nu\rho}(q,k) \,,
\eeq
where
\beq
\Delta_{\mu\nu\rho}(q,k)_{\mbox{\rm\tiny anomaly}}=
\sum_f g_{A,f}^{Z'}e^2 Q_f^2 a_n\frac{(q-k)_{\nu}}{(q-k)^2}
\epsilon_{\mu\rho\alpha\beta}q^{\alpha}k^{\beta}\,,
\eeq
with $a_n=-i/(2\pi^2)$.
The function $\tilde{\Delta}^f_{\mu\nu\rho}(q,k)$ is transverse with respect to
the momenta $q^{\mu}$ and $k^{\rho}$
\beq
q^{\mu}\tilde{\Delta}^f_{\mu\nu\rho}(q,k)=0\,,\nonumber\\ \;
k^{\rho}\tilde{\Delta}^f_{\mu\nu\rho}(q,k)=0\,,
\label{wi}
\eeq
but in the presence of massive fermions we isolate the longitudinal components of the
corresponding broken Ward identity
\beq
\tilde{\Delta}^f_{\mu\nu\rho}(q,k)=
\tilde{\Delta}_{\mu\nu\rho}^{f,\mbox{\tiny long}}(q,k)
+\tilde{\Delta}_{\mu\nu\rho}^{f,\mbox{\tiny trans}}(q,k)\,.
\eeq
Differentiating the 2nd expression in Eq. (\ref{wi}) with respect to $k_{\rho}$ we obtain
\ba
\tilde{\Delta}_{\mu\nu\rho}(q,k)=-k^{\sigma}\frac{\partial}{\partial k^{\rho}}\tilde{\Delta}_{\mu\nu\sigma}(q,k)
\ea
where we have suppressed the flavor index $f$ for simplicity.
Since we are interested in the soft photon limit, the relevant contributions are those linear in
$k$.  In \cite{Knecht:2002hr} these are extracted in the form
\ba
\tilde{\Delta}_{\mu\nu\rho}^{\mbox{\tiny trans}}(q,k)=k^{\sigma}\Delta_{\mu\nu\rho\sigma}(q) + ...
\label{rosi}
\ea
where the tensor $\Delta_{\mu\nu\rho\sigma}(q)$ is obtained by using the projection operator
$\Pi^{\nu\nu'}$ as follows
\ba
&&\Pi^{\nu\nu'}(q,k)=\left( g^{\nu\nu'}-\frac{(q-k)^{\nu}}{(q-k)^2}(q-k)^{\nu'}\right),
\nonumber\\
&&\Delta_{\mu\nu\rho\sigma}(q)=-\frac{\partial}{\partial k^{\rho}}\left(\Pi^{\nu\nu'}\Delta_{\mu\nu'\rho}\right)\vert_{\lim k\rightarrow 0}.
\label{proj}
\ea
It is not difficult to notice that
\ba
\Pi^{\nu\nu'}\Delta_{\mu\nu'\rho}=\Pi^{\nu\nu'}\tilde{\Delta}_{\mu\nu'\rho}.
\ea
As we have already mentioned, the action of $\Pi^{\nu\nu'}$ is to remove all the longitudinal parts from
the $\Delta_{\mu\nu\rho}$ tensor, including the anomalous term.
 $\Delta_{\mu\nu\rho\sigma}(q)$ in (\ref{rosi}) has the form
\ba
\Delta_{\mu\nu\rho\sigma}(q)=i \Delta(Q^2)\left[ q_{\rho}\varepsilon_{\mu\nu\alpha\sigma}q^{\alpha}
- q_{\sigma}\varepsilon_{\mu\nu\alpha\rho}q^{\alpha}\right],
\ea
where $q^2=-Q^2$.
We can now try to apply this formalism to the anomalous
triangle diagrams. We use the generic parametrization of a $AVV$ triangle given in \cite{Kniehl:1989qu},
\ba
&&\Delta^f_{\mu\nu\rho}(q,k)=-\frac{g_2}{\pi^2 \cos\theta_W} g_{A,f}^{Z'}e^2Q_f^2
\left[
A(k,-q,m_f)(k_{\rho}\varepsilon_{\nu\mu\beta\sigma}k^{\beta}
-k^2\varepsilon_{\nu\mu\rho\sigma})(-q)^{\sigma}
\right.\nonumber\\
&&\hspace{1.5cm}\left.+A(-q,k,m_f)
(q_{\mu}\varepsilon_{\nu\rho\beta\sigma}q^{\beta}
-q^2\varepsilon_{\nu\rho\mu\sigma})k^{\sigma}
\right.\nonumber\\
&&\hspace{1.5cm}\left.
-B(k,-q,m_f)(q-k)_{\nu}\varepsilon_{\rho\mu\alpha\beta}k^{\alpha}(-q)^{\beta}
\right],
\ea
where the functions $A(k,-q,m_f)$ and $B(k,-q,m_f)$ are given by in terms of the
tensor-redution coefficients $C_{ij}$ as follows
\ba
&&A(k,-q,m_f)=(C_{11} -C_{12}+C_{21}-C_{23})(k,-q,m_f)
\nonumber\\
&&B(k,-q,m_f)=(C_{12}+C_{23})(k,-q,m_f),
\ea
and are defined in Eq.(A.2),(A.3) of ref. \cite{Kniehl:1989qu}.
The Ward identity on the axial-vector current is given by
\ba
(q-k)^{\nu}\Delta^f_{\mu\nu\rho}(q,k)=
-\frac{g_2}{\pi^2 \cos\theta_W} g_{A,f}^{Z'}e^2Q_f^2\left[\frac{1}{2}-2m_f^2 C_0\right]\varepsilon_{\rho\mu\alpha\beta}k^{\alpha}(-q)^{\beta}
\label{gen}
\ea
and the most general expression of the coefficient $C_0$ is given
in Eq.(A.8) of ref. \cite{Kniehl:1989qu}. $C_0$ is the scalar 3-point function with a fermion of
mass $m_f$ circulating in the loop. In the soft photon limit the invariant amplitude defined by the right-hand-side of (\ref{gen}) reduces to (\ref{carn}).

The purely transverse part (for $m_f\neq 0$) is obtained by applying the projection operator given in
(\ref{proj})
\ba
\Delta^{trans}_{\mu\nu\rho}(q,k)&=&-k^{\sigma}\frac{\partial}{\partial k^{\rho}}\left(\Pi^{\nu\nu'}(q,k)\Delta^f_{\mu\nu\sigma}(q,k)\right)\vert_{\lim k\rightarrow 0}
\nonumber\\
&=&-k^{\sigma}\frac{\partial}{\partial k^{\rho}}
\Delta^T_{\mu\nu\sigma}(q,k)\vert_{\lim k\rightarrow 0},
\ea
where
\ba
&&\Delta^T_{\mu\nu\sigma}(q,k)=A(k,-q,m_f)\left[
q^{\alpha}\varepsilon_{\alpha\mu\nu\sigma}k^2 -\frac{k_{\nu}\,k^2}{(q-k)^2}\varepsilon_{\mu\sigma\alpha\beta}k^{\alpha}q^{\beta}
+\frac{q_{\nu}\,k^2}{(q-k)^2}\varepsilon_{\mu\sigma\alpha\beta}k^{\alpha}q^{\beta}
+k_{\sigma}\varepsilon_{\mu\nu\alpha\beta}k^{\alpha}q^{\beta}
\right]
\nonumber\\
&&\hspace{1.5cm}+A(-q,k,m_f)\left[
k^{\alpha}\varepsilon_{\alpha\mu\nu\sigma}q^2
-\frac{k_{\nu}\,q^2}{(q-k)^2}\varepsilon_{\mu\sigma\alpha\beta}k^{\alpha}q^{\beta}
+\frac{q_{\nu}\,q^2}{(q-k)^2}\varepsilon_{\mu\sigma\alpha\beta}k^{\alpha}q^{\beta}
-q_{\mu}\varepsilon_{\nu\sigma\alpha\beta}k^{\alpha}q^{\beta}\right],
\nonumber\\
\ea
Differentiating with respect to $k_{\rho}$ and taking the $\lim k \rightarrow 0$ we obtain
\ba
\frac{\partial}{\partial k^{\rho}}\Delta^T_{\mu\nu\sigma}(q,k)\vert_{\lim k \rightarrow 0}&=&
A(Q^2)\left[ \varepsilon_{\rho\mu\nu\sigma}q^2
+q_{\nu}\varepsilon_{\mu\sigma\rho\beta}q^{\beta}
-q_{\mu}\varepsilon_{\nu\sigma\rho\beta}q^{\beta}\right]
\nonumber\\
&=&A(Q^2)\left[q_{\rho}\varepsilon_{\mu\nu\alpha\sigma}q^{\alpha}
- q_{\sigma}\varepsilon_{\mu\nu\alpha\rho}q^{\alpha}\right].
\ea
where $A(Q^2)$ denotes the soft limit of the $A(-q,k,m_f)$ amplitude.
The intermediate steps to simplify the contribution to $a_\mu$  are those of \cite{deRafael:1993za}.
In our case, with the modifications discussed above, the Pauli form-factor for a circulating fermion of flavor $f$
we obtain

\ba
&&F_{2}(0)|_{Z'}=(-e^2)\frac{g_2^2}{16\cos^2\theta_{W}}
\frac{1}{M_{Z'}^2}
\lim_{k^2\rightarrow 0}\int\frac{d^4q}{(2\pi)^4}\frac{1}{q^2}
\left(\frac{M_{Z'}^2}{q^2 -M_{Z'}^2}\right)\times
\nonumber\\
&&\hspace{2cm}\frac{1}{4k^2}\tr\left\{(\pslsh+m_{\mu})
\left[\gamma^{\rho}\ksls-\left(k^{\rho}+
\frac{p^{\rho}}{m_{\mu}}\ksls\right)\right]
\right.\nonumber\\
&&\left.\hspace{2cm}\left[\gamma^{\mu}\frac{(\pslsh\ -\qsls+m_{\mu})}
{q^2-2q\dd p}g_{A,\mu}^{Z'}
\gamma^{\nu}\gamma_{5}+g_{A,\mu}^{Z'}\gamma^{\nu}\gamma_{5}
\frac{(\pslsh\ +\qsls+m_{\mu})}{q^2+2q\dd p}\gamma^{\mu}
\right]
\right\}\times
\nonumber\\
&&\hspace{2cm}g_{A,f}^{Z'} Q_f \left[
-i\tilde{\Delta}_{\mu\nu\rho}^{f,\mbox{\rm\tiny long  }}(q,k)+
k^{\sigma}\left[q_{\rho}\varepsilon_{\mu\nu\alpha\sigma}q^{\alpha}-q_{\sigma}
\varepsilon_{\mu\nu\alpha\rho}q^{\alpha}\right]A(Q^2)\right]\,,\nonumber\\
\label{TOT}
\ea
where
$\tilde{\Delta}_{\mu\nu\rho}^{\mbox{\rm\tiny f,long}}$ is not anomalous
and in the soft photon limit it is given by
\ba
\tilde{\Delta}_{\mu\nu\rho}^{\mbox{\rm\tiny f,long}}(q,k)= \frac{q_{\nu}}{q^2}\tilde{w}_L^f \varepsilon_{\mu\rho\alpha\sigma}q^{\alpha}k^{\sigma}
\ea
where
\ba
\tilde{w}_L^f= -\left[-\frac{2}{\pi^2}\frac{m_f^2}{q^4}
\log{\frac{(-q)^2}{m_f^2}}+O\left(\frac{1}{q^6}\right)\right].
\ea
It is obvious from this analysis that in presence of the Green-Schwarz mechanism there is a 2-loop counterterm which removes the pure anomalous contribution in Eq.~(\ref{TOT}) and is given by see Fig.~(\ref{magn_GS}b. Diagram c) in the same figure is the longitudinal part of the diagram and appears in the broken Ward identity that we will discuss in the last section. Finally, after some manipulations, similar to those performed in \cite{Knecht:2003xy, Knecht:2002hr, Peris:1995bb, Czarnecki:2002nt}, the final result for the anomalous contributions to $a_\mu$ takes the form
\ba
F_{2}(0)|_{Z'}=(-e^2 Q_f)\frac{g_2^2 g_{A,\mu}^{Z'}
g_{A,f}^{Z'}}{16\cos^2\theta_{W}}\left(\frac{m_{\mu}^2}
{M_{Z'}^2}\right)\frac{1}{4\pi^2}\int_{m_{\mu}}^{\infty}dQ^2
\left(\tilde{w}_L^{f}(Q^2)+\frac{M_{Z'}^2}{Q^2+M_{Z'}^2}A(Q^2)\right)\,,
\ea
where $-q^2=Q^2$ and $\tilde{w}_L^{f}(Q^2)$ vanishes in the chiral limit.

\subsubsection{The Wess-Zumino Counterterm}
A similar analysis can be performed in the case of the Wess Zumino mechanism.
The leading non anomalous 1-loop contributions Fig.~(\ref{magn_LO}) have been calculated in \cite{Kiritsis:2002aj} for a specific D-brane model. These are due to the coupling of the axi-Higgs to the fermions. The organization of the perturbative expansion for a theory with an axion-like particle has been discussed in \cite{Coriano:2007fw, Coriano:2006xh}, where the explicit cancellation of the gauge dependence has been discussed on general grounds. We show in Fig.~\ref{magn_WZ} the contribution coming from the $Z^\prime$ propagator (graph a) in the anomalous exchange, the additional graphs b) and c) represent the axion counterterm due to the WZ interaction (b) and the correction due to the coupling of the axion to the massive fermions (c). We have omitted a graph similar to (c) in which the exchanged pseudoscalar is a goldstone and cancels the gauge dependence of the $Z^\prime$ propagator.

The computation of graph a) follows exactly the analysis of \cite{Knecht:2002hr} and can be performed in dimensional regularization in the unitary gauge, to give
\ba
&&\lambda_{\mbox{\tiny{$\overline{\mbox{MS}}$}}}\,\equiv\,
\frac{\mu^{d-4}}{16\pi^2}\,\Big[ \,\frac{1}{d-4}\,-\,
\frac{1}{2}\,\big(\log 4\pi +2+ \Gamma '(1)\big)\Big]\,,
\nonumber\\
&&F_2(0)\Big\vert^{(f)}_{\mbox{\tiny anom}} = \frac{g_2^2}{16\pi^2\cos^2\theta_W}
\frac{m_{\mu}^2}{M_{Z'}^2}\frac{1}{4\pi^2}Q_f^2 g_A^{Z',f}\times
\left[ \log\left(\frac{\mu_R^2}{M_{Z'}^2}\right)\,
-\,32\pi^2\lambda_{\mbox{\tiny{$\overline{\mbox{MS}}$}}}\,+
\log\left(\frac{M_{Z'}^2}{m_{\mu}^2}\right)\,+\, \frac{1}{2}
\right]\,.
\nonumber\\
\ea
The expression of the extra contributions when a physical axion is exchanged are given by
\ba
&&F_{2}(k^2)\big\vert_{(c)\mbox{\rm\tiny long}}=
e Q_{\mu}c^{\chi}_{\mu} \lim_{k^2\rightarrow 0}\int\frac{d^4q}{(2\pi)^4}\,
\frac{1}{q^2}\,\frac{1}{(p'-p-q)^2-M_{Z'}^2}\times
\nonumber\\
&&\bar{u}(p')\left[\gamma^{\mu}\frac{1}{\pslsh'\ -\qsls-m_{\mu}}\gamma_{5}+
\gamma_{5}\frac{1}{\pslsh\ +\qsls-m_{\mu}}\gamma^{\mu}\right]u(p)\times
\nonumber\\
&&\left[2 \frac{m_{\mu}}{M_{Z'}^2} \sum_f e Q_f c^{\chi}_f \frac{1}{(q-k)^2-M_{\chi}^2}\right]
\Delta_{\mu \rho}(m_f,q,k,q-k),
\ea
where $c^{\chi}_f$ is coupling of the axi-Higgs to the fermions and
\ba
&&\Delta_{\mu \rho}(m_f,q,k,q-k)=\varepsilon_{\mu\rho\alpha\beta}q^{\alpha}k^{\beta}
\left(-\frac{1}{2\pi^2}\right)I(m_f)
\nonumber\\\nonumber\\
&&I(m_f)\equiv -\int_0^1\int_0^{1-x}dx dy \frac{1}{m_f^2+(x-1)x q^2+(y-1)y k^2-2 x y q\cdot k}\,.
\ea
Using the projection operator we get
\ba
&&F_{2}(0)\big\vert_{(c)\mbox{\rm\tiny long}}=
e Q_{\mu}c^{\chi}_{\mu} \lim_{k^2\rightarrow 0}\int\frac{d^4q}{(2\pi)^4}\,
\frac{1}{q^2}\,\frac{1}{(p'-p-q)^2-M_{Z'}^2}\times
\nonumber\\
&&\frac{1}{4k^2}\tr\left\{(\pslsh+m_{\mu})
\left[\gamma^{\rho}\ksls-\left(k^{\rho}+
\frac{p^{\rho}}{m_{\mu}}\ksls\right)\right]\times
\right.\nonumber\\
&&\left.\left[\gamma^{\mu}\frac{1}{\pslsh'\ -\qsls-m_{\mu}}\gamma_{5}+
\gamma_{5}\frac{1}{\pslsh\ +\qsls-m_{\mu}}\gamma^{\mu}\right]\right\}\times
\nonumber\\
&&\left[2 \frac{m_{\mu}}{M_{Z'}^2} \sum_f e Q_f c^{\chi}_f \frac{1}{(q-k)^2-M_{\chi}^2}\right]
\Delta_{\mu \rho}(m_f,q,k,q-k).
\ea
The contribution coming from the diagram in Fig.~\ref{magn_WZ}b is similar to Fig.~\ref{magn_WZ}c and we obtain
\ba
&&F_{2}(0)\big\vert_{(b)\mbox{\rm\tiny long}}=
e Q_{\mu}c^{\chi}_{\mu} \lim_{k^2\rightarrow 0}\int\frac{d^4q}{(2\pi)^4}\,
\frac{1}{q^2}\,\frac{1}{(p'-p-q)^2-M_{Z'}^2}\times
\nonumber\\
&&\frac{1}{4k^2}\tr\left\{(\pslsh+m_{\mu})
\left[\gamma^{\rho}\ksls-\left(k^{\rho}+
\frac{p^{\rho}}{m_{\mu}}\ksls\right)\right]\times
\right.\nonumber\\
&&\left.\left[\gamma^{\mu}\frac{1}{\pslsh'\ -\qsls-m_{\mu}}\gamma_{5}+
\gamma_{5}\frac{1}{\pslsh\ +\qsls-m_{\mu}}\gamma^{\mu}\right]\right\}\times
\nonumber\\
&&\left[-2 \frac{m_{\mu}}{M_{Z'}^2} g^{\chi}_{\gamma\gamma} \frac{1}{(q-k)^2-M_{\chi}^2}\right]
\varepsilon_{\mu\rho\alpha\beta}q^{\alpha}k^{\beta}.
\ea
where the coefficient $g^{\chi}_{\gamma\gamma}$ is the coupling of the axi-Higgs to the photons
and it will be given explicitly in the next section together with the coefficient $c^{\chi}$.

\subsection{Corrections to muonium}

\begin{figure}[t]
\begin{center}
\includegraphics[scale=0.7]{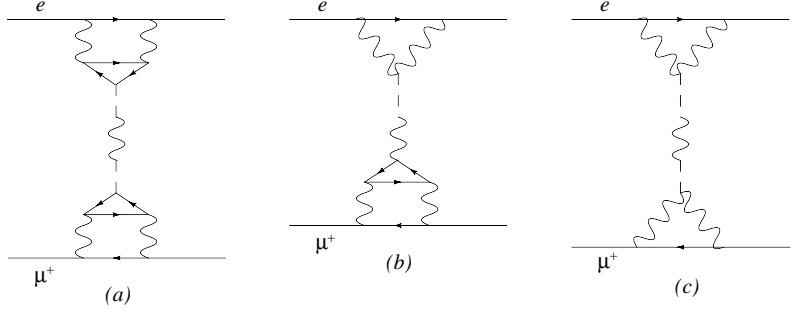}
\caption{\small As in Fig.~\ref{LBL_WZ} for a GS vertex.}
\label{LBL_GS}
\end{center}
\end{figure}

A similar analysis of the role played by both mechanisms in anomalous processes at higher orders can be
done in the case of muonium. A recent analysis of the hadronic effects in this type of systems can be found
in \cite{Vainshtein:2007zz}. One of the typical contributions is given by virtual light-by-light scattering, shown in Fig.~\ref{magn2}. In the presence of anomalous gauge interactions a dominant contribution for the GS case is given by diagram (b). Diagram (c) is subdominant. This is expanded in terms single and double counterterms, typically given in
Fig.~\ref{LBL_GS}. In the WZ case we report some of the corresponding contributions in Figs.~\ref{muonium_others} and \ref{LBL_WZ}, where we allow a coupling of the axi-Higgs to the fermions.  The leading contribution are diagrams
(b) and (e) of Fig.~\ref{muonium_others}, which are the analogue of (a) and (b) of Fig.~\ref{magn_LO}.

The diagrams involving light-by-light scattering in the presence of a WZ vertex with a physical axi-Higgs $\chi$ coupled to fermions are shown in Fig.~\ref{LBL_WZ}; their expression can be easily obtained by taking into account some recent results  on two-loop QCD corrections \cite{Bernreuther:2005rw} and a specific choice of parameters for an anomalous model developed and fully described in
\cite{Coriano:2007fw, Coriano:2007xg}. So we have
\ba
{\mathcal M}_a &=&
\bar u (p_2)  \Bigg[ e^4 \sum_f g_{Z^\prime} a^f_{Z^\prime} Q^2_f \, \Lambda_\mu(s,m_f,m_e) \Bigg]
u(p_1) \frac{-i}{k^2-M_{Z^\prime}^2}  \left( g^{\mu \nu}-\frac{k^\mu k^\nu}{M^2_{Z^\prime}}  \right)    \nonumber\\
&&\bar v( p^\prime_1)
 \Bigg[ e^4 \sum_{f^\prime} g_{Z^\prime} a^{f^\prime}_{Z^\prime} Q^2_{f^\prime} \Lambda_\nu(s,m_{f^\prime},m_\mu) \Bigg]
 v(p^\prime_2),
\ea
where the axial-vector vertex function $\Lambda_\mu(s,m_f,m_{ext})$ is given \cite{Bernreuther:2005rw} in terms of some coefficients named $G_1$ and $G_2$ as
\ba
\Lambda_\mu(s,m_f,m_{ext})  = \gamma_\mu \gamma_5 G_1(s,m_f,m_{ext})
+ \frac{1}{2 m_{ext}} k_\mu \gamma_5 G_2(s,m_f,m_{ext})
\ea
and $m_{ext}$ refers to the electron or the muon. Their explicit expression can be found in \cite{Bernreuther:2005rw}.
For $\mathcal M_b$, with an axi-Higgs exchanged in the t-channel we obtain
\ba
{\mathcal M}_b &=&
\bar u (p_2) \Bigg[ e^2  \sum_f c^{\chi, f}_{\g\g}  \Lambda(s,m_f,m_e) \Bigg]
u(p_1) \frac{i}{k^2-m_\chi^2}
\bar v( p^\prime_1)
 \Bigg[ e^2 \sum_{f^\prime} c^{\chi, f^\prime}_{\g\g}  \Lambda(s,m_{f^\prime},m_\mu)  \Bigg]    v(p^\prime_2), \nn \\
\ea
with the general coupling of the physical axion
\ba
c^{\chi,f}_{\gamma\gamma}= e^{2} Q^2_f c^{\chi,f},   \qquad f=u,d,\nu, e.
\label{chi_coupling}
\ea
and  the pseudoscalar vertex function $\Lambda(s,m_f,m_{ext})$ \cite{Bernreuther:2005rw}
\ba
\Lambda(s,m_f,m_{ext}) = \gamma_5 A(s,m_f ,m_{ext}).
\ea
We have used a condensed notation for the flavors in Eq.~\ref{chi_coupling}
with u = \{u, c, t\}, d = \{d, s, b\}, $\nu$ = \{$\nu_{e}$, $\nu_{\mu}$, $\nu_{\tau}$\}
and e = \{ e, $\mu$, $\tau$\}, whose expansion yields
\ba
c^{\chi, u} &=& \Gamma^{u}  \frac{i}{\sqrt 2} O^{\chi}_{11} = \frac{m^{}_{u}}{v_u} i O^{\chi}_{11} ,  \qquad
c^{\chi, d} = - \Gamma^{d} \frac{i}{\sqrt 2} O^{\chi}_{21} = - \frac{m^{}_{d}}{v^{}_{d}} i O^{\chi}_{21},    \nonumber\\
c^{\chi, \nu} &=& \Gamma^{\nu}  \frac{i}{\sqrt 2} O^{\chi}_{11} = \frac{m^{}_{\nu}}{v^{}_u} i O^{\chi}_{11} ,  \qquad
c^{\chi, e} = - \Gamma^{e} \frac{i}{\sqrt 2} O^{\chi}_{21} = - \frac{m^{}_{e}}{v^{}_{d}} i O^{\chi}_{21},
\ea
where the elements of the $O^{\chi}$ rotation matrix from the interaction to the mass eigenstate basis are given in \cite{Coriano:2007xg}.\\
The most difficult to analyze are those of higher order, shown in Fig.~\ref{LBL_WZ}
\ba
{\mathcal M}_c &=&
\bar u (p_2) \Bigg[ e^2  \sum_f c^{\chi, f}_{\g\g}  \Lambda(s,m_f,m_e) \Bigg]
u(p_1) \frac{i}{k^2-m_\chi^2} \bar v( p^\prime_1)  F(s,m_\mu) g^\chi_{\g\g}  e^2  v(p^\prime_2) \\
{\mathcal M}_d &=& \bar u(p_2) F(s,m_e) g^\chi_{\g\g} e^2
u(p_1) \frac{i}{k^2-m_\chi^2} \bar v( p^\prime_1)  F(s,m_\mu) g^\chi_{\g\g}  e^2  v(p^\prime_2),
\ea
with the one-loop anomalous vertex function $F(s, m_{ext})$  \cite{Bernreuther:2005rw}
\ba
F(s,m_{ext})= 2 i m_{ext} f(s,m_{ext})  \,\gamma_5
\ea
where $m_{ext}$ is the mass of the external fermion, in this case the muon mass,
and the specific choice of coupling given by
\ba
\label{GS_coeffs}
g^{\chi}_{\g\g}=\left[\frac{F}{M_1}(O^{A}_{W\g})^2
+\frac{C_{YY}}{M_1}(O^{A}_{Y\g})^2\right]O^{\chi}_{31}
\ea
in terms of model dependent parameters defined in \cite{Coriano:2007xg}. \\
The simplest corrections are those of Fig.~\ref{muonium_others} and can be written as
\ba
{\mathcal N}_a &=&
\bar u(p_2) ( g_{Z^\prime} a^e_{Z^\prime} ) \g^\mu \gamma^5 u(p_1)  \frac{-i}{k^2-M_{Z^\prime}^2}
\left( g^{\mu \nu} - \frac{k^\mu k^\nu}{M^2_{Z^\prime}}  \right)    
\bar v( p^\prime_1) \Bigg[ e^4 \sum_f g_{Z^\prime} a^f_{Z^\prime} Q^2_f \Lambda_\nu(s,m_f,m_\mu) \Bigg] v(p^\prime_2), \nn \\ \\
{\mathcal N}_b &=&
\bar u(p_2)c^{\chi, e} \gamma^5 u(p_1) \frac{i}{k^2-m_\chi^2}\bar v( p^\prime_1) c^{\chi, \mu} \gamma^5 v(p^\prime_2), \\
{\mathcal N}_c &=&
\bar u(p_2) c^{\chi, e} \gamma^5 u(p_1) \frac{i}{k^2-m_\chi^2}
\bar v( p^\prime_1) F(s,m_\mu) g^{\chi}_{\g\g} e^2  v(p^\prime_2), \\
{\mathcal N}_d &=&
\bar u(p_2) c^{\chi, e} \gamma^5 u(p_1) \frac{i}{k^2-m_\chi^2}
\bar v( p^\prime_1)  \Bigg[ e^2  \sum_f c^{\chi,f}_{\g\g} \Lambda(s,m_f,m_\mu) \Bigg] v(p^\prime_2),
\ea
where the one-loop functions  $\Lambda_\nu$, $F$ and $\Lambda$ have been given above and can be found in the literature.
\section{The longitudinal subtraction and the broken Ward identities of the GS vertex}
There is one last important point that we will address in this final section which concerns the correct interpretation of the anomaly counterterm in both (chiral) phases of theory. The GS vertex satisfies a broken Ward identity, which is easy to derive diagrammatically. The identity is similar to that of the ordinary triangle diagram, but with a subtraction of the (massless) anomaly pole. In the massless fermion case, the GS counterterm restores the Ward identity on the anomaly vertex; in the massive case the mass-dependent terms are a signal of chiral symmetry breaking, but are not counterterms. The only counterterm is the anomaly pole. We briefly clarify this point.

We recall that, for on-shell photons (analogously for gluons) in an anomalous theory,
 the pole contribution to the $AVV$ triangle is given by
\ba
&&T^{\lambda\mu\nu}=g_{V_{KK}}\theta_f^{KK} g_s^2 Tr[t^a t^b] k^{\lambda}\varepsilon[k_1,k_2,\mu,\nu]
\left[\frac{1}{2\pi^2 s}-\frac{m_f^2}{ 2\pi^2 s^2}\log^2 \left(\frac{\rho_f+1}{\rho_f-1}\right) \right]
\nonumber\\
&&\rho_f=\sqrt{1-\frac{4 m_f^2}{s}}
\ea
and the anomalous Ward identity on the axial-vector line gives
\ba
&&k_{\lambda}T^{\lambda\mu\nu}=g_{V_{KK}}\theta_f^{KK} g_s^2 Tr[t^a t^b]\left(\varepsilon[k_1,k_2,\mu,\nu]\frac{1}{2\pi^2}
+2 m_f T^{\mu\nu}\right)
\nonumber\\
&&T^{\mu\nu}=\frac{m_f}{2\pi^2}
\int_0^1\int_0^{1-x}d x d y \frac{\varepsilon[k_1,k_2,\mu,\nu]}{m_f^2 -2 x y k_1\cdot k_2}.
\label{reT}
\ea
The second term in the Ward identity above, or $T_{\mu\nu}$, in a local gauge theory with  spontaneous symmetry breaking (in an anomly-free theory), is determined by the BRS invariance of the correlator. In an anomalous theory the first term is the anomaly, while the second term comes from chiral symmetry breaking. If we use  a WZ counterterm to restore the gauge symmetry, the Ward identity is modified with the addition of the $bF\tilde{F}$ graph, and the analysis can be found  in \cite{Armillis:2007tb}. Eq. \ref{reT} takes a more general form for off-shell gauge lines.
The general corrections to the anomaly pole are of the form
\ba
&&\Delta^{\lambda\mu\nu}=g_{V_{KK}}\theta_f^{KK} g_s^2 Tr[t^a t^b] k^{\lambda}\varepsilon[k_1,k_2,\mu,\nu]
\left[\frac{1}{2\pi^2 s}-2 {m_f^2} C_0(t, k_1^2,k_2^2,m_f)  \right]\nonumber\\
\label{TT}
\ea
where $C_0(t, k_1^2,k_2^2,m_f) $ is the scalar triangle diagram. Also in this case the $C_0$ terms are not counterterms. We don't need to add any mass-dependent term to the GS vertex to restore the Ward identity of the non-local theory. These longitudinal contributions, following the analysis of $g-2$ \cite{Knecht:2002hr} and the discussion of the previous sections,
are easily interpreted as the longitudinal parts of the non-anomalous components of the vertex, generated by the breaking of the chiral symmetry.

There are two ways to write the broken Ward identity in GS case. The first form is given by
\ba
k_{\lambda} \left(\Delta^{\lambda\mu\nu} + \Gamma_{GS}^{\lambda\mu\nu}\right) + {{T}}^{\mu\nu}=0,
\ea
in which the $T_{\mu\nu}$ term, which is of the form \ref{TT}, is derived simply by acting with the Ward
identity on the GS vertex (anomaly plus massless pole term) and bringing the result to the first member. The chiral symmetry breaking corrections to the pole term are then obtained from the decomposition

\beq
\Delta^{GS\,\lambda\mu\nu}=
\tilde{\Delta}_{long}^{\lambda\mu\nu} + \tilde{\Delta}_{trans}^{\lambda\mu\nu}
\eeq
where
\beq
k_{\lambda}\tilde{\Delta}_{long}^{\lambda\mu\nu}=\frac{1}{4 \pi^2} {m_f^2} C_0(t, k_1^2,k_2^2,m_f)
\epsilon[\mu,\nu,k_1,k_2]\equiv -T^{\mu\nu} \eeq
and
\beq
k_{\lambda}\tilde{\Delta}_{trans}^{\lambda\mu\nu}=0.
\eeq
A second form of the same equation is obtained by extracting a masless pole from $T_{\mu\nu}$

\ba
k_{\lambda} \left(\Delta^{\lambda\mu\nu} + \Gamma_{GS}^{\lambda\mu\nu} + \frac{k^{\lambda}}{k^2} {{T}}^{\mu\nu}\right) =0,
\ea
\begin{figure}[t]
\begin{center}
\includegraphics[scale=0.8]{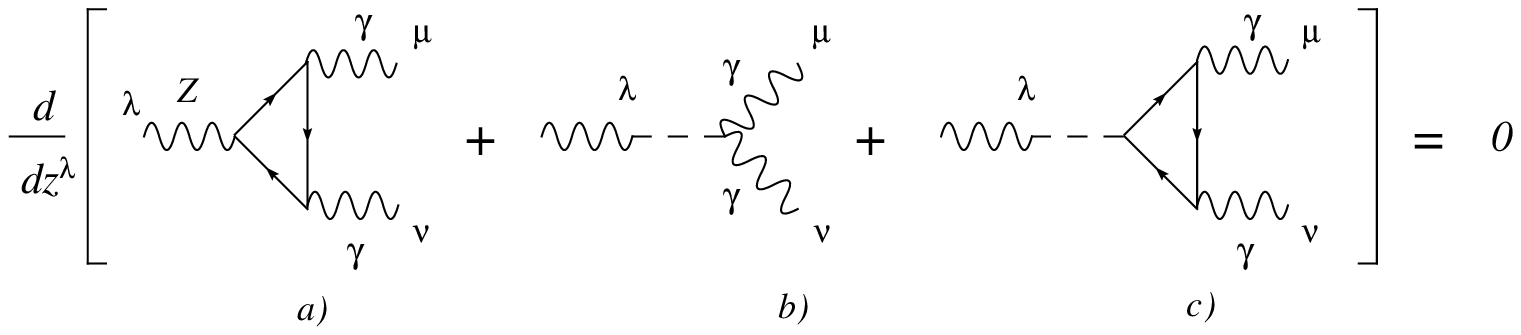}
\caption{\small Broken Ward identity in the presence of a GS interaction.}
\label{GS1}
\end{center}
\end{figure}
whose explicit form is shown in Fig.~\ref{GS1}. This result is in disagreement with \cite{Djouadi:2007eg}, where the authors write down an {\em exact} Ward identity for the GS vertex in the chirally broken phase, identity which clearly does not exist, since the pole counterterm and the effects due to chiral symmetry breaking should be kept separate.

There are other issues concerning the use of this vertex to describe the mixing of the Kaluza-Klein excitations of gauge bosons to an axion, claimed to be relevant in $t\bar{t}$ production,
which also point toward an inconsistency of these types of formulations in theories with extra dimensions and chiral delocalization on the brane. These quietly assume that the GS vertex is generated by sewing together local-interactions ($b F\tilde F$ vertex and $B\,b$ mixing), which are claimed to be obtained from extra dimensional theories \cite{Djouadi:2007eg}. The bilinear mixing is assumed to be physical (i.e. no gauge fixing condition can remove it). If these constructions were consistent, this would imply that the anomaly can be removed by adding a finite number of local interactions. Instead, the anomaly pole can be removed, but at the expense of building a non-local theory.
This result does not contradict the use of the WZ mechanism for the "cancellation" of the anomaly, since the WZ theory, being local, generates an effective theory which is unitary only below a certain  scale, while remaining gauge invariant at all scales. The theory, in fact, needs to be amended by higher dimensional operators for the restoration of unitarity, and therefore the description of axion-like particle, involving St\"uckelberg axions and PQ interactions, are not unitary at all scales. In fact, the number of local interactions needed to obtain an anomaly-free theory is infinite, which is the price to pay for not having a pole counterterm as in the GS case. The presence of BIM amplitudes for the WZ mechanism provides a clear example of processes with a non-unitary growth at high energy. We refer to \cite{Coriano:2008pg} for more details on some of these issues.

\section{Conclusions}
We have investigated the consistency of the subtraction of pole counterterm  in an anomalous theory,
re-analysing the problem of the generation of double poles in the perturbative expansion due to the extra subtractions, and in particular, in some s-channel exchanges. Having the anomaly diagram a natural separation into longitudinal and transverse contributions, the subtraction of the longitudinal component can be viewed simply as the remotion of one of its {\em independent} invariant amplitudes. If the structure of a given graph does not render the anomaly vertex harmless, the longitudinal subtraction is explicit, otherwise
the subtraction vanishes by itself, as does the longitudinal component of the anomaly in that case.

In principle, the perturbative expansion for the GS vertex can be formulated directly in terms of its transverse components. Away from the chiral limit there is still no anomaly pole, and the decoupling of the anomaly should hold at all orders and also in the broken chiral phase.

We have argued by explicit examples that the organization of the perturbative expansion in terms of anomaly diagrams and pole counterterms
(or DZ counterterms) is just  a matter of convenience, especially if a given computation has to be carried out to
higher orders. In this case, the double poles due to the counterterms have to be interpreted as genuine contributions which are embedded in 2-loop graphs.  We have pointed out that
the emergence of double poles is not an isolated case, but a standard
result, common to a specific way to address the tensor decomposition of a Feynman graph.

In this approach the computation of tensor integrals is performed using scalar integrals with higher power of the denominators and then re-formulated in terms of suitable sets of master integrals. Therefore, an ordinary
perturbative expansion at 2-loop level - after integration on one of the loop momenta -  gives -with no surprise-  a theory with propagators of second order and higher.

A final comment goes to the high energy behavior od the GS vertex.
The good high energy behavior of the vertex is related to its gauge invariance, with BIM amplitudes which are identically vanishing in the chiral limit. A similar feature is absent in the WZ case, which violates unitarity at high energy.
Finally, we have investigated the emergence of GS and of Wess-Zumino vertices in $g-2$ of the muon and in muonium, describing the differences between the SM case and its anomalous extensions, involving one axion-like particle and an extra anomalous $Z^\prime$, concentrating our attention, in particular, on the anomalous contributions, which can be studied accurately in the future in view of the planned experiments on $g-2$  at BNL.  In general, in the WZ case, the leading contributions to $g-2$ come from the exchange of an axi-Higgs and an anomalous $Z^\prime$, while in the GS case they involve directly the transverse components of the GS vertex. We hope to return with a quantitative analysis of some of these contributions in the near future.

\centerline{\bf Acknowledgements}
We thank Nikos Irges, George Sterman and Alan White for discussions and C. Anastasiou for discussions and help with the use of his program ${AIR}$ for the analysis of the tensor reductions. The work of C.C. was supported (in part)
by the European Union through the Marie Curie Research and Training Network ``Universenet'' (MRTN-CT-2006-035863).

\section{Appendix. Some features of the GS and WZ vertices} 
We comment on the relation between the WZ and GS formulation. 

There are several ways to parameterize an anomaly vertex (AVV), the most well known being the one due to Rosenberg \cite{Rosenberg:1962pp} which involves 6 invariant amplitudes $(A_1,A_2,...A_6)$, two of which are ill-defined and determined by the Ward identities of the theory in terms of the finite ones. The presence of an anomaly pole is not obvious in this formulation, although its structure was clearly established by Dolgov and Zakharov in their work \cite{Dolgov:1971ri} using dispersion relations. The basic interpretation of this result is that the anomaly is not just an ultraviolet but also an infrared effect.

The extraction of the anomaly pole from the rest of the amplitude is not so evident from the Rosenberg parameterization, but is quite obvious from the L/T formulation of this vertex, discussed in section 2.4.
As we have discussed in the same section, the GS mechanism corresponds to a redefinition of the anomaly vertex. In plain words it means that whenever we encounter an anomaly diagram we replace it with 
another vertex in which the DZ pole has been explicitly removed. In a lagrangean formulation this operation is equivalent to the addition of the counterterm shown in diagram c) of 
Fig. \ref{GS_AVV}. We stress once more that there is no direct coupling of the axion to the fermion, since in this approach the axion is not an asymptotic state. As we have extensively discussed in the previous 
sections this subtraction can be understood in a local version of the effective action by using Federbush's formulation of the GS mechanism with two pseudoscalars (Eq. \ref{fedeq}), one of them being actually a ghost, with negative kinetic energy. This formulation could, in principle, be extended so to describe a coupling of one of these two axions to the fermions. 

In the WZ case the local counterterm  $b F \wedge F$ introduces the axion as an asymptotic state of the corresponding S-matrix. Therefore, the axion takes an important role in the mechanisms of symmetry breaking, being this due either to a Higgs sector or to the St\"uckelberg mechanism, or to both. 
 For this reason, in the presence of electroweak symmetry breaking, there is a direct coupling of the axion to the fermions via the corresponding Yukawa couplings. One point which is worth to stress is that the WZ mechanism guarantees the gauge invariance of the 1-loop effective action but not of the trilinear gauge vertex. 
The differences between the two mechanism can be seen rather clearly, for instance, by comparing Fig. \ref{LBL_WZ} and \ref{LBL_GS}. Notice that fermion mass effects, in the WZ case, induced either by chiral symmetry breaking and/or electroweak symmetry breaking cause a direct interaction of the axion to the fermion. If they are both absent, then those diagrams in which the axion couples to the fermions are trivially vanishing. 

\subsection{Gauge choices} 
The cancellation of the gauge dependence in the perturbative expansion is rather trivial in the GS case while it is less straightforward in the WZ case. In the first case, the redefinition of the trilinear gauge vertex is sufficient to obtain from the beginning a gauge invariant result. For this purpose we may work directly in the 
$R_\xi$ gauge, denoting with $\xi_B$ the gauge-fixing parameter. The gauge dependent propagator for the gauge field is given by
\beqa
\frac{- i}{k^2}\left[  g^{\, \lambda\, \lambda^{\prime}} - \frac{k^\lambda \, k^{\lambda^\prime}}{k^2 }
(1 - \xi_B) \right] 
\eeqa 
and the longitudinal components disappear whenever they are attached to a GS vertex, due to the Ward identities satisfied on all the gauge lines. In the WZ case the cancellation of the gauge dependence is more subtle and has been discussed extensively in \cite{Coriano:2007fw}. We briefly summarize here the way this cancellation is achieved. The analysis is rather technical but can grasped more easily using a simple 
model. 

\begin{figure}[t]
{\centering \resizebox*{9cm}{!}{\rotatebox{0}
{\includegraphics{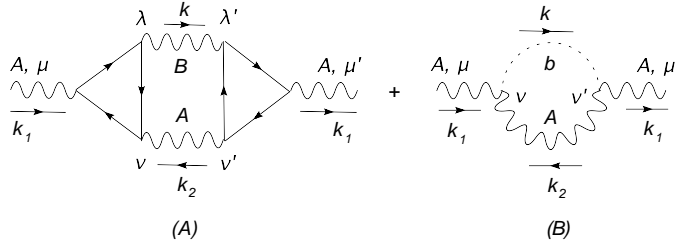}}}\par}
\caption{Cancellations of the gauge dependence in a self-energy graph}
\label{box}
\end{figure}

\begin{figure}[h]
{\centering \resizebox*{14cm}{!}{\rotatebox{0}
{\includegraphics{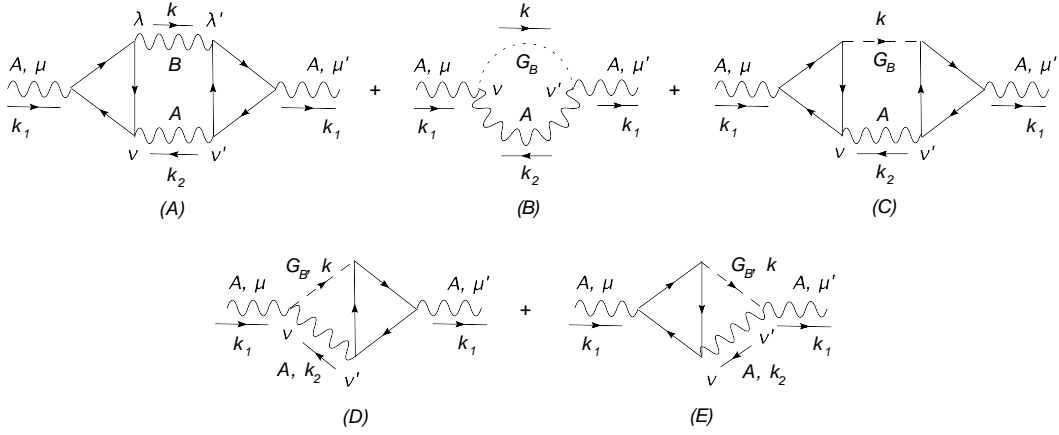}}}\par}
\caption{The complete set of diagrams in the broken phase.}
\label{boxmass}
\end{figure}

One of the working example is provided by the self-energy graph discussed in the main section 
(see Fig. \ref{self_terms}). There we have analyzed this diagram using the GS vertex and, as we have just mentioned, it is straightforward to verify gauge invariance if we use this vertex as a replacement for any anomalous triangle diagram, as shown in Fig. \ref{self}. In the WZ case, instead, the cancellation of the gauge dependence involves the exchange of the St\"uckelberg axion. This is shown in Fig. \ref{box} in the case of a simple $U(1)_A\times U(1)_B$ model with  $A$ vector-like and $B$ axial-vector like (anomalous). The counterterm diagram has interactions which are fixed by the requirement of gauge invariance of the anomalous action by the inclusion of suitable axion counterterms such as $b F\wedge F$. Before symmetry breaking the axion is a goldstone mode and diagram (B) is necessary in order to cancel the gauge dependence of diagram (A). After symmetry breaking, the $b$ field has to be decomposed into a goldstone mode $G_B$ of the gauge field $B$ and a physical axion $\chi$. This decomposition is discussed in 
 \cite{Coriano:2007fw}. The complete set of diagrams, in this case, is shown in Fig. \ref{boxmass}. We reproduce in this figure only the gauge dependent contributions, omitting the (gauge independent) contributions due to the exchange of the physical axion. In particular we assume here that $B$ 
becomes massive via a combination of the Higgs and the St\"uckelberg mechanisms. Notice that the set of graphs include also the coupling of the goldstone to the massive fermions. The derivation of the normalization for the counterterm and direct proofs of gauge invariance for this and other similar graphs can be found in 
the same work. 
 
\section{Appendix. Simplifications in some of the integrands on higher point functions}
\subsection{Computation of the diagrams in Fig.~\ref{null2}}
We show the vanishing of the counterterms in Fig.~\ref{null2}.
We have
\bea
C_1^\la &=& \int  \frac{d^4 k_1}{(2 \pi)^4} \bar v (p_2) \gamma_\nu \frac{1}{\ds p_1 - \ds k_1} \gamma_\mu u(p_1) \frac{1}{k_1^2} \frac{1}{k_2^2} \, C^{\mu\nu\la}_{AVV}(k_1, -k_2, k) \nn \\
&=& \int  \frac{d^4 k_1}{(2 \pi)^4} \bar v (p_2) \gamma_\nu \frac{1}{\ds p_1 - \ds k_1} \gamma_\mu u(p_1)
\frac{1}{k_2^2} \frac{k_1^\mu}{k_1^4}\frac{a_n}{3}\epsilon[\la,\nu,k,k_2],
\eea
with $k_2=k-k_1$ so that
\bea
C_1^\la&=& \bar v (p_2) \gamma_\nu u(p_1)
 \frac{a_n}{3}   \varepsilon[ \lambda, \nu, k, \rho]
 \int  \frac{d^4 k_1}{(2 \pi)^4} \frac{k_1^\rho}{(k-k_1)^2 k_1^4 },
\eea
where the expansion of the integrand function yields a result proportional to the independent momentum $k^\rho$ and finally $C_1^\la=0$.
The $C_2$ counterterm vanishes in an analogous way as $C_1^\la$, so we take into account the last diagram in Fig.~\ref{null2}
\bea
C_3^{\la} &=&
\int  \frac{d^4 k_1}{(2 \pi)^4}
\bar v (p_2) \gamma_\nu \frac{\ds p_1 - \ds k_1}{(p_1 -k_1)^2} \gamma_\mu u(p_1) \frac{1}{k_1^2}
\frac{1}{k_2^2} C^{\la\mu\nu}_{AVV}(-k,-k_1,-k_2),
\label{C3}
\eea
with $k_2=k-k_1$ and $C^{\la\mu\nu}_{AVV}(-k,-k_1,-k_2)=\frac{a_n}{3} \frac{k^\la}{k^2} \epsilon[\mu, \nu, k_1, k_2]$.
$C^\la_3$ is given by the sum of a first-rank and a second rank tensor integral which can be further reduced with the well-known tensor-reduction technique. The general expansion for the two integrands is
\bea
\int \frac{d^4 k_1}{(2 \pi)^4} \frac{k_1^\alpha}{(p_1 -k_1)^2 k_1^2 (k-k_1)^2} &=& C_1 p_1^\alpha + C_2 k^\alpha  \\
\int \frac{d^4 k_1}{(2 \pi)^4} \frac{k_1^\alpha k_1^\beta}{(p_1 -k_1)^2 k_1^2 (k-k_1)^2} &=&
C_{00} g^{\alpha \beta} + C_{12}(p_1^\alpha k^\beta + p_1^\beta k^\alpha)+ C_{11} p_1^\alpha p_1^\beta + C_{22} k^\alpha k^\beta;
\eea
first we notice all the terms proportional to $k^\alpha$ trivially vanish after the contraction with the antisymmetric Levi-Civita tensor in Eq.\ref{C3} and then we conclude $C^\la_3=0$ by using the following relations in eq.\ref{C3}
\bea
\bar v(p_2) \gamma_\nu \ds p_1 \gamma_\mu \epsilon[\mu, \nu, p_1, k] u(p_1)
&=& 2 i \, \bar v(p_2) (p_1^2 \ds k  - k\cdot p_1\ds p_1 ) \, \gamma^5 u(p_1) = 0,\\
\bar v(p_2) \gamma_\nu \gamma_\beta \gamma_\mu  \epsilon[\mu, \nu, \beta, k] u(p_1)
&=& 6 i\,  \bar v(p_2) \ds k \, \gamma^5 u(p_1)=0.
\label{AM}
\eea
for massless external fermions with momenta $p_1$ and $p_2$ and $k=p_1+p_2$.
\subsection{Simplifications of the integrand in section 2.7}
The third amplitude ${\mathcal S}_C$ does not contribute to ${\mathcal S}$,
in fact we have
\bea
{\mathcal S}_C &=&  \int  \frac{d^4 k_1}{(2 \pi)^4} \Bigg( \bar v (p_2) \gamma_\nu \frac{1}{\ds p_1 - \ds k_1} \ds k_1 u(p_1) \frac{1}{k_2^2}
 \frac{1}{k_1^2}  \frac{a_n}{3} \frac{1}{k_1^2}  \epsilon[\nu, \lambda, k_2, k]   \Bigg) \frac{1}{k^2}
 \mathcal{ BT}^{ \lambda}_{AAA}    \nonumber\\
 &=&-   \bar v (p_2) \gamma_\nu u(p_1)
 \frac{a_n}{3}  \int  \frac{d^4 k_1}{(2 \pi)^4}   \Bigg(  \frac{1}{(k - k_1)^2}
 \frac{1}{k_1^4}  \epsilon[\nu, \lambda, k - k_1, k]   \Bigg) \frac{1}{k^2}\mathcal{ BT}^{ \lambda}_{AAA} \nonumber\\
&=& \bar v (p_2) \gamma_\nu u(p_1)
 \frac{a_n}{3}   \epsilon[\nu, \lambda, \rho, \sigma] k^\sigma \int  \frac{d^4 k_1}{(2 \pi)^4}   \Bigg(  \frac{k_1^\rho}{(k-k_1)^2 k_1^4 }
  \Bigg) \frac{1}{k^2}\mathcal{ BT}^{ \lambda}_{AAA}   \nonumber\\
 &\propto& \bar v (p_2) \gamma_\nu u(p_1)
 \frac{a_n}{3}   \epsilon[\nu, \lambda, k, k] \frac{1}{k^2}  \mathcal{ BT}^{ \lambda}_{AAA}  = 0,
\eea
where by the tensor integral decomposition we obtain the following result
\ba
\epsilon[\nu, \lambda, \rho, \sigma] k^\sigma \int  \frac{d^4 k_1}{(2 \pi)^4}
\Bigg(  \frac{k_1^\rho}{(k-k_1)^2 k_1^4 }\Bigg)
=\epsilon[\nu, \lambda, \rho, \sigma] k^\sigma B k_{\rho}=0.
\ea
Here we omit the explicit form of the coefficient of the rank-1 tensor decomposition $B$,
since it is not essential for the calculation.
We can apply the same arguments to prove that ${\mathcal S}_C =0$.

\section{Tensor reductions with higher order poles}

Trading tensor decompositions in favour of propagators with higher powers is a standard result
\cite{Anastasiou:2004vj}. Our aim, here is to illustrate how the computations with explicit GS counterterms proceed. The perturbative expansion that follows, as we have mentioned above, is the one typical of a computation at higher order
(2-loop and higher). Here we outline the procedure for the counterterms of Fig.~\ref{twoblobs1} starting from rank-1.
\subsection{Rank-1}
We consider diagrams with products of generic propagators in the general form
\bea
\frac{1}{A^{\nu_1}_1 A^{\nu_2}_2 A^{\nu_3}_3 A^{\nu_4}_4}
\eea
for the specific choice $\nu_1=\nu_2=\nu_3=1$ and $\nu_4=2$.
\bea
A_1=k^2 \qquad A_2 = (k+p_1)^2 \qquad A_3 =(k+p_1+p_2)^2 \qquad A_4=(k-p_4)^2.
\eea
In the amplitude we can isolate the following different tensor structures
\bea
{\bf J}^{D} &=& \int Dx  \,\, {\mathcal I},  \\
{\bf J}^{D}_\alpha &=& \int Dx  \,\, \chi^\alpha {\mathcal I},  \\
{\bf J}^D_{\alpha \beta} &=&  \int Dx  \,\,  \left(   \chi^\alpha \chi^\beta - \frac{1}{2 P} g^{\alpha \beta}    \right) {\mathcal I},  \\
{\bf J}^D_{\alpha \beta \rho} &=&  \int Dx  \,\,  \left(   \chi^\alpha \chi^\beta \chi^\rho - \frac{1}{2 P}
\left\{  g^{\alpha \beta} \chi^\rho +  g^{\alpha \rho } \chi^\beta  +  g^{\beta \rho } \chi^\alpha \right\}  \right) {\mathcal I},
\eea
with
\bea
\int Dx &=& \prod_{i=1}^{4} \frac{(-1)^{\nu_i}}{\Gamma(\nu_i)} \int^\infty_0  d x_i x_i^{\nu_i - 1}    \\
\chi^\alpha &=& - \frac{d^\alpha}{P} = - \frac{\left[  p^\alpha_1 x_2 + (p_1 + p_2)^\alpha x_3 - p_4^\alpha x_4 \right]}{P},
\qquad P=x_1 +x_2 +x_3 +x_4,     \\
\mathcal I &=& \frac{1}{P^{D/2}} \exp\left( Q/P \right), \qquad  \mbox{in the massless case} \,\,\,\,  Q= x_1x_3 s + x_2 x_4 t,
\eea
where the usual Mandelstam variables are $s = (p_1 +p_2)^2=2 p_1 \cdot p_2$ and $t=(p_1+p_4)^2 = 2 p_1 \cdot p_4$.
For the the rank-1 integral $J^D_\a$ we have
\bea
{\bf J}^{D}_{\alpha} = \int Dx  \,\, \chi^\alpha {\mathcal I} =  - p_1^\alpha {\bf J}^{D}_2  - ( p_1 + p_2 )^\alpha {\bf J}^{D}_3 + p_4^\alpha {\bf J}^{D}_4,
\eea
where, because of the antisymmetry of the Levi-Civita symbol $\varepsilon[ \mu^\prime, \mu, - p_4, \alpha ]$ in Eq.(\ref{SDelta3}), we don't need to compute explicitly the integral $ p_4^\alpha {\bf J}^{D}_4$ as
\bea
\varepsilon[ \mu^\prime, \mu, - p_4, \alpha ] {\bf J}^{D}_{\alpha} =  \varepsilon[ \mu^\prime, \mu, - p_4, \alpha ]
\,\, \left\{ \,- p_1^\alpha {\bf J}^{D}_2 -  ( p_1 + p_2 )^\alpha {\bf J}^{D}_3  \, \right\}.
\eea
Therefore we have
\bea
{\bf J}^{D}_2 \equiv {\bf J}^{D}_2(s,t)&=&  \prod_{i=1}^{4} \frac{(-1)^{\nu_i}}{\Gamma(\nu_i)} \int^\infty_0 dx_1 dx_2  dx_3  dx_4 x_1^{\nu_1 - 1}   x_2^{\nu_2 - 1}
x_3^{\nu_3 - 1}  x_4^{\nu_4 - 1}    \frac{1}{P}  x_2   \frac{1}{P^{D/2}} \exp \left( Q/P \right)   \nonumber\\
&=& -  \nu_2 {\bf J}^{D+2}(\nu_1,\nu_2 + 1, \nu_3, \nu_4; s , t ) = -   {\bf J}^{D+2}(1,2, 1,2; s , t )
\label{JD2}
\eea
\bea
{\bf J}^{D}_3 \equiv {\bf J}^{D}_3(s,t)  &=&  \prod_{i=1}^{4} \frac{(-1)^{\nu_i}}{\Gamma(\nu_i)} \int^\infty_0 dx_1 dx_2  dx_3  dx_4 x_1^{\nu_1 - 1}   x_2^{\nu_2 - 1}
x_3^{\nu_3 - 1}  x_4^{\nu_4 - 1}    \frac{1}{P}  x_3   \frac{1}{P^{D/2}} \exp \left( Q/P \right)   \nonumber\\
&=&  - \nu_3 {\bf J}^{D+2}(\nu_1,\nu_2 , \nu_3 + 1, \nu_4; s , t )    =  -  {\bf J}^{D+2}(1,1,2,2;s,t),
\label{JD3}
\eea

where we have used the following identity
\bea
\frac{   (-1)^{\nu_i} x^{\nu_i -1}  }{\Gamma(\nu_i)} x_i = - \nu_i  \frac{ (-1)^{\nu_i+1} x^{\nu_i}  }{\Gamma(\nu_i + 1 )}.
\label{index}
\eea
We start now the direct evaluation of the two scalar integrals involved in the reduction, that is ${\bf J}^{D+2}(1,2, 1,2; s , t )$ and ${\bf J}^{D+2}(1,1,2,2;s,t)$. \\
For both the integrals  we have $D+2=6-2\eps$ and  $N=\sum_i \nu_i =6$, while the specific choices of the indices are  $\nu_1=\nu_3= 1, \nu_2=\nu_4=2$ for ${\bf J}^{D+2}(1,2, 1,2; s , t )$ and $\nu_1=\nu_2= 1, \nu_3=\nu_4=2$ for ${\bf J}^{D+2}(1,2, 1,2; s , t )$; their complete expressions in terms of few master integrals after the dimensional regularization and the analytic continuation in the physical region $s>0$ and $t<0$ are
\bea
{\bf J}^{D+2}(1,2, 1,2; s , t)=
&-& \frac{4 Bub^{D+2}(s) (D-6) (D-3) (D-1)}{(D-4) s^2 t^2}+ \frac{8 Bub^{D+2}(t) (D-3) (D-1)}{(D-4) s t^3} \nn \\
&+& \frac{Box^{D+2}(s,t) (D-3) ((D-6) s-2 t)}{s t^2}, \\
{\bf J}^{D+2}(1,1,2,2;s,t) =
&-& \frac{4 Bub^{D+2}(s) (D-1) (D-3)}{s^3 t}- \frac{4 Bub^{D+2}(t) (D-1) (D-3)}{s t^3} \nn \\
&+& \frac {Box^{D+2}(s,t) (D-4) (D-3)}{s t}
\eea
The explicit expressions of the scalar box and self-energy in severalm dimensions are given below.

\subsection{Rank-2}
In this section we deal with the two-rank tensor integral of the form
\bea
{\bf J}^D_{\alpha \beta} &=&  \int Dx  \,\,  \left(   \chi^\alpha \chi^\beta - \frac{1}{2 P} g^{\alpha \beta}    \right) {\mathcal I}   \nonumber\\
&=&  p_1^\alpha p_1^\beta  {\bf J}^D_{22}
+ (2 p_1^\alpha p_1^\beta +  p_1^\beta p_2^\alpha +  p_2^\beta p_1^\alpha )  {\bf J}^{D}_{23}
- (p_4^\alpha p_1^\beta + p_1^\alpha p_4^\beta)   {\bf J}^{D}_{24}   \nonumber\\
& +&  ( p_1^\alpha p_1^\beta +  p_1^\alpha p_2^\beta +  p_1^\beta p_2^\alpha +  p_2^\beta p_2^\alpha  )  {\bf J}^{D}_{33}
-   ( p_1^\alpha p_4^\beta +  p_2^\alpha p_4^\beta +  p_1^\beta p_4^\alpha +  p_2^\beta p_4^\alpha  )  {\bf J}^{D}_{34}   \nonumber\\
&+& p_4^\alpha p_4^\beta   {\bf J}^{D}_{44}   - \frac{g^{\alpha \beta}}{2} {\bf J}^{D+2},
\eea
where as in the previous case we don't need to compute explicitly
the contributions proportional to the momentum $p_4^\alpha$, because of  the antisymmetry of the Levi-Civita tensor $\varepsilon[ \mu^\prime, \mu, - p_4, \alpha ]$  so that we're left with the following contributions in Eq.(\ref{SDelta3})
\bea
&&\varepsilon[ \mu^\prime, \mu, - p_4, \alpha ] {\bf J}^{D}_{\alpha \beta}  \nonumber\\
&=&\varepsilon[ \mu^\prime, \mu, - p_4, \alpha ] \,\, \left\{ \,   p_1^\alpha p_1^\beta  {\bf J}^D_{22}
   + (2 p_1^\alpha p_1^\beta +  p_1^\beta p_2^\alpha +  p_2^\beta p_1^\alpha )  {\bf J}^{D}_{23}    -  p_1^\alpha p_4^\beta {\bf J}^{D}_{24}   \right.  \nonumber\\
 &+&  \left.  ( p_1^\alpha p_1^\beta +  p_1^\alpha p_2^\beta +  p_1^\beta p_2^\alpha +  p_2^\beta p_2^\alpha  )  {\bf J}^{D}_{33}
  -   ( p_1^\alpha p_4^\beta +  p_2^\alpha p_4^\beta ) {\bf J}^{D}_{34}  - \frac{g^{\alpha \beta}}{2} {\bf J}^{D+2}  \, \right\}.
\eea
For the integrals
\bea
{\bf J}^D_{22} &\equiv& {\bf J}^D_{22}(s,t)
= \,\nu_2 \,(\nu_2 + 1) \, {\bf J}^{D+4}\, (\nu_1, \nu_2 +2 ,\nu_3, \nu_4; s , t )
=  2  \, {\bf J}^{D+4}(1,3,1,2;s,t),\\
{\bf J}^D_{23} &\equiv& {\bf J}^D_{23}(s,t)
=\,\nu_2 \,\nu_3 \, {\bf J}^{D+4}\,(\nu_1, \nu_2 +1 ,\nu_3 + 1, \nu_4; s , t )
= {\bf J}^{D+4}(1,2,2,2;s,t),   \\
{\bf J}^D_{24} &\equiv& {\bf J}^D_{24}(s,t)
=\, \nu_2 \, \nu_4 \,{\bf J}^{D+4}(\nu_1, \nu_2 +1 ,\nu_3 , \nu_4 + 1; s , t )
= 2 \, {\bf J}^{D+4}(1,2,1,3;s,t),\\
{\bf J}^D_{33} &\equiv& {\bf J}^D_{33}(s,t)
= \, \nu_3 \, (\nu_3 + 1) \, {\bf J}^{D+4}(\nu_1, \nu_2 ,\nu_3 + 2, \nu_4; s , t )
= \,2  \, {\bf J}^{D+4}(1,1,3,2;s,t), \\
{\bf J}^D_{34} &\equiv& {\bf J}^D_{34}(s,t)
= \, \nu_3 \, \nu_4 \,{\bf J}^{D+4}(\nu_1, \nu_2 ,\nu_3 + 1, \nu_4+1; s , t )
= \,2  \,{\bf J}^{D+4}(1,1,2,3;s,t), \\
{\bf J}^{D+2} &\equiv& {\bf J}^{D+2}(s,t)
= {\bf J}^{D+2}(1,1,1,2;s,t),
\eea
where we have used the identity in Eq.(\ref{index}) and its extension of the form
\bea
\frac{(-1)^{\nu_i} x^{\nu_i -1} }{\Gamma(\nu_i)} x^2_i =  \nu_i (\nu_i +1)  \frac{ (-1)^{\nu_i +  2 } x^{\nu_i + 1}  }{\Gamma(\nu_i + 2 )}.
\eea
The reduction in terms of bubble and box master integrals is
\bea
{\bf J}^{D+4}(1,3,1,2;s,t)&=&
\frac{2 Bub^{D+4}(s) (D-3) (D-1) (D+1)}  {(D-4) (D-2) s^3 t^3} \left[s (D+4)^2 \right.\nn \\
 &+& \left. (-18 s-2 t) (D+4)+80 s+12 t\right]  - \frac{8 Bub^{D+4}(t) (D-3) (D-1) (D+1)}{(D-2) s t^4} \nn \\
&-&  \frac{Box^{D+4}(s,t) (D-3) (D-1) ((D-6) s-4 t)}{2 s t^3}  \\
{\bf J}^{D+4}(1,2,2,2;s,t)&=&
\frac{4 Bub^{D+4}(t) (D-3) (D-1) (D+1)} {(D-4) (D-2) s^2 t^4}  \left[s (D+4)^2 \right. \nn \\
&+&  \left. (-16 s-2 t) (D+4)+60 s+16 t\right]
+ \frac{4   Bub^{D+4}(s) (D-4) (D-3) (D-1) (D+1)}{(D-2) s^3 t^2} \nn \\
&-& \frac{Box^{D+4}(s,t) (D-3) (D-1) ((D-4) s-2 t)}{s^2 t^2} \\
{\bf J}^{D+4}(1,2,1,3;s,t)&=&
\frac{2 Bub^{D+4}(s) (D-3) (D-1) (D+1)}{(D-4) (D-2) s^3 t^3} \left[s (D+4)^2\right. \nn \\
&+& \left. (-18 s-2 t) (D+4)+80 s+12 t\right]
- \frac{8 Bub^{D+4}(t) (D-3) (D-1) (D+1)}{(D-2) s t^4} \nn \\
&-& \frac{Box^{D+4}(s,t) (D-3) (D-1) [(D-6) s-4 t]}{2 s t^3} \\
{\bf J}^{D+4}(1,1,3,2;s,t)&=&  \frac{2 Bub^{D+4}(t) (D-3) (D-1)(D+1) \left[t (D+4)^2+2 s-8 t\right] }{(D-4) s^2 t^4} \nn \\ &+& \frac{2 Bub^{D+4}(s) (D-3) (D+1) (D-1)}{s^4 t} \nn \\
&-& \frac{Box^{D+4}(s,t) (D-3) (D-2)(D-1)}{2 s^2 t} \\
{\bf J}^{D+4}(1,1,2,3;s,t)&=& \frac{2 Bub^{D+4}(s) (D-3) (D-1)(D+1) [(D+4) s-8 s+2 t] }{(D-4) s^4 t^2} \nn \\
&+& \frac{2 Bub^{D+4}(t) (D-3) (D+1) (D-1)}{s t^4} \nn \\
&-& \frac{Box^{D+4}(s,t) (D-3) (D-2) (D-1)}{2 s   t^2} \\
{\bf J}^{D+2}(1,1,1,2;s,t)&=& \frac{4 Bub^{D+2}(s) (D-3) (D-1)} {(D-4) s^2 t} + \frac{Box^{D+2}(s,t) (3-D)}{t},
\eea
with $D=4-2 \eps$ and $\epsilon>0$.
A similar approach can be followed also for the lengthier integrals of rank-3. Details are left to an appendix.
It is evident, from this analysis, that a theory with GS counterterms has all the characteristics of a typical
higher order perturbative expansion. This should not be so surprising since the pole counterterm is a 1-loop effect and our expansion therefore is essentially composed of 2-loop graphs.

\subsection{Rank-3}
The complete third rank tensor integral has the form
\bea
{\bf J}^D_{\alpha \beta \rho} = \int Dx  \,\,  \left(   \chi^\alpha \chi^\beta \chi^\rho - \frac{1}{2 P}
\left\{  g^{\alpha \beta} \chi^\rho +  g^{\alpha \rho } \chi^\beta  +  g^{\beta \rho } \chi^\alpha \right\}  \right) {\mathcal I},
\eea
where
\bea
&& \int Dx  \,\,   \left(   \chi^\alpha \chi^\beta \chi^\rho   \right) {\mathcal I}      \nonumber\\
&=& -  p_1^\alpha p_1^\beta p_1^\rho {\bf J}^D_{222} - \left( 3 p_1^\alpha p_1^\beta p_1^\rho
+  p_1^\beta p_2^\alpha p_1^\rho + p_1^\alpha p_2^\beta  p_1^\rho
+ p_1^\alpha p_1^\beta p_2^\rho \right) {\bf J}^D_{322}
+ \left(  p_1^\alpha p_1^\rho p_4^\beta +  p_1^\alpha p_1^\beta p_4^\rho \right) {\bf J}^D_{422}  \nonumber\\
&-& \left(  3 p_1^\alpha p_1^\beta p_1^\rho + 2 p_1^\beta p_2^\alpha p_1^\rho + 2 p_1^\alpha p_2^\beta  p_1^\rho
+  p_2^\alpha p_2^\beta p_1^\rho +2 p_1^\alpha p_1^\beta p_2^\rho + p_1^\beta p_2^\alpha p_2^\rho
+ p_1^\alpha p_2^\beta p_2^\rho  \right)  {\bf J}^D_{332}    \nonumber\\
&+& \left(  2 p_1^\alpha p_1^\rho p_4^\beta + p_1^\rho p_2^\alpha p_4^\beta
+ p_1^\alpha p_2^\rho p_4^\beta + 2 p_1^\alpha p_1^\beta p_4^\rho
+ p_1^\beta p_2^\alpha p_4^\rho + p_1^\alpha p_2^\beta p_4^\rho \right){\bf J}^D_{342}
- p_1^\alpha p_4^\rho p_4^\beta {\bf J}^D_{442}  \nonumber\\
&-& \left(  p_1^\alpha p_1^\beta p_1^\rho + p_1^\beta p_2^\alpha p_1^\rho
+ p_1^\alpha p_2^\beta  p_1^\rho + p_2^\alpha p_2^\beta p_1^\rho + p_1^\alpha p_1^\beta p_2^\rho + p_1^\beta p_2^\alpha p_2^\rho
+ p_1^\alpha p_2^\beta p_2^\rho + p_2^\alpha p_2^\beta p_2^\rho \right){\bf J}^D_{333}    \nonumber\\
&+& \left( p_1^\alpha p_1^\rho p_4^\beta + p_1^\rho p_2^\alpha p_4^\beta
+ p_1^\alpha p_2^\rho p_4^\beta + p_2^\alpha p_2^\rho p_4^\beta
+ p_1^\alpha p_1^\beta p_4^\rho + p_1^\beta p_2^\alpha p_4^\rho
+ p_1^\alpha p_2^\beta p_4^\rho + p^\alpha_2 p_2^\beta p_4^\rho   \right){\bf J}^D_{334}    \nonumber\\
&-& \left(  p_1^\alpha p_4^\rho p_4^\beta + p_2^\alpha p_4^\rho p_4^\beta  \right){\bf J}^D_{344},
\label{threechi}
\eea
and
\bea
&&\int Dx  \,\,  \left(  - \frac{1}{2 P}     \right)
\left\{  g^{\alpha \beta} \chi^\rho +  g^{\alpha \rho } \chi^\beta  +  g^{\beta \rho } \chi^\alpha \right\}   {\mathcal I}   \nn\\
&=&  \frac{g^{\alpha \beta}}{2} \left[   - p_1^\rho {\bf J}^D_2 - ( p_1 + p_2 )^\rho {\bf J}^D_3
+ p_4^\rho {\bf J}^D_4  \right]   +  \frac{g^{\alpha \rho}}{2} \left[   - p_1^\beta {\bf J}^D_3 - (p_1 + p_2 )^\beta {\bf J}^D_3
+ p_4^\beta {\bf J}^D_4  \right]   \nn \\
&+& \frac{g^{\beta \rho}}{2} \left[   - p_1^\alpha {\bf J}^D_2 - (p_1 + p_2 )^\alpha {\bf J}^D_3  \right],
\label{onechi}
\eea
recalling that  in the last term we have omitted the contribution coming from $ p_4^\alpha {\bf J}^{D+2}(1,1,1,3;s,t)$ thanks to the antisymmetry of the tensor $\varepsilon[ \mu^\prime, \mu, - p_4, \alpha ]$.

The $J^D$ integrals with three indices in Eq.(\ref{threechi}) are defined as
\bea
{\bf J}^D_{222} \equiv  {\bf J}^D_{222}(s,t)
&=& - \nu_2 (\nu_2 +1)(\nu_2 +2) {\bf J}^{D+6}(\nu_1,\nu_2 +3, \nu_3, \nu_4;s,t) \nn \\
&=& - 6 \, {\bf J}^{D+6}(1,4, 1, 2;s,t)   \\
{\bf J}^D_{322}     \equiv  {\bf J}^D_{322}
&=& -  \nu_3 \nu_2 (\nu_2 +1) {\bf J}^{D+6}(\nu_1,\nu_2 +2, \nu_3+1, \nu_4;s,t)\nn \\
&=& - 2 \,  {\bf J}^{D+6}(1,3, 2, 2;s,t),    \\
{\bf J}^D_{422}   \equiv    {\bf J}^D_{422} (s,t)
&=& -  \nu_4 \nu_2 (\nu_2 + 1)     {\bf J}^{D+6}(\nu_1,\nu_2 +2, \nu_3, \nu_4+1;s,t) \nn \\
&=& -  4 \,  {\bf J}^{D+6}(1,3, 1, 3;s,t),   \\
{\bf J}^D_{332}   \equiv   {\bf J}^D_{332}(s,t)
&=& -  \nu_3 (\nu_3 + 1) \nu_2    {\bf J}^{D+6}(\nu_1,\nu_2 +1, \nu_3+2, \nu_4;s,t) \nn \\
&=& - 2  \,   {\bf J}^{D+6}(1,2, 3, 2;s,t), \\
{\bf J}^D_{342}  \equiv  {\bf J}^D_{342}(s,t)
&=& -  \nu_3 \nu_4 \nu_2   {\bf J}^{D+6}(\nu_1,\nu_2 +1, \nu_3+1, \nu_4+1;s,t) \nn \\
&=& - 2 \,  {\bf J}^{D+6}(1,2, 2, 3;s,t),  \\
{\bf J}^D_{442}   \equiv {\bf J}^D_{442}(s,t)
&=& -  \nu_4 (\nu_4 +1) \nu_2    {\bf J}^{D+6}(\nu_1,\nu_2 +1, \nu_3, \nu_4+2;s,t) \nn \\
&=& - 6 \,{\bf J}^{D+6}(1,2 , 1, 4;s,t),  \\
{\bf J}^D_{333}   \equiv  {\bf J}^D_{333}(s,t)
&=& - \nu_3 (\nu_3 + 1)(\nu_3 +2)   {\bf J}^{D+6}(\nu_1,\nu_2 , \nu_3, \nu_4+3;s,t) \nn \\
&=& - 6 \,    {\bf J}^{D+6}(1,1 ,1, 5;s,t), \\
{\bf J}^D_{334}   \equiv  {\bf J}^D_{334}(s,t)
&=& -   \nu_3 (\nu_3 + 1)   \nu_4   {\bf J}^{D+6}(\nu_1,\nu_2 , \nu_3+2, \nu_4+1;s,t) \nn \\
&=& -  4 \,   {\bf J}^{D+6}(1,1 , 3, 3;s,t), \\
{\bf J}^D_{344}   \equiv   {\bf J}^D_{344}(s,t)
&=&  - \nu_3 \nu_4 (\nu_4 +1)     {\bf J}^{D+6}(\nu_1,\nu_2 , \nu_3+1, \nu_4+2;s,t) \nn \\
&=& - 6 \,     {\bf J}^{D+6}(1,1 , 2, 4;s,t),
\eea
where we have used the following property
\bea
\frac{   (-1)^{\nu_i} x^{\nu_i -1}  }{\Gamma(\nu_i)} x^3_i =  -\nu_i (\nu_i +1)(\nu_i +2)
  \frac{ (-1)^{\nu_i +  3 } x^{\nu_i + 2}  }{\Gamma(\nu_i + 3 )}.
\eea
The integrals appearing in Eq.(\ref{onechi}) have been partially computed in the section relative to the one-rank tensor integral decomposition. From Eqs.(\ref{JD2}) and (\ref{JD3}) we have indeed
\bea
{\bf J}^D_2 &\equiv & - {\bf J}^{D+2}(1,2,1,2;s,t)         \\
{\bf J}^D_3 &\equiv & - {\bf J}^{D+2}(1,1,2,2;s,t)
\eea
and the remaining one is
\bea
{\bf J}^D_4  &=& \prod_{i=1}^{4} \frac{(-1)^{\nu_i}}{\Gamma(\nu_i)} \int^\infty_0 dx_1 dx_2  dx_3  dx_4 x_1^{\nu_1 - 1}   x_2^{\nu_2 - 1}
x_3^{\nu_3 - 1}  x_4^{\nu_4 - 1}    \frac{1}{P}  x_4    \frac{1}{P^{D/2}} \exp \left( Q/P \right)     \nonumber\\
&=&  -  {\bf J}^{D+2}(1,1,1,3;s,t)
\eea
The last step to be accomplished refers to the reduction in terms of master integrals for all the $J^D$ involved in the computation
\bea
{\bf J}^{D+6}(1,4, 1, 2;s,t) &=&
-\frac{2 Bub^{D+6}(s) (D-1) (D+1) (D+3)} {3 (D-4) D s^4 t^4}\left[s^2 (D+6)^3+\left(-30 s^2-4 t s\right) (D+6)^2 \right. \nn \\
&& \left. + \left(296 s^2+64 t s-4 t^2\right)  (D+6)-960 s^2+24 t^2-240 s t\right] \nn \\
   &+& \frac{4 Bub^{D+6}(t) (D-2) (D-1) (D+1) (D+3)}{D s   t^5} \nn \\
   &+& \frac{Box^{D+6}(s,t) (D-2) (D-1) (D+1) ((D+6) s-12 s-6 t)}{6 s t^4}, \\
{\bf J}^{D+6}(1,3,2,2;s,t) &=&
- \frac{2 Bub^{D+6}(s) (D-1) (D+1) (D+3) \left[s (D+6)^2+(-18 s-2 t) (D+6)+80 s+12 t\right]}{D s^4 t^3} \nn \\
&-& \frac{2 Bub^{D+6}(t) (D-1) (D+1) (D+3) \left[s (D+6)^2+(-18 s-4 t) (D+6)+72 s+32 t\right]}{D s^2 t^5} \nn \\
&+& \frac{Box^{D+6}(s,t) (D-2) (D-1) (D+1) [(D+6) s-10 s-4 t]}{2 s^2 t^3}, \\
{\bf J}^{D+6}(1,3,1,3;s,t) &=&
- \frac{Bub^{D+6}(s) (D-4) (D-1) (D+1) (D+3)} {(D-2) D s^3 t^4} \left[s (D+6)^2 \right. \nn \\
&& \left. +(-20 s-6 t) (D+6)+96 s+40 t\right]  \nn \\
&+& \frac{8 Bub^{D+6}(t) (D-1) (D+1) (D+3) \left[s (D+6)^2+(-18 s-t) (D+6)+78 s+8 t\right]} {(D-2) D s^2 t^5} \nn \\
&+& \frac{Box^{D+6}(s,t) (D-1) (D+1) \left[ \left((D+6)^2-22 (D+6)+120\right) s^2-8 (D-4) s t + 8 t^2\right]}{4 s^2 t^4}, \nn \\ \\
{\bf J}^{D+6}(1,2,3,2;s,t) &=&
-\frac{2 Bub^{D+6}(s) (D-1) (D+1) (D+3) (D-2)^2}{D s^4 t^2}   \nn \\
&-& \frac{2 Bub^{D+6}(t) (D-1) (D+1) (D+3) } {(D-4) D s^3 t^5} \left[s t (D+6)^3 + \left(2 s^2-26 t s-2 t^2\right) (D+6)^2 \right. \nn \\
&&+ \left. \left(-36 s^2 + 220   t s+36 t^2\right) (D+6)+144 s^2-160 t^2-600 s t\right] \nn \\
&+& \frac{Box^{D+6}(s,t) (D-1) (D+1) ((D+6) s-8 s-2 t) (D-2)}{2 s^3 t^2}, \\
{\bf J}^{D+6}(1,2,2,3;s,t) &=& {\bf J}^{D+6}(1,3,2,2;s,t) , \\
{\bf J}^{D+6}(1,2,1,4;s,t) &=& {\bf J}^{D+6}(1,4,1,2;s,t) , \\
{\bf J}^{D+6}(1,1,1,5;s,t) &=&
- \frac{Bub^{D+6}(s) (D-1) (D+1) (D+3) }{6 (D-6) (D-4) s^5   t^4}
\left[(D+6)^3 s^3+960 s^3+240 t s^2-96 t^2 s+48 t^3 \right. \nn \\
&& +  \left. (D+6)^2 \left(2 s^2 t-30 s^3\right)+(D+6) \left(296 s^3-44 t s^2+8 t^2 s\right)\right] \nn \\
&+& \frac{Box^{D+6}(s,t) (D-2) (D-1) D (D+1)}{24 t^4} , \\
{\bf J}^{D+6}(1,1,3,3;s,t) &=&
-\frac{Bub^{D+6}(s) (D-1) (D+3) [(D+6) s-8 s+2 t] (D+1)}{s^5 t^2} \nn \\
&-& \frac{Bub^{D+6}(t) (D-1) (D+3) [2 s+(D+6) t-8 t](D+1)}{s^2 t^5} \nn \\
&+& \frac{Box^{D+6}(s,t) (D-2) (D-1) D (D+1)}{4 s^2 t^2}, \\
{\bf J}^{D+6}(1,1,2,4;s,t) &=&
-\frac{2  Bub^{D+6}(s)(D-1) (D+3)(D+1)}{3 (D-4) s^5  t^3}
\left[(D+6)^2 s^2-18 (D+6) s^2+80 s^2 \right. \nn \\
&& \left. +2 (D+6)  s t -20 t s+8 t^2\right]
- \frac{2 Bub^{D+6}(t) (D-2) (D-1) (D+3) (D+1)} {3 s t^5} \nn \\
&+& \frac{ Box^{D+6}(s,t) (D-2) (D-1) D (D+1)}{6 s t^3}, \\
{\bf J}^{D+2}(1,1,1,3;s,t) &=&  \frac{Box^{D+2}(s,t) (D-4) (D-3)}{2 t^2}
- \frac{2 Bub^{D+2}(s) (D-3) (D-1) [(D+2) s-8 s+2 t]}{(D-6) s^3 t^2} \nn \\.
\eea
\subsection{Bubble and box master integrals in generic dimensions}
The analytic continuation in the physical region  $s>0$ and  $t<0$ of the $D=4-2\epsilon$ one-loop bubble yields \cite{Anastasiou:2000kg}
\bea
Bub^{D}(s) &=& \frac{i \pi^{D/2}}{(2 \pi)^D}\, \mu^{2 \eps} \, \biggl(\frac{e^{\g}}{4 \pi}\biggr)^\eps\frac{c_\Gamma}{ \eps (1-2 \eps)} (s)^{- \eps} (-1)^{\eps},
\label{BubDs}\\
Bub^{D}(t) &=& \frac{i \pi^{D/2}}{(2 \pi)^D}\, \mu^{2 \eps} \, \biggl(\frac{e^{\g}}{4 \pi}\biggr)^\eps \frac{c_\Gamma}{\eps (1-2 \eps)}(-t)^{- \eps}.
\label{BubDt}
\eea
The bubble master integral in $D+2$, $D+4$ and $D+6$ dimensions can be obtained in a straightforward way starting from $Bub^D(s)$ and performing first a $\eps\rightarrow(4-D)/2$ shift and then another shift to bring $D$ to the desired number of dimensions; we have for instance
\bea
Bub^{D+2}(s) &=&  \frac{i \pi^{(D+2)/2}}{(2 \pi)^{D+2}}\, \mu^{2 \eps} \, \biggl(\frac{e^{\g}}{4 \pi}\biggr)^\eps \frac{c_\Gamma \, s }{ 2 \eps (1-2 \eps)(3-2 \eps)} (s)^{- \eps} (-1)^{\eps}
\label{BubDp2s}
\eea
for the bubble in $6-2\eps$ dimensions.\\
The basis of master integrals we have used includes the one-loop bubble and the one-loop box in $6-2\eps$ dimensions \cite{Bern:2001df}
\bea
Box^{D+2}(s,t) &=& \frac{i \pi^{(D+2)/2}}{(2 \pi)^{D+2}}\, \mu^{2 \eps} \, \biggl(\frac{e^{\g}}{4 \pi}\biggr)^\eps
\frac{c_ \Gamma \, |s|^{-\epsilon }} {u (1-2 \epsilon )}
\left\{\frac{X^2}{2} \right. \nn \\
&& \left. +\epsilon  \left(-\frac{X^3}{3}+\frac{Y X^2}{2}+Li_2(-x) X - \frac{\pi ^2 X}{2}- Li_3(-x)+ \zeta   (3)\right) \right.\nn \\
&&-  \epsilon ^2 \left(\frac{Y X^3}{3} + \frac{Y^2 X^2}{4}-\frac{Y^3
   X}{6}+\frac{1}{3} \pi ^2 Y X + Li_3(-y) X - \frac{1}{8} \left(X^2+\pi ^2\right)^2 \right. \nn \\
&&  \left.  + \frac{1}{24} \left(Y^2 + \pi^2\right)^2 + \frac{1}{2} \left(X^2+\pi ^2\right) Li_2(-x)+ Li_4\left(-\frac{x}{y}\right)- Li_4(-y)+ \frac{7 \pi ^4}{360}\right) \nn \\
&& + i \pi  \left[\left(\frac{X^3}{6}-\frac{Y X^2}{2}- Li_3(-x)- Li_3(-y)+\zeta (3)\right) \epsilon^2 \right.\nn \\
   && + \left. \left. \left(-\frac{X^2}{2}+Y X+Li_2(-x)-\frac{\pi ^2}{6}\right) \epsilon +X\right]\right\} + \mathcal{O}(\eps^3),
\eea
where $c_\Gamma$, $x$, $y$, $X$ and $Y$ are defined as
\bea
c_\Gamma= \frac{\Gamma(1+\eps) \Gamma^2(1-\eps)}{\Gamma(1-2\eps)} \qquad \qquad
x=\frac{t}{s} \qquad  y=\frac{u}{s} \qquad  \qquad X=\ln\left(- \frac{t}{s}\right)
\qquad Y=\ln \left(- \frac{u}{s}\right).
\eea
Since  the scalar box integral in $D+2=6-2\eps$ is completely finite as $\eps \rightarrow 0$, it is convenient to perform a dimensional shift of the box integrals in higher dimensions like $Box^{D+4}(s,t)$ and  $Box^{D+6}(s,t)$ in terms of it with these formulas
\bea
Box^{D+4}(s,t) &=& -\frac{1}{u} \biggl[
\frac{st} {2(D-1)}Box^{D+2}(s,t) + \frac{2}{D-2}( Bub^{D+2}(s)+Bub^{D+2}(t))  \biggr]\\
Box^{D+6}(s,t) &=& -\frac{1}{u} \biggl[
\frac{st} {2(D+1)}Box^{D+4}(s,t) + \frac{2}{D}( Bub^{D+4}(s)+Bub^{D+4}(t))  \biggr].
\eea
\bibliographystyle{h-elsevier3}
\bibliography{Zero_Augfinalissimo}
\end{document}